\documentclass[12pt,twoside,a4paper,openright]{report}

\usepackage{amsmath}
\usepackage{graphicx,lineno}
\usepackage{graphics}   
\usepackage{dcolumn,enumerate}    
\usepackage{bm,bbm}         
\usepackage{mathrsfs}
\usepackage{mathtools}
\usepackage{color}
\usepackage[T1]{fontenc}
\usepackage{rotating}
\usepackage{txfonts}
\numberwithin{equation}{chapter}

\usepackage{geometry}
\usepackage{fancyhdr}
\usepackage[small,bf]{caption}
\usepackage{setspace}
\newcommand{\HRule}{\rule{\linewidth}{0.3mm}} 

\begin{document}

\begin{titlepage}
\begin{center}

{\LARGE \bfseries Nuclear Spins as Quantum Testbeds:}\\[0.3cm]
{\LARGE \bfseries Singlet States, Quantum Correlations, and}\\[0.3cm]
{\LARGE \bfseries Delayed-choice Experiments}\\[1.5cm]

A thesis\\
Submitted in partial fulfillment of the requirements\\
Of the degree of\\
Doctor of Philosophy\\[0.8cm]
By\\[1cm]
{\large Soumya Singha Roy}\\
20083009 \\[2.5cm]

\includegraphics[width=3cm]{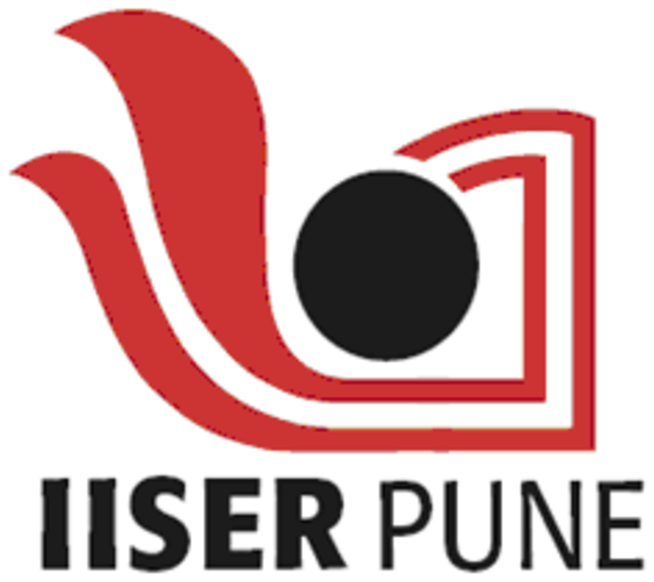}\\    

{INDIAN INSTITUTE OF SCIENCE EDUCATION AND RESEARCH PUNE}\\[1cm]

\vfill
{August, 2012}

\end{center}
\end{titlepage}
\thispagestyle{empty}
\cleardoublepage

\pagenumbering{roman}
\chapter*{Certificate}\addcontentsline{toc}{section}{\textbf{Certificate}}
Certified that the work incorporated in the thesis entitled 
``\textit{Nuclear Spins as Quantum Testbeds: Singlet States, Quantum Correlations, and Delayed-choice Experiments}'',
submitted by \textit{Soumya Singha Roy} was carried out by the candidate, under my supervision. The work presented here or any part of it has not been included in any other thesis submitted previously for the award of any degree or diploma from any other University or institution.\\[3cm]
\textit{Date} \hspace{10cm} \textbf{Dr. T. S. Mahesh}
\cleardoublepage

\chapter*{Declaration}\addcontentsline{toc}{section}{\textbf{Declaration}}
I declare that this written submission represents my ideas in my own words and where others' ideas have been included, I have adequately cited and referenced the original sources. I also declare that I have adhered to all principles of academic honesty and integrity and have not misrepresented or fabricated or falsified any idea/data/fact/source in my submission. I understand that violation of the above will be cause for disciplinary action by the Institute and can also evoke penal action from the sources which have thus not been properly cited or from whom proper permission has not been taken when needed.\\[3cm]
\textit{Date} \hspace{10.2cm} \textbf{Soumya Singha Roy}
\begin{flushright}
Roll No.- 20083009 
\end{flushright}
\cleardoublepage

\begin{flushright}
\textbf{\LARGE Acknowledgement}\\
\HRule \\[0.4cm]
\end{flushright}\addcontentsline{toc}{section}{\textbf{Acknowledgement}} 

This thesis would have not been possible without the help and support by many individuals whom I met during my PhD life at IISER Pune. I have been very privileged to have so many wonderful friends and collaborators. \

First of all, I am grateful to my advisor Dr. T. S. Mahesh for teaching me everything about NMR and Quantum Information Processing starting from the scratch. His cheerful guidance and deep understanding in NMR-QIP are the driving forces behind this thesis. Being his first PhD student, I consider myself to be very fortunate to get all of his support, care and affection.\

I would like to thank Dr. Vikram Athalye for his valuable set of lectures which ultimately led to a fruitful collaborative work. I thank Prof. G. S. Agarwal for some constructive discussions with him and visiting us in spite of his busy schedule. I am thankful to Prof. Apoorva Patel for useful discussions and collaboration. Very special thanks to Prof. Anil Kumar for his great support and insightful conversations on various scientific problems and my future career path. \

I thank all the members of NMR Research Center -past and present- with whom I have worked. It's been a great pleasure working with Abhishek (whom we fondly call `Shuklaji') for all his `complicated' queries and thoughtful discussions that I had with him. His famous `I am not saying this, but I am saying that' is really unforgettable. It was highly exciting to work with Hemant who became an integral part of the NMR center ever since he joined the lab. Discussions on `Quantum weirdness' were always intriguing with Manvendra. Working with Swathi was interesting because of her thorough theoretical understandings. Short stay of Philipp left many sweet memories of the discussions that I had with him. Discussions with Sheetal and Abhijeet were interesting and that actually made me learn many aspects of NMR and Quantum Computing. I also thank Pooja for her organized way of managing the spectrometers for years. Sachin was always there to look after the spectrometers and he made sure that it's up all the time.\

I thank my RAC members- Dr. A. Bhattacharyay, Dr. T. G. Ajitkumar, and Dr. K. Gopalakrishnan for all their support. I also thank Dr. R. G. Bhat, Dr. V. G. Anand, Dr. H. N. Gopi, for all those help and affection. I am thankful to Prof. K. N. Ganesh for providing all the necessary experimental facilities in the lab. I thank Dr. V. S. Rao for all his help whenever I needed, be it academic or non-academic. I would like to thank CSIR and IISER for the graduate scholarships that I received during my PhD.\

Life outside lab was also highly enjoyable for many many friends and I thank them all. I thank all my friends in Physics department, especially my 2008 batch mates for all their support and good times we had together. 
Arthur's casual approach and Murthy's `mass' approach created a funny contrast that I had thoroughly enjoyed all these years. Mayur always surprises me with his witty comments and jokes. Arun, Kanika, Padma, Ramya, and Resmi provided jovial company always. I thank all my friends in Chemistry department for helping me with many chemical compounds whenever I needed.\

I thank Anurag, Harsha, JP, and Amar for all those adventurous weekend treks and cheerful company. Dinner table was always full of noise, argument, and fun because of Biplab, Abhigyan, and Dada and I thank them all for making it so fascinating. Dada also has been my very close friend and room-mate for all these years.\

I thank all my long-time and long-distance friends- Sudipta, Diganta, Kalyan, Nandan, Swarup and Souravda for their support and encouragement. Conversations were always cheerful and motivating with them.\

My research career would have not been possible without the active support of my family. I thank my mother, father and sister for their love, support and encouragement throughout all the time.

\fancyhf{}
\fancyhead[RO]{\nouppercase{\emph{\rightmark}}}
\fancyhead[LE]{\nouppercase{\emph{\leftmark}}}
\fancyfoot[C]{\thepage}
\setlength{\headheight}{15pt} 
\pagestyle{fancy}

\tableofcontents
\clearpage{\pagestyle{empty}\cleardoublepage}
\addcontentsline{toc}{section}{\textbf{List of Figures}}
\listoffigures
\clearpage{\pagestyle{empty}\cleardoublepage}
\thispagestyle{plain}
\addcontentsline{toc}{section}{\textbf{List of Publications}} 
\begin{singlespace}
\begin{flushright}
\textbf{\LARGE List of Publications}\\
\HRule \\[0.4cm]
\end{flushright}
\begin{enumerate}
              
	\item S. S. Roy and T. S. Mahesh,\\
	\textit{Density Matrix Tomography of Singlet States},\\
	{{J. Magn. Reson. \textbf{206}, 127 (2010)}}.\\
	
	\item S. S. Roy and T. S. Mahesh,\\
	\textit{Initialization of NMR Quantum Registers using Long-Lived Singlet States},\\
	{{Phys. Rev. A \textbf{82}, 052302 (2010)}}.\\
	
	
	\item S. S. Roy, T. S. Mahesh, and G. S. Agarwal,\\
	\textit{Storing Entanglement of Nuclear Spins via Uhrig Dynamical Decoupling},\\
	{{Phys. Rev. A \textbf{83}, 062326 (2011)}}.\\
	
	\item V. Athalye, S. S. Roy, and T. S. Mahesh,\\
	\textit{Investigation of Leggett-Garg Inequality for Precessing Nuclear Spins},\\
	{{Phys. Rev. Lett. \textbf{107}, 130402 (2011)}}.\\
	
	\item S. S. Roy, A. Shukla, and T. S. Mahesh,\\
	\textit{NMR Implementation of Quantum Delayed-Choice Experiment},\\
	{{Phys. Rev. A \textbf{85}, 022109 (2012)}} \\
	
	\item S. S. Roy and T. S. Mahesh,\\
	\textit{Study of Electromagnetically Induced Transparency using Long-Lived Singlet States},\\
  arXiv : quant-ph/1103.3386\\

	\item H. Katiyar, S. S. Roy, T. S. Mahesh, and A. Patel,\\
	\textit{Evolution of Quantum Discord and its Stability in Two-qubit NMR Systems},\\
	{{Phys. Rev. A \textbf{86}, 012309 (2012)}} \\
	
	\item S. S. Roy, M. Sharma, V. Athalye, and T. S. Mahesh,\\
	\textit{Experimental Test of Quantum Contextuality in Nuclear Spin Ensembles},\\
	(in preparation).\\

\end{enumerate}
\end{singlespace}
\clearpage{\pagestyle{empty}\cleardoublepage}

\addcontentsline{toc}{section}{\textbf{Abstract}} 
\chapter*{}
\vspace{-4.5cm}
\begin{center}
\textbf{\large Abstract}\\[0cm]
\end{center}

Nuclear Magnetic Resonance (NMR) forms a natural test-bed to perform quantum information processing (QIP) and has so far proven to be one of the most successful quantum information processors. The nuclear spins in a molecule are treated as quantum bits or qubits which are the basic building blocks of a quantum computer. 

The long lived singlet state (LLS) has found wide range of applications ever since it was discovered by Carravetta, Johannessen, and Levitt in 2004. Under suitable conditions, singlet states can live up to minutes or about many times of longitudinal relaxation time constant (T1). For the first time, we have exploited the long lifetime of singlet states in NMR to execute several potentially important QIP problems. We were able to prepare high fidelity pseudopure states (PPS) in multi-qubit systems starting from LLS. We developed an efficient scheme of density matrix tomography to study all these quantum states. The tomographic study on LLS shows some interesting results. We performed experiments, where we created all the four Bell states from LLS and then studied the effect of various dynamical decoupling sequences on preserving these states. We found that Uhrig dynamical decoupling sequence is better than CPMG sequence in preserving Bell states for longer duration under suitable conditions. 

Nuclear spin systems form convenient platforms for studying various quantum phenomena. We used violation of Leggett-Garg Inequality (LGI) in a two-qubit system to study the transition from quantum to macrorealistic behavior.  We observed perfect violation of LGI for time scales which are much small compared to the spin-spin relaxation time scales. However, with the increasing time scales, we notice a gradual transition of spin-states from quantum to classical behavior. This steady arrival of classicality can be attributed to the decoherence process. In a separate experiment we performed quantum delayed choice experiment in nuclear spin ensembles to study the wave-particle duality of quantum states. These set of experiments clearly demonstrate a continuous morphing of the target qubit between particle-like and wave-like behaviors, thus supporting the theoreticians' demand to reinterpret Bohr's complementary principles.

\clearpage{\pagestyle{plain}\cleardoublepage}

\pagenumbering{arabic}

\chapter{Introduction}
\begin{quotation}
\textit{``If we cannot possibly reach our desired destination, what is the point
in setting out? This question might seem reasonable, but its premise is too
restrictive: sometimes one walks not to reach a destination, but to observe the
scenery along the way, and the pursuit of NMR quantum computation has
thrown up some surprising sights.''}\
\begin{flushright}
- Jonathan A. Jones, 2010\\
\end{flushright}
\end{quotation}
	
\section{Nuclear Magnetic Resonance}

There are four main physical properties in an atomic nucleus: mass, electric charge, magnetism, and spin \cite{Abragam, LevBook}. Most of the macroscopic physical or chemical properties of matter depend on the mass and charge characteristics of nucleus. Though it is less evident, most of the nuclei are magnetic and behave like a tiny bar magnet \cite{Abragam}. However, this nuclear magnetism is very weak and may have little consequence on the matter's property. The dynamics of a nuclear spin can not be understood fully under classical physics and one has to invoke quantum mechanics. The spin and the associated nuclear magnetism provide us the tool to look not only inside the atom but also its microscopic world \cite{Abragam, LevBook}.\  

The first direct evidence of nuclear magnetism was given by Stern and Gerlach in 1922 \cite{feynmanlect}. The Stern-Gerlach experiment involves sending a beam of particles through an inhomogeneous magnetic field and observing their deflection \cite{feynmanlect}. Much to the astonishment of classical physics, the beam splits into only two parts depending on the parallel and anti parallel alignment of their respective magnetic moment in the magnetic field. The exact measure of proton's magnetic moment was done by a series of experiments performed by Frisch, Estermenn, and Stern during 1933-1937 \cite{frisch, estermann, estermann2}. Almost around the same time Isidor Rabi was working on the nuclear magnetism using the extended version of the Stern-Garlach apparatus. Rabi and co-workers showed the first indication of `nuclear magnetic resonance' in molecular beams \cite{rabi}. Soon after, this resonance effect achieved its spectroscopic importance after Bloch \cite{bloch1, bloch2, bloch3} and Purcell \cite{purcell1, purcell2, purcell3} independently observed nuclear magnetization in a bulk matter in 1946.\

Since then, NMR has been studied extensively and has found wide field of applications in physical, chemical, biological, medical and material sciences. In this section, we give a brief overview of the basic principles of NMR. Later sections will describe the field of quantum information processing and its physical realization through NMR.

\subsection{A nuclear spin under a static magnetic field}
Let us consider the simplest situation where we have a single nucleus (an isolated spin) placed in an external magnetic field $\bm{B_0}$. The magnetic nuclei will have a characteristic `Larmor' frequency of $\bm{\omega_0} = -\gamma \bm{B_0}$. Here $\gamma$ represents gyromagnetic ratio of the particular nuclear isotope. The Zeeman Hamiltonian can be written as
\begin{eqnarray} 
\mathcal{H}_z &=& -\bm{\mu \cdot B} \nonumber \\
            &=& -\hbar \bm{I_z} \gamma B_0 = \hbar \omega_0 \bm{I_z},
\label{zH1}
\end{eqnarray}
where $\bm{\mu}$ is the nuclear magnetic moment operator and the external magnetic field $\bm{B_0}$ is taken along $\hat{z}$  direction. 
$\bm{I_z}$ denoting the z component of the nuclear spin operator and the relation between the spin operator and magnetic moment can be written as
$\bm{\mu} = \gamma\hbar \bm{I_z}$.
Below is a table showing comparative study of different properties relevant in NMR for various nuclei \cite{LevBook}.
\begin{table}[h]
\centering
\begin{tabular}{c c c c c}
\hline
Nucleus & Spin & Natural & Gyromagnetic ratio & NMR frequency at 11.7 T \\
& & abundance(\%) & $\gamma/10^6$ rad s$^{-1}$T$^{-1}$ &($\omega_0/2\pi$) in MHZ\\
\hline
\hline\\
$^1$H & 1/2 & $\sim$100 & 267.522 & -500.000\\[1mm]
$^{13}$C & 1/2 & 1.1 & 67.283 & -125.725\\[1mm]
$^{19}$F & 1/2 & $\sim$100 & 251.815 & -470.470\\[1mm]
$^{31}$P & 1/2 & $\sim$100 & 10.394 & -202.606\\[3mm]
\hline
\end{tabular}
\caption{Table of most commonly used nuclear isotopes in NMR}
\label{tab:1}
\end{table}

The eigenvalues of the Zeeman Hamiltonian (Eq. \ref{zH1}) represent the energy levels of the nucleus and are given by
\begin{equation}
E_m = -m\hbar\omega_0.
\label{Em}
\end{equation}
Here $m$ represents the magnetic quantum number and it can take certain discrete values $m = -I, -I+1,.....,I-1, I$, where $I$ can be integer or half-integer and is known as the spin quantum number. $I(I+1)\hbar^2$ is the eigenvalues of total spin operator $I^2$.

While $\bm{I_z}$ represents a stationary state under the Zeeman Hamiltonian, $\langle I_x\rangle$ and $\langle I_y\rangle$ show out of phase oscillations at Larmor frequency ($\omega_0$). Higher (positive) $m$ values have lower energy state (Eq. \ref{Em}) and thus the ground state is the state with $m = I$. In a semiclassical picture, it can be seen as the nuclear spin that is aligned along the static magnetic field direction. On the other hand, highest excited state corresponds to a spin-alignment against the magnetic field. 

For an ensemble of nuclear spins at thermal equilibrium , the population distribution can be represented by Boltzmann statistics. For spin-1/2 ensemble, there will be only two possible energy levels with $m = -1/2$ and $m = +1/2$. The population ratio of these two levels is determined by Boltzmann distribution
\begin{equation}
\frac{p_-}{P_+} = e^{-\hbar\omega_0/k_BT},
\end{equation}
where $k_B$ is the Boltzmann constant and $T$ is the absolute temperature of the ensemble. In the case of $^1H$ nuclei at a 10 T magnetic field strength, $\hbar\omega_0 \approx 10^{-6} eV$. Whereas at room temperature $k_BT \approx 2.5\times 10^{-2} eV$, hence the ratio $\hbar\omega_0/k_BT \approx 10^{-5}$. So the Boltzmann factor $e^{-\hbar\omega_0/k_BT}$ is almost close to unity. This can naively be interpreted as, there are slightly more spins in the parallel direction (lower state) than in the anti-parallel direction (upper state) and this slight imbalance in the populations is responsible for the `net' nuclear magnetization along the $z$-direction. This also reveals the fact that, NMR is a very low sensitive technique. 

The nuclear magnetization for an ensemble of spin-1/2 nuclei at thermal equilibrium is given by \cite{Slichterbook}
\begin{equation}
M_0 = \frac{n_0\gamma^2\hbar^2 B_0}{4k_BT},
\end{equation}  
where $n_0$ is the number of nuclei per unit volume. From above equation it is clearly seen that the magnetization increases linearly with the external field strength, whereas it is inversely proportional to the temperature. Hence nuclear magnetism is paramagnetic in nature and follows Curie's law \cite{Slichterbook}.	Also, the demand for higher field strength can be understood from the above equation. However, the temperature of the ensemble can not be reduced as per wish, since it is related to the `state' of the matter and hence on its dynamics. Here it can be noted that, electrons also posses paramagnetism and the magnitude of electron paramagnetism is three order of magnitude higher than the nuclear magnetism.   
  
\subsection{Radiofrequency field}
The application of static magnetic field will create a Zeeman splitting according to the Eq. \ref{zH1}. Now the transitions between the energy levels can be induced by the application of suitable oscillatory magnetic fields with appropriate frequencies. From the Table \ref{tab:1}, it is seen that the Larmor frequencies are of the order of MHz in present days' magnet of a few Tesla and resonance can be achieved by the application of RF fields. In Comparison, typical electron Larmor frequencies are of the order of GHz range.\

The dynamics of nuclear spin excitation due to the application of oscillatory magnetic field can be well understood by considering a time dependent magnetic field, $\bm{B}_1(t)$ applied perpendicular to the static magnetic field $\bm{B}_0$. The RF interaction Hamiltonian ($\mathcal{H}_{RF}$), can be written in a similar way as the Zeeman Hamiltonian. 
\begin{eqnarray}
\mathcal{H}_{RF} &=& -\bm{\mu}.\bm{B}_1(t) = -\gamma\hbar \bm{I}_x [2B_1 \cos(\Omega t + \phi)]\\
\text{where,}\quad \bm{B}_1(t) &=& 2B_1 \cos(\Omega t + \phi)\hat{x}
\label{Hrf}
\end{eqnarray}
Here $\Omega$ and $\phi$ are respectively the frequency and phase of the RF field which is along the $\hat{x}$ direction.
The strength of the oscillatory magnetic field $\left(\bm{B}_1(t)\right)$ is much smaller than the Zeeman field strength ($\bm{B}_0$) an hence it is reasonable to treat the RF Hamiltonian ($\mathcal{H}_{RF}$) as a perturbation to the Zeeman Hamiltonian ($\mathcal{H}_z$). The dynamics can be described by the standard time dependent perturbation theory \cite{Sakuraibook}. The result shows that, at resonance condition ($\Omega \cong \omega_0$), there will be induced transitions between the eigenstates of $\mathcal{H}_z$ with a transition rate given by the Fermi golden rule \cite{Olinmrqip}
\begin{equation}
p_{m_1\rightarrow m_2} = p_{m_2\rightarrow m_1} \propto \gamma^2\hbar^2 B_1^2\; \Big\vert\langle m_1\vert I_x\vert m_2\rangle\Big\vert^2,
\end{equation}
where $m_1$ and $m_2$ are two energy eigenstates of the system. It can be seen from the above equation that, the transition probability on either way depends on square root of gyromagnetic ratio of the nucleus and the magnitude of RF field. The selection rule for the allowed transition should be, $\Delta m = \pm1$.\

Now we will discuss the logic behind choosing the RF magnetic field similar to Eq. \ref{Hrf}. We can think of a linearly polarized magnetic field $\bm{B}_1(t)$ as composed of two circularly polarized fields with same frequency and amplitude but precessing in opposite directions about $z$-axis.
\begin{eqnarray}
\bm{B}_1(t) &=& 2B_1\cos(\Omega t+\phi)\hat{x} = \bm{B}_1^+(t) + \bm{B}_1^-(t)\\[.2cm]
\text{where,}\quad \bm{B}_1^+(t) &=& B_1\big[\cos(\Omega t+\phi)\hat{x} + \sin(\Omega t+\phi)\hat{y}\big]\\[0.1cm]
\bm{B}_1^-(t) &=& B_1\big[\cos(\Omega t+\phi)\hat{x} - \sin(\Omega t+\phi)\hat{y}\big]
\end{eqnarray}
The RF field interactions can be better described in rotating frame formalism. At resonance condition, (i. e. $\Omega=\omega_0$) the field $\bm{B}_1^-(t)$ rotates coherently with the nuclear Larmor precession along z-axis. Whereas, the field $\bm{B}_1^+(t)$ rotates exactly in opposite sense. In a frame which is rotating along with the Larmor frequency, the field $\bm{B}_1^-(t)$ is stationary along with nuclear spin, whereas the field $\bm{B}_1^+(t)$ rotates with a frequency twice the Larmor frequency. Therefore, at high static fields it can safely be assumed that only the field $\bm{B}_1^-(t)$ has effect on the nuclear spins.\

Let us assume the on-resonance condition i.e., $\Omega=\omega_0$. In a frame that is rotating with $\bm{B}_1^-(t)$ with same frequency and direction, the magnetic moment sees a static field, say along direction $\hat{x'}$, and precesses about it. In the case of off-resonance conditions (i.e. $\Omega\neq\omega_0$), the precession of magnetic moments in the rotating frame is around an axis defined by an effective magnetic field given by,
\begin{equation}
\bm{B}_{eff} = \left(B_0 - \frac{\Omega}{\gamma}\right)\hat{z} + B_1\hat{x'}.
\end{equation}
The relation between laboratory frame and rotating frame is given by,
\begin{eqnarray}
\hat{x'}=\cos(\Omega t+\phi)\hat{x}-\sin(\Omega t+\phi)\hat{y}
\end{eqnarray}
At on-resonance condition, the precession frequency about $\bm{B}_{eff}$ is also known as nutation frequency $\omega_{nut} = -\gamma\bm{B}_{eff}$ in analogy with the Larmor frequency. Application of an RF pulse for the time duration $t_P$, makes the magnetization shift from its initial z-direction by a nutation angle given by,
\begin{equation}
\theta_p = \gamma B_1t_p.
\end{equation}  
Hence, a $\pi/2$ pulse is defined as a pulse which can take the magnetization from longitudinal direction to transverse plane. One must remember that in laboratory frame the magnetization is always precessing around the $z$- axis in addition to nutating about the RF axis. 
 
\subsection{Nuclear spin interactions}
So far we have described the nuclear spins in isolated situation without any kind of interactions. In practice, nuclear spins are interacting with each other as well as with the environment. The interaction of nuclear spins with each other makes NMR a very sophisticated tool with versatile applications. However interaction of nuclear spins with environment remains a challenge in the field of NMR-QIP and we will discuss this case in detail in a later chapter. Here we describe the main interactions involving in the nuclear spins under normal conditions \cite{Olinmrqip, haeberlen}.

The total nuclear Hamiltonian is given by
\begin{equation}
\mathcal{H}_{total} = \mathcal{H}_{RF} + \mathcal{H}_{int},
\end{equation}
where $\mathcal{H}_{int}$ represents the internal interactions of the nuclei.
Here we will concentrate on the $\mathcal{H}_{int}$ part of the total Hamiltonian. There are several contributors to the internal Hamiltonian part based on its physical and chemical characters. In most of the case the material in study under NMR is a diamagnetic insulating substance. For this the internal Hamiltonian is given by
\begin{equation}
\mathcal{H}_{int} = \mathcal{H}_{CS} + \mathcal{H}_{D} + \mathcal{H}_{J} + \mathcal{H}_{Q}
\end{equation}
where, $\mathcal{H}_{CS}$ is the chemical shift interaction, $\mathcal{H}_{D}$ is the direct dipolar interaction, $\mathcal{H}_{J}$ is the indirect spin-spin interaction, and $\mathcal{H}_{Q}$ is the quadrupolar interaction.
\subsubsection{Chemical Shift}
Though the external magnetic field applied is same for all the nuclei, it is not even exactly same for a same type of nuclei in a molecule. The slight change in the magnetic field is due to the modified chemical environment created by the electron density surrounding it. The modified magnetic field is given by
\begin{equation}
\bm{B}_{loc} = (1-\widetilde{\sigma})\bm{B}_0,
\end{equation} 
where $\widetilde\sigma$ is known as chemical shielding tensor allied to that particular nuclear site. Hence the chemical shift Hamiltonian can be written as,
\begin{equation}
\mathcal{H}_{CS} = -\bm{\mu}\;.\; (-\widetilde{\sigma}B_0) \cong \gamma\hbar\sigma_{zz}(\theta, \phi)B_0\bm{I}_z.
\end{equation} 
The approximation is known as secular approximation. Now,
\begin{equation}
\sigma_{zz}(\theta, \phi) = \sigma_{11} \sin^2\theta \cos^2\phi + \sigma_{22} \sin^2\theta \sin^2\phi + \sigma_{33} \cos^2\theta,
\end{equation} 
where, $\sigma_{11}, \sigma_{22}$, and $\sigma_{33}$ are the principle values of the chemical shielding tensor $\widetilde{\sigma}$. Here $\theta$ and $\phi$ are the azimuthal and polar angle respectively, describing the magnetic field $\bm{B_0}$ in the principle axis system. In isotropic liquid, due to rapid molecular motions, shielding tensor get averaged. Hence, the time averaged shielding constant for isotropic liquid can be written as,
\begin{equation}
\sigma_{iso} = \frac{1}{3}\left(\sigma_{xx} + \sigma_{yy} + \sigma_{zz}\right).
\end{equation} 
The consequence of the above calculation is the introduction of a shift in resonance frequency,
\begin{equation}
\omega = \omega_0(1-\sigma_{iso}).
\end{equation}
For a monocrystalline material, the above equation will be modified just by replacing $\sigma_{iso}$ with $\sigma_{zz}$. In the case of polycrystalline material or powder samples, the continuous distribution of orientations of the several crystallites causes an anisotropic broadening, known as chemical shift anisotropy (CSA)\cite{LevBook},
\begin{equation}
\Delta\sigma = \sigma_{zz} - \frac{1}{2}(\sigma_{xx}+\sigma_{yy}).
\end{equation}
The resonance frequency is conventionally expressed by the relative shift from the reference resonance frequency ($\omega_{ref}$),
\begin{equation}
\delta = \frac{\omega-\omega_{ref}}{\omega_{ref}}.
\end{equation} 
Here $\delta$ represents the chemical shift of the resonance lines and normally expressed in terms of parts per million (ppm).
\subsubsection{Direct dipolar coupling}
Any two magnetic dipole moments interact directly with each other through the magnetic fields created by each one for the others. It provides rich structural information about the materials. The dipolar Hamiltonian is defined as,
\begin{eqnarray}
\mathcal{H}_D &=& \sum_{k<l}\bm{I}_k\bm{\widetilde{D}}_{kl}\bm{I}_l\\
              &=& \sum_{k<l}\frac{\mu_0}{4\pi}\frac{\gamma_k\gamma_l\hbar^2}{r_{kl}^{3}}\left[ \bm{I}_k\;.\;\bm{I}_l - 3\frac{1}{r_{kl}^2}(\bm{I}_k\;.\;r_{kl})(\bm{I}_l\;.\;r_{kl})\right],
\end{eqnarray}
where $\bm{\widetilde{D}}_{kl}$ is the dipole-dipole interaction tensor, $r_{kl}$ is the radius vector connecting the two spins. Under secular approximation, the Hamiltonian can be rewritten as,
\begin{equation}
\mathcal{H}_D^{trunc} = -\sum_{k<l}\frac{\mu_0}{4\pi}\frac{\gamma_k\gamma_l\hbar^2}{r_{kl}^{3}}\frac{1}{2}\left(3\cos^2\theta_{kl} - 1 \right)\left[ 3I_{kz}I_{lz} - \bm{I}_k\;.\;\bm{I}_l\right],
\end{equation}
where $\theta_{kl}$ is the angle between $r_{kl}$ and $\hat{z}$. In case of heteronuclear spin systems (i.e. $\gamma_k\neq\gamma_l$), further simplification is possible,
\begin{equation}
\mathcal{H}_D^{IS} = \frac{\mu_0}{4\pi}\frac{\gamma_k\gamma_l\hbar^2}{r_{kl}^{3}}\left(1-3\cos^2\theta_{kl} \right)\;I_{kz}I_{lz}.
\end{equation}
\subsubsection{Indirect spin-spin coupling}
Indirect spin-spin coupling (also called J-coupling or scalar coupling) is also an interaction between the nuclear magnetic dipole moments. This type of coupling is not direct and being mediated by the electron cloud involved in the chemical bonds between the atoms. The J-coupling Hamiltonian is defined as,
\begin{equation}
\mathcal{H}_J = 2\pi\hbar\sum_{k<l}\bm{I}_k \bm{\widetilde{J}} \bm{I}_l,
\end{equation}
where $\bm{\widetilde{J}}$ is the J-coupling tensor. J-coupling posses an isotropic part which survives under random molecular motion in an isotropic substance (e.g. liquid samples), whereas direct dipolar coupling is averaged out under similar situation. In the case of solid samples, the J-coupling is generally overwhelmed by the strong direct dipolar couplings. Under secular approximation, the simplified J-coupling term is written as,
\begin{equation}
\mathcal{H}_J^{kl} = 2\pi\hbar J_{kl}\bm{I}_{kz}\bm{I}_{lz}.
\end{equation}
The approximation can be carried out when $2\pi J_{ij} << \vert\omega_i - \omega_j\vert$. It can be seen that this approximation holds for all heteronuclear pairs.
\subsubsection{Quadrupolar coupling}
All the nuclei with spin, I > 1/2 are subjected to electrostatic interaction with the neighboring electrons, ions due to the non-spherical charge distribution of nuclei \cite{ErnstBook}. The Hamiltonian form of quadrupolar interaction is defined as,
\begin{equation}
\mathcal{H}_{Q} = \sum_{k=1}^{N}\bm{I}_k \bm{\widetilde{Q}}_k\bm{I}_k,
\end{equation}
where, $\bm{\widetilde{Q}}_k$ is the quadrupolar coupling tensor and it can be expressed in terms of electric field gradient tensor $\bm{V}_k$ at the $k^{th}$ nuclear site,
\begin{equation}
\bm{Q}_k = \frac{eQ_k}{2I_k(2I_k-1)}\bm{V}_k.
\end{equation}
Here $Q_k$ is the nuclear quadrupolar moment of the $k^{th}$ nucleus. In the partial axis coordinate system, the quadrupolar Hamiltonian for the $k^{th}$ nucleus can be written as,
\begin{equation}
\mathcal{H}_Q = \frac{3e^2q_kQ_k}{4I_k(2I_k-1)}\left[ \left( I_{kz}^2-\frac{1}{3}\bm{I}_{k}^{2}\right) + \frac{\eta}{3}\left(I_{kx}^2 - I_{ky}^2 \right)\right],
\end{equation}
where $eq_k = V_{kzz}$ and $\eta_k$ defines the assymetry parameter,
\begin{equation}
\eta_k = \frac{V_{kxx}-V_{kyy}}{V_{kzz}}.
\end{equation}

\subsection{Systems of spin-1/2 nuclei}  
The Hamiltonian for N coupled spin-1/2 nuclear spins in an isotropic medium is given by,
\begin{equation}
\mathcal{H} = \sum_{k=1}^{N}\omega_kI_{kz} + \sum_{k < l}2\pi J_{kl}\bm{I}_{k}.\bm{I}_{l},
\end{equation}
where $\omega_k$ is the chemical shift for the $k^{th}$ nucleus and $J_{kl}$ is the J-coupling constant between the two spins. Considering weak coupling condition i.e. $|2\pi J_{kl}| << |\omega_k - \omega_l|$, the Hamiltonian can be written as,
\begin{equation}
\mathcal{H} = \sum_{k=1}^{N}\omega_kI_{kz} + \sum_{k < l}2\pi J_{kl}I_{kz}I_{lz}.
\end{equation}
The coupling part of the above equation actually commutes with the Zeeman part and hence both the part will share common eigenbasis. All the $2^N$ eigenstates of $\mathcal{H}$ can be expressed as tensor products of the single spin eigenstates, namely $\vert\alpha\alpha\ldots\alpha\rangle,\vert\alpha\alpha\ldots\beta\rangle$, $\ldots,\vert\beta\beta\ldots\beta\rangle$. Here $\vert\alpha\rangle$ and $\vert\beta\rangle$ denote $\vert+1/2\rangle$ and $\vert-1/2\rangle$ single-spin eigenstates, which are labeled as $\vert 0\rangle$ and  $\vert 1\rangle$ in QIP terminology. The NMR spectrum displays $N$ set of $2^{N-1}$ spectral lines of equal intensity.\

The Hamiltonian for a pair of spin-1/2 system in an isotropic liquid environment can be written as,
\begin{equation}
\mathcal{H}_{kl} = \omega_kI_{kz} + \omega_lI_{lz} + 2\pi J_{kl}I_{k}I_{l},
\end{equation}
where we have considered the weak coupling condition. This Hamiltonian will have four eigenstates and corresponding four eigenenergy values. The four probable transition will reflect as four transition line in an NMR spectra. The eigenstates and eigenenergy are:
\begin{equation}
\begin{aligned}
\vert00\rangle\quad ::\quad E_{00} &=& -\left(-\omega_k - \omega_l + \pi J \right)/2\\
\vert01\rangle\quad ::\quad E_{01} &=& -\left(-\omega_k + \omega_l - \pi J \right)/2\\
\vert10\rangle\quad ::\quad E_{10} &=& -\left(\omega_k - \omega_l - \pi J \right)/2\\
\vert11\rangle\quad ::\quad E_{11} &=& -\left(\omega_k + \omega_l + \pi J \right)/2
\end{aligned}
\end{equation}
\subsection{NMR Relaxation}  
In equilibrium, the population distribution of the spins follow Boltzmann statistics with off diagonal elements are zero for the density matrix of the system. The NMR mechanism depends on the perturbation of the system from equilibrium situation. For example, application of a single $\pi/2$ pulse on equilibrium equalizes the populations and also creates the coherences. Now, this is clearly a non-equilibrium situation and the disturbed state tends to go back to the original equilibrium state through relaxation mechanism of the spins. There are two different processes, occurring simultaneously but in general independently that can be identified for this relaxation. These two relaxation mechanisms known as transverse relaxation and longitudinal relaxation \cite{proteinnmr, Olinmrqip}.\

Just after the RF pulse, the magnetization is on the transverse plane perpendicular to the static magnetic field $\bm{B}_0$. The transverse relaxation mechanism makes the magnetization along transverse plane to disappear. The result of the transverse relaxation is the loss of coherences among the spins. This happens due to the spread in nuclear precession frequencies of the spin ensemble. As shown earlier, the Larmor frequency of each spin depends on the external magnetic field as well as locally created magnetic field for various reasons. Hence due to this slight variance in the Larmor frequency, after some time these spins are oriented in a complete random direction on transverse plane and the vector sum of all this magnetization will be zero. The decay of coherences due to the inhomogeneous fields is one part of the transverse relaxation process. The other important part occurs due to the fluctuations in the local magnetic field \cite{LevBook}.\

Under normal conditions, the decaying of the transverse component of the magnetization of the nuclear spins ensemble in the rotating frame can be described by a phenomenological differential equation given by Bloch \cite{bloch2, LevBook}.
\begin{equation}
\frac{dM_{x,y}}{dt} = -\frac{M_{x,y}}{T_2},
\end{equation} 
where $T_2$ is known as transverse or spin-spin relaxation constant. The solution of the above equation is simple and can be written as,
\begin{equation}
M_{x,y} = M_0e^{-t/T_2},
\end{equation} 
where $M_0$ represents the initial value of the transverse magnetization. Hence from the above equation it is seen that the transverse magnetization decays with time in exponential fashion. The exact value of $T_2$ depends on the detail of each particular nuclear spin system and its environment.\

The longitudinal part of the nuclear magnetization also goes under relaxation simultaneously with transverse relaxation. The mechanism can be understood as follows. Just after the $\pi/2$ pulse, the longitudinal magnetization, $M_z = 0$ and the population of a two level system is equalized. Since this condition is non-equilibrium, the system will tend to go back to its equilibrium condition that is supported by Boltzmann distribution. The preferable way towards the equilibrium is by giving up its excess populations in upper to lower energy level till the Boltzmann distribution is reestablished. Since this mechanism involves energy exchange and that happens with the lattice part of the system, this relaxation mechanism is also termed as spin-lattice relaxation process \cite{LevBook}.\

Similar to the transverse case, the longitudinal relaxation mechanism is also described by a phenomenological differential equation given by Bloch \cite{bloch2, LevBook}.
\begin{equation}
\frac{dM_{z}}{dt} = \frac{M_0 - M_z}{T_1},
\end{equation} 
where $T_1$ representing the longitudinal or spin-lattice relaxation constant. The solution of the above equation is given by,
\begin{equation}
M_z = M_0\left(1-e^{-t/T_1}\right).
\end{equation}
As it can be seen from the above solution, the longitudinal magnetization is gaining with time beginning from zero and reaches to the stable magnetization $M_0$ after certain time. The exact values of $T_1$ and $T_2$ time constants depend on various factors such as physical state of matter (liquid or solid), temperature, molecular mobility, viscosity, concentration, external magnetic field etc \cite{LevBook}. In most of the cases it is found that $T_1 \geq T_2$. In case of liquids, $T_2$ values are comparable with $T_1$ and in many cases both are almost equal. However in case of solids, $T_1$ is much larger than $T_2$. \

It is worth noting that the above simplistic approach of relaxation formalism in nuclear spins is not straightforward in many complicated situations. The relaxation phenomenon can be best understood by the elaborative mechanism of Redfield theory \cite{Slichterbook}.  
 
\section{Quantum Information Processing}
Quantum information processing (QIP) is the study of the information processing tasks that can be accomplished using quantum mechanical systems \cite{chuangbook}. The idea of utilizing quantum systems for the information processing was first introduced by Benioff in early 1980s \cite{benioffjstat, benioffprl}. The exponential time required for simulating the dynamics of quantum systems using classical computers inspired Feynman to propose exploiting quantum systems for such purpose \cite{feynman82}. He was rather skeptical whether a classical computer is capable enough to simulate a quantum system and advocated for building a quantum computer for this purpose. In 1985 Deutsch gave a decisively important step towards quantum computers by presenting the first example of quantum algorithm which utilizes the fact of quantum superposition in speeding up computational process \cite{deutsch85}. He is also the pioneer of quantum computer history for introducing the notion of quantum logic gate in 1989 \cite{deutsch89}. Since then there has been a good theoretical progress in the field of quantum computation and quantum information. Classically intractable problems were reduced to tractable regime by treating it in quantum way. It was 1994, when a major breakthrough happened, calling the attention of scientific community for the potential practical importance of quantum computation and its direct consequence on our society. Peter Shor discovered a quantum algorithm which is capable of factorization of prime numbers in polynomial time instead of exponential time \cite{shor94, shor99}. Prime factorization being the heart of computational security, draws tremendous attention from computer scientists and cryptographers as well. The successful experimental tests of Shor's algorithm have been performed using a liquid state NMR systems \cite{vandersypen}. A few years after that, in 1997, another important discovery had been made by Lov Grover by introducing a quantum search algorithm for searching an unsorted database \cite{grover97}. Grover's algorithm makes use of quantum superposition and quantum phase interference to find an item in an unsorted database, faster than any other classical algorithms. Various schemes on error correction has also being developed to counter the faulty outcomes \cite{preskill}. In the meantime, other branches of QIP, namely quantum teleportation, quantum key distribution and quantum cryptography are also being developed. Many of these techniques have actually making commercial success and continue to be better \cite{BB84}. Considering the extreme difficulty in controlling a quantum system, there has been modest development towards a practical quantum computer. Nonetheless, commercialization of quantum computer has been taken very seriously and till date it has already arrived (arguably) in the markets \cite{dwave}. This section intend to give a brief theoretical understanding on QIP and later its physical realization by various experimental schemes.\

\subsection{Computational science}
Today, we can not even think a society without the machine called computer. The impact of a computer is such that, there hardly any field left where we are not using a computer directly or indirectly. There is a long history of development of computers and the theoretical notion of computation. As put by David Deutsch \cite{deutschbook}, `Computers are physical objects, and computations are physical processes. What computers can or cannot compute is determined by the laws of physics alone, and not by pure mathematics'. Computation is carried out through a procedure called algorithm and it needs three basic resources (space, time, energy) for it \cite{chuangbook}. Space refers to the the computer hardware, i.e. the number of logic gates used. Time refers to the computational time required and energy refers to the energy spent for the computational work. The basic model of a modern day computer was mainly given by Alonzo Church and Alan Turing in early 20th century. Later it became famous as Turing machine \cite{turing36}. A Turing machine is a hypothetical, idealized theoretical model of an actual computer. There is not a single computation work which can be done by an actual computer but not by a Turing machine. In that sense, a real computer is a physical realization of a Turing machine. It consisted of a program, a finite state of control, a memory tape, and a read-write head \cite{chuangbook}. The Church-Turing thesis calls a problem `computable' only if it can be done by a Turing machine. Quantum computation also obeys the ideology of the Church-Turing thesis and hence the notion of `computable' has not changed, only efficient algorithms could be possible. The efficiency of an algorithm is studied by its asymptotic behavior as the size of the input increases \cite{chuangbook}. Consider the time taken by an algorithm varies as $f(N)$, where $N$ is the number of input bits. Now, if $f(N)$ is polynomial, then the highest power in $f(N)$, say $g(N)$ is known as order of algorithm denoted by $O\left(g(N)\right)$. Depending on these requirements, computational problems are classified into various classes known as `complexity classes' as shown in table \ref{complex}. 
\begin{table}[h]
\centering
\begin{tabular}{p{4cm} p{4cm} p{2cm}}
\hline
\textbf{Class} & \textbf{Time} & \textbf{Space} \\
\hline
\hline\\
EXP & exponential & unlimited \\[1mm]
PSPACE & unlimited & polynomial\\[1mm]
NP & exponential & polynomial\\[1mm]
P & polynomial & polynomial\\[1mm]
L & logarithmic & polynomial\\[3mm]
\hline
\end{tabular}
\caption{Complexity classes in computational science. These classes are related as : L $\subseteq$ P $\subseteq$ NP $\subseteq$ PSPACE $\subseteq$ EXP}
\label{complex}
\end{table}
A simple addition, subtraction, or multiplication are in class L, while division comes under class P. Prime factorization is believed to be a class NP problem, however not proven till date. Many of the complexity classes are unclear even today. In fact it is a great source of debate whether N = NP or N $\neq$ NP and nobody has come up with a concrete prove so far. \

The relationship of energy with information processing has an important physical significance \cite{preskill}. Erasure of information is a dissipative process, as pointed out by Rolf Landauer in 1961 \cite{rolf}. Erasure of each bit increases the amount of entropy by $k \ln 2$ and the energy dissipates at least by an amount $kT \ln 2$. However, this amount is negligible compared with the energy dissipated in a modern computer which is of the order of $500kT\ln 2$. All the irreversible gates involve in loss of information and hence dissipates energy. Interestingly in 1973, it was found by Charles Bennett that the dissipation of energy can be made vanishingly small by making all the gates reversible \cite{bennett73}.

\subsection{Quantum Information}
Information always exists as encoded with an physical system and therefore it should obey the physical laws. In other words, `Information is Physical' \cite{preskill, rolf} and physical systems obey quantum mechanics. Hence the information encoded in such system is `Quantum Information'. Treating some problems in quantum mechanical way can actually make it much more efficient than classical way. For example, prime factorization is a `NP' class problem classically (require exponential time), whereas solving it in quantum mechanical way can make it a `P' class problem (require polynomial time). 

\subsection{Quantum Bits}
\begin{figure}
	\centering
		\includegraphics[width=6cm]{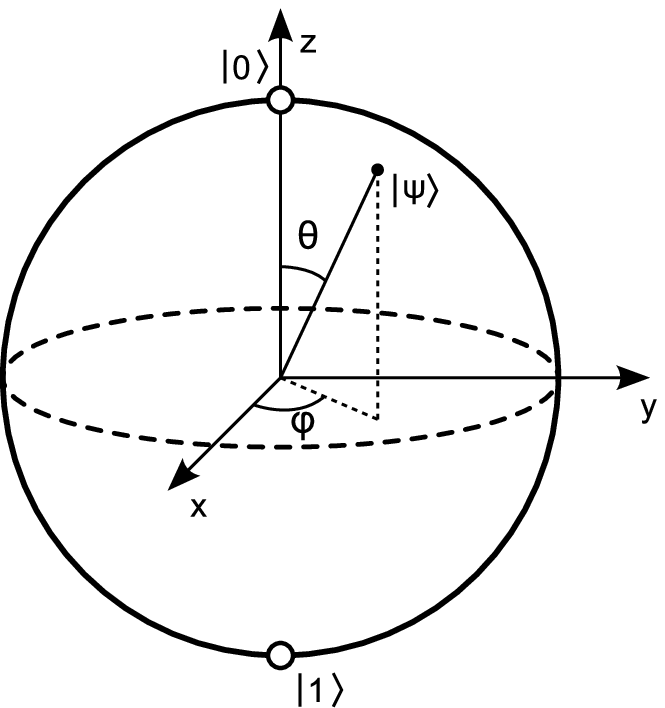}
		\caption[Bloch sphere representation]{Bloch sphere representation of a two level quantum system.}
	\label{fig:bloch0}
\end{figure}

The unit of information in quantum computation and quantum information is known as quantum bit or `qubit'. A qubit can assume a logical values `0' and `1' along with a state that is a linear combination of them. Physically a qubit can be represented by any well defined distinct eigenstates. For example, qubits can be the polarization states of a photon or nuclear spins inside a static magnetic field. Let us consider a two level quantum system, where the eigenstates are represented by $\vert0\rangle$ and $\vert1\rangle$. The general form of a quantum state under this condition can be written as,
\begin{equation}
\vert\psi\rangle = \cos\left(\frac{\theta}{2}\right)\vert0\rangle + e^{i\phi}\sin\left(\frac{\theta}{2}\right)\vert1\rangle
\end{equation}  
where $0\leq\theta\leq\pi$ and $0\leq\phi\leq2\pi$, neglecting the global phase factor. On measurement in ${\vert0\rangle, \vert1\rangle}$ basis, the probability of getting the state $\vert0\rangle$ is $\cos^2(\theta/2)$ and for $\vert1\rangle$ it is $\sin^2(\theta/2)$. Also this kind of representation allows one to visualize this complex quantum state geometrically. The qubit states are designated as some geometrical point on the surface of a `Bloch sphere' (Fig. \ref{fig:bloch0}). Any surface point on the Bloch sphere is a `pure' state while any non-surface point represents a `mixed' state. A more detailed description about pure and mixed states is given in chapter 3. The power of quantum computation comes from the quantum mechanical laws such as superposition of states of qubits and the ability to manipulate the quantum states through unitary transformations as will be seen in next subsections.

\subsection{Quantum Gates}
A classical gate can transform a string of $n$ bits into a string of $m$ bits,
\begin{equation}
f : \{0,1\}^n \stackrel{f}{\longrightarrow} \{0,1\}^m .
\end{equation}
Now for $f$ to be a reversible classical gate, it should be one to one (each inout is mapped to a unique output). In general $n$ and $m$ are not equal and hence a classical gate is a irreversible gate. Quantum gates on the other hand transform a state of quantum system from one point in the Hilbert space to another point. A single qubit can be expressed by $\vert\psi\rangle = a\vert0\rangle + b\vert1\rangle$, where $a$ and $b$ are coefficients having a relationship $\vert a\vert^2 + \vert b\vert^2 = 1$. Quantum gates on a qubit must preserve this normalization condition and thus it can be described by a $2\times2$ unitary matrices \cite{chuangbook}. Since all the quantum operations are unitary operators, quantum gates must also be reversible.\

Some important unitary transformations for one qubit are the Pauli matrices,
\begin{align}
\sigma_x &= \begin{pmatrix}
    0 & 1 \\
    1 & 0
\end{pmatrix};
&
\sigma_y &= \begin{pmatrix}
    0 & -i \\
    i & 0
\end{pmatrix};
&
\sigma_z &= \begin{pmatrix}
    1 & 0 \\
    0 & -1
\end{pmatrix}.
\end{align}  
These three Pauli matrices along with the $2\times 2$ identity matrix form a $2\times 2$ basis matrix space. Hence any one qubit operation can be decomposed as a linear combination of the four matrices. A NOT gate is nothing but the Pauli-x matrix and it flips the $\vert 0\rangle$ to $\vert 1\rangle$ and vice versa.
\begin{eqnarray}
U_{NOT} &=& \begin{pmatrix}
     0 & 1\\
     1 & 0
\end{pmatrix};
\label{notgate}\\
U_{NOT}\vert0\rangle &=& \begin{pmatrix}
     0 & 1\\
     1 & 0
\end{pmatrix}
\begin{pmatrix}
1\\
0
\end{pmatrix} = 
\begin{pmatrix}
0\\
1
\end{pmatrix} = \vert1\rangle;\\  
U_{NOT}\vert1\rangle &=& \begin{pmatrix}
     0 & 1\\
     1 & 0
\end{pmatrix}
\begin{pmatrix}
0\\
1
\end{pmatrix} = 
\begin{pmatrix}
1\\
0
\end{pmatrix} = \vert0\rangle.
\end{eqnarray}
Another very important one qubit gate is Hadamard gate which has no classical analogue and it is used for the creation of superposition states as shown below. One important property of Hadamard operator is its self-reversibility, i.e. H$^2$ = $\mathbbm{1}$.
\begin{eqnarray}
U_H = \frac{1}{\sqrt{2}}\begin{pmatrix}
      1 & 1\\
      1 & -1
\end{pmatrix};
\label{Hgate}\\
\vert0\rangle \stackrel{H}{\longrightarrow}\frac{1}{\sqrt{2}}\:\Big(\vert0\rangle + \vert 1\rangle\Big);\\
\vert1\rangle \stackrel{H}{\longrightarrow}\frac{1}{\sqrt{2}}\:\Big(\vert0\rangle - \vert 1\rangle\Big).
\end{eqnarray}
A phase shift gate P selectively introduces a phase to either of the qubit of a superposition state,
\begin{eqnarray}
&&U_P = \begin{pmatrix}
    1 & 0\\
    0 & e^{i\phi}
\end{pmatrix};
\label{Pgate}\\
&&\Big(a\vert0\rangle + b\vert1\rangle\Big)\stackrel{P}{\longrightarrow}\Big(a\vert0\rangle + b.e^{i\phi}\vert1\rangle\Big).
\end{eqnarray}\
For a two qubit system the dimension of Hilbert space is $4\times 4$ and can be realized by tensor products among the one qubit states,
\begin{eqnarray}
\{\vert0\rangle,\vert1\rangle\} \otimes \{\vert0\rangle,\vert1\rangle\} = \{\vert00\rangle,\vert01\rangle,\vert10\rangle,\vert11\rangle\},
\end{eqnarray}
where,
\begin{align}
\vert00\rangle &= \begin{pmatrix}
                 1\\
                 0\\
                 0\\
                 0
                 \end{pmatrix};
&
\vert01\rangle &= \begin{pmatrix}
                 0\\
                 1\\
                 0\\
                 0
                 \end{pmatrix};
&
\vert10\rangle &= \begin{pmatrix}
                 0\\
                 0\\
                 1\\
                 0
                 \end{pmatrix};
&
\vert11\rangle &= \begin{pmatrix}
                 0\\
                 0\\
                 0\\
                 1
                 \end{pmatrix}.                            
\end{align}
The matrix representation for operators that act only on one of the qubits of a system of two qubit can be constructed by tensor product between one qubit operator and $2\times2$ identity operator. 
\begin{equation}
\mathcal{O}_a = \mathcal{O}\otimes\mathbbm{1}; \;\;\;\; \mathcal{O}_b = \mathbbm{1}\otimes\mathcal{O}
\end{equation}
Here $\mathcal{O}$ denoting the Pauli matrix operators. The above given scheme can be worked out for any number of qubits in a similar fashion.
\begin{figure}
	\centering
		\includegraphics[width=10cm]{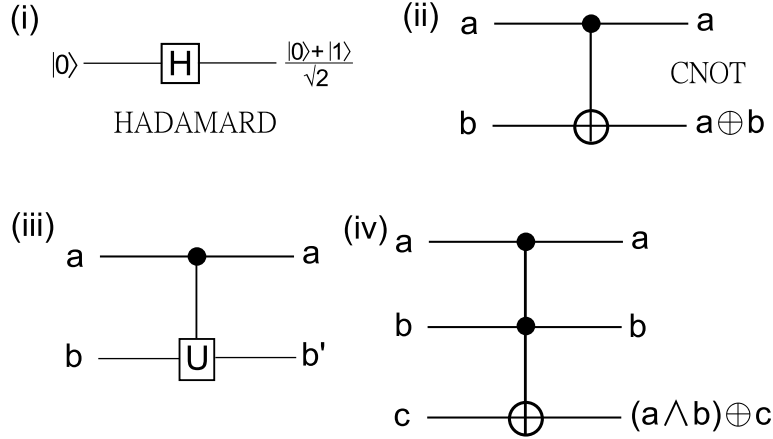}
	\caption[Quantum gates]{Quantum gates. (i) Hadamard gate acting on $\vert0\rangle$ qubit, (ii) CNOT gate, (iii) A general two qubit controlled- gate, where U can be any one qubit operator, and (iv) Toffoli gate. In the above circuits, the inputs are assumed to be individual basis states. If on the otherhand, inputs are in superposition, output may be entangled.}
	\label{cnotfig}
\end{figure}
The most important two qubit gate is definitely the CNOT (or controlled not) gate. It can be proved that all the quantum operations necessary for quantum computation can be achieved using only CNOT and set of one qubit gates \cite{chuangbook}. In that sense CNOT is the universal quantum gate similar to NAND gate in classical counterpart. A CNOT gate has a control qubit and a target qubit. Depending on the state of control qubit, the status of target qubit get flipped while the control qubit remaining same. The operator form of CNOT gate whose control is `a' and target is `b' (and vice versa) can be written as,
\begin{align}
CNOT_a &= \begin{pmatrix}
         1 & 0 & 0 & 0\\
         0 & 1 & 0 & 0\\
         0 & 0 & 0 & 1\\
         0 & 0 & 1 & 0\\
         \end{pmatrix};
         &
CNOT_b &= \begin{pmatrix}
         1 & 0 & 0 & 0\\
         0 & 0 & 0 & 1\\
         0 & 0 & 1 & 0\\
         0 & 1 & 0 & 0\\
         \end{pmatrix}; 
         &       
\end{align} 
CNOT can also be represented by binary addition of two qubits, i.e. CNOT$_a\vert a,b\rangle = \vert a,a\oplus b\rangle$ and CNOT$_b\vert a,b\rangle = \vert a\oplus b,b\rangle$. Here the symbol $\oplus$ represents the addition modulo 2, for which $0\oplus 0=0, 0\oplus 1=1, 1\oplus 0=1,$ and $1\oplus 1=1$. The application of CNOT gate on two qubit states has the following results,
\begin{eqnarray}
CNOT_a\vert00\rangle = \vert00\rangle,\;\;\;\;\;
CNOT_a\vert01\rangle = \vert01\rangle,\\
CNOT_a\vert10\rangle = \vert11\rangle,\;\;\;\;\;
CNOT_a\vert11\rangle = \vert10\rangle.
\end{eqnarray}
The circuit diagram of a CNOT gate is shown in Fig. \ref{cnotfig}. For a three qubit system, TOFOLLI gate is the universal gate which is nothing but a controlled-CNOT gate. 
\subsection{Quantum Algorithms}
Quantum algorithm solves problems by exploiting the properties of quantum mechanics. An efficient algorithm will require minimum resources. Many quantum algorithms are much more efficient than any classical algorithms by exploiting the features like superposition and entanglement. The first quantum algorithm was given by David Deutsch and is known as Deutsch algorithm. This algorithm is capable of finding out whether a binary function of one qubit is `constant' or `balanced' in one go \cite{deutsch85}. The most powerful quantum algorithm till date is the prime factorization given by Peter Shor. Shor's algorithm can factor a number by exponentially faster than its classical version. Grover's search algorithm can search an unsorted database in polynomially faster than classical algorithm.

Adiabatic quantum algorithm gains much attention due to its universality\cite{farhi}. In most cases a quantum algorithm begins with a uniform superposition and ends with an eigenstate which is the desired result. Often it is found that the ground state of the final Hamiltonian ($\mathcal{H}_f$) is the desired answer, however it is not easy to find the answer. Now suppose we have a Hamiltonian $\mathcal{H}_i$ whose ground state can easily be found. Hence by evolving the system adiabatically from $\mathcal{H}_i$ to $\mathcal{H}_f$, one can reach the ground state of $\mathcal{H}_f$ and hence the desired result. One has to make sure that there is no crossover of the ground state with any other state and the evolution process is slow enough that there won't be any possible transition. 

\subsection{Experimental implementations of QIP}
While there is a good amount of progress in the theoretical understanding of QIP, the physical realization of a quantum computer is proving extremely challenging. DiVincenzo laid out five criteria which must be fulfilled for a successful quantum computer architecture \cite{Div5}.
\begin{enumerate}
\itemsep-0.5em 
	\item Well defined qubits
	\item Ability to initialize
	\item Universal set of quantum gates
	\item Qubit-specific measurement
	\item Long coherence time
\end{enumerate}
Also there are two more criteria that will be needed for quantum communication. Meeting all these criteria in a single experimental setup is a highly challenging task. Nonetheless, various techniques have been proposed and is being explored for QIP tasks. All the techniques have there own advantages and some disadvantages. The major techniques available today are :
\begin{enumerate}
\itemsep-0.5em 
   \item Nuclear spins in NMR
   \item Trapped ions / atoms
   \item Photons
   \item Spins in semiconductor :\\[-.2cm]--- Quantum dots\\[-.2cm]--- NV centers in diamond
   \item Superconducting circuits  
\end{enumerate}
The first two techniques deal with the mutual interaction of quantum particles (atomic nucleus, atoms, ions) and controlled by electromagnetic field. Polarization of photons can be treated as qubits and it is controlled by optical means. Quantum dots and NV center in diamond techniques utilize the much developed semiconductor field in miniaturization scale. The well defined `phase' and `flux' parameters can serve as qubits in a superconducting circuits. Apart from these techniques, there are few more interesting techniques which might get much attention in future due to its hybrid approach. These methods are exploiting the best features among the available techniques and intend to make out a optimized experimental setup. For example, nuclear spins have much larger coherence time, but nuclear magnetization is very faint. By transferring magnetization from electrons to nuclei, the above problem can be solved and thus integrating the NMR with the ESR technique \cite{baugh11}. Another approach is integrating NMR with Atomic Force Microscopy (AFM) which is capable of measuring a single atom compare to bulk sample measurement done by NMR \cite{nmrafm}. A comparative study of all the key techniques is given in Table \ref{tab:exptech} \cite{nori2011, photon10, qcnature2010, photoncnot2008}.
\begin{sidewaystable}
\begin{tabular}{| p{3.5cm} | p{4cm} | p{3cm} | p{2.5cm} | p{3cm} | p{3cm} |}
\hline
& {\textbf{NMR}}& {\textbf{Trapped ions}}& {\textbf{Photons}}& {\textbf{Semiconductors}}& {\textbf{Superconductors}}\\[5mm] \hline \hline
\textbf{\textit{System}} & Nucleus & Atom & Photon & Atom, Vacancy & Phase, Flux, Charge \\[2mm] \hline
\textbf{Maximum Qubits demonstrated} & 12 (entangled) in liquids, >100 (correlated) in solids & 10-10$^3$ stored, 14 (entangled) & 10 (entangled) & 1 (QDs), 3(NV centers) & 128 (fabricated), 3 (entangled) \\[2mm] \hline

\textbf{Coherence time} & >1s (liquids), $\sim$100ms (solids) & >1s & $\sim 100\mu$s & 1-10 $\mu$s (QDs), 1-10 ms (NV) & $\sim 10\mu$s \\[2mm] \hline

\textbf{Two qubit gates (highest fidelity)} & CNOT (>99\%) & CNOT (>99\%) & CNOT (>94\%) & $\sim$90\% (NV centers) & >90\% \\[2mm] \hline

\textbf{Measurement} & Bulk magnetization & Fluorescence: `quantum jump' technique & Optical & Electric, optical & SQUID\\[2mm] \hline

\textbf{Controls} & RF pulses & Optical, MW, electrical & Optical & RF,  electrical, optical pulses & MW, voltages, currents \\[2mm] \hline
\end{tabular}
\caption{Comparison of main features for different available techniques in QIP}
\label{tab:exptech}
\end{sidewaystable}

\section{NMR QIP}
Application of NMR for the physical realization of QIP adds one more feather to the much colorful NMR application field. Implementation of QIP by NMR was independently proposed by Cory et al \cite{corypps} and Gershenfeld et al \cite{chuangpps} in 1997. Since the criteria laid by DiVincenzo was fulfilled more or less by NMR, it became an automatic choice whilst all other techniques were slowly coming up. One more thing fueled the NMR-QIP initiative was the fact that many of the QIP experimental basics are routinely done in conventional NMR experiments \cite{jones2000}. For example, the selective inversion of populations achieved in 1973 is described as a CNOT gate \cite{wessels}. The inversion of zero quantum coherence takes the name as SWAP gate \cite{kavita95, kavita00}. However, NMR-QIP gains much of its attention after Cory et al and Gershenfeld et al independently showed the preparation of `pseudopure state' in a liquid state NMR at room temperature. NMR-QIP in liquids containing small number of spins (preferably spin-1/2) have been studied extensively and its proven to be an excellent testbed for a small scale quantum information processor. Many complicated algorithms have been tested and verified. For example, Shor's factorizing algorithm has been tested till date only by liquid state NMR \cite{vandersypen}. However, scalability of liquid state NMR is an issue which hurdling the possibility of being an `useful' quantum information processor in long-run. It is unlikely to get more than 15-20 qubits unless some technological breakthrough occurs \cite{jones2000}. On the other hand Solid state NMR has the potential to become a reliable QIP architecture in future, since scalability issue and preparing `true' ground state seems more realistic. Some aspects of NMR-QIP are discussed in the following.\

\subsection{NMR- A suitable candidate for QIP}
A small scale NMR system is an attractive candidate for QIP for several reasons.
\begin{itemize}
	\item The fast reorientation of nuclear spins in liquids makes them fairly well isolated from
the environment and therefore provide a good source of qubits. Since they are weakly coupled with the environment, nuclear spins have long coherence times of the order of seconds.
   \item Nuclear spins can easily be manipulated and controlled using RF pulses. It is fairly simple to construct
quantum logic gates using RF pulses and evolution of couplings. 
    \item The development of NMR over half a century itself makes huge difference in developing and optimizing so many experiments. NMR is a perfect experimental tool for a quantum mechanical theorist.
    \item Modern NMR spectrometers are well developed. Though they are extremely sophisticated, they can be easily controlled.
\end{itemize}
\subsection{NMR Qubits}
A spin-1/2 nuclei ($^1$H, $^{13}$C, $^{19}$F, $^{31}$P) in an external magnetic field will be placed either `parallel' or `antiparallel' to the applied magnetic field. These two orthogonal states can be labeled as $\vert0\rangle$ and $\vert 1\rangle$ state of a qubit. A spin-1/2 nuclei is most preferable since it is naturally equivalent to a qubit. A multi qubit system should have the individual addressing capability and strong enough coupling constant with the farthest qubit. Since NMR is an ensemble system, it can address qubits separately depending on the slight variance in Larmor frequencies. Coupling is provided by J-interaction (in liquids) or dipole-dipole
interaction or both (in liquid crystals, solids). A stronger coupling constant is always welcome since it reduces the time duration of gates operation. Nuclei with spin> 1/2 has both advantages and disadvantages. Each nucleus can be treated as multiple qubits. Quadrupolar nuclei having spin $I=(2^N-1)/2$ (for $N\geq2$) will have $2^N$ states. Thus, a spin 3/2 nucleus has 4 states and can be treated as a 2-qubit system provided individual transitions are selectively addressable. This can be achieved by introducing first order quadrupolar coupling. Quadrupole systems partially oriented in liquid crystals can form an ideal multiqubit system. 

Another possibility is that, one can take advantage of higher number of `base' ($N\neq2$) into computation work. For example, a spin-1 system has three orthogonal states, which makes a `qutrit' system \cite{chuangbook}. 

\subsection{Initialization of NMR Qubits}
NMR is an ensemble system which deals with a large number ($10^{18}$) of identical spin-systems. At room temperature the nuclear energy levels are overwhelmed by the Boltzmann energy distribution. In order to achieve a state like $\vert000...0\rangle$, all the spins should be brought to the ground state which needs extremely low temperatures or extremely high magnetic fields. It is almost a impossible task to prepare a `pure' initial state at room temperature NMR. On the other hand for high resolution NMR, we need to have the systems to be in solution state. It is however pointed out by Cory et al \cite{corypps} and Chuang et al \cite{chuangpps} that the problem of preparing a pure initial state can be alleviated by preparing a pseudopure state which is a specially `mixed' state mimicking a pure state. We will have a thorough discussion on preparing pseudopure states by various methods in Chapter 3. 
\subsection{NMR Quantum Gates}
Single qubit gates can be thought of as a rotation in a Bloch sphere and can be implemented by the simple RF pulses \cite{cory98, moscajmr, review2002}. In NMR, it is convenient to understand the effect of RF pulses in rotating frame.\

\begin{figure}
	\centering
		\includegraphics[width=14cm]{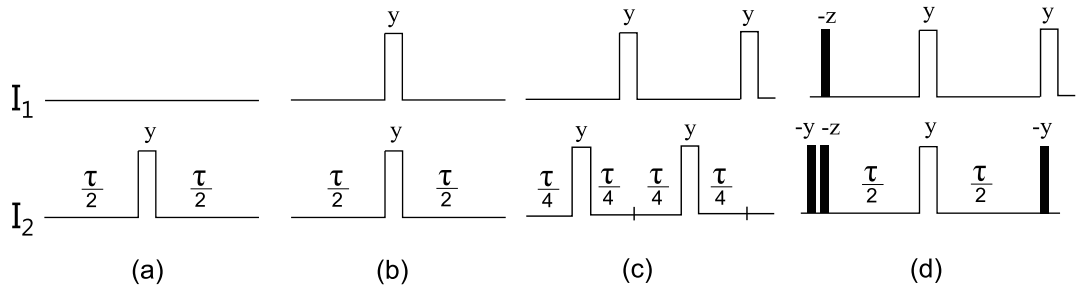}
	\caption[Switching ON and OFF of interactions for the first qubit]{Switching ON and OFF of interactions for the first qubit (I$_1$) over a time $\tau$. (a) Chemical shift ON, J-coupling OFF, (b) chemical shift OFF, J-coupling ON, (c) both chemical shift and J-coupling OFF. (d) CNOT gate ($\tau = 1/2J$). Filled rectangles indicates $\pi/2$ pulses and rest are representing a $\pi$ pulses with phase mentioning on each pulse.}
	\label{nmrgates1}
\end{figure}

The NOT gate as given in expression (\ref{notgate}) can be realized by a $\pi$ pulse.
\begin{equation}
\Big[\pi_x\Big] = e^{-i\pi I_x} = -i\begin{pmatrix} 0 & 1\\ 1 & 0 \end{pmatrix} = U_{NOT}.
\end{equation} 
The factor $-i$ can be ignored since it produces an undetectable global phase \cite{chuangbook}. Here the $\pi$ rotation is achieved by using a RF pulse of power $\omega_p$ for duration $\tau$ and phase $x$ such that $\omega_p\tau = \pi$. Similarly a Hadamard gate (see expression \ref{Hgate}) can be realized by applying two RF pulses.
\begin{equation}
\Big[(\pi/2)_y\:.\pi_x\Big] = e^{-i\pi I_x}\:e^{-i(\pi/2) I_y} = \frac{-i}{\sqrt 2}\begin{pmatrix} 1 & 1\\ 1 & -1 \end{pmatrix} = U_{H}.
\end{equation} 
A phase shift gate (Eq. \ref{Pgate}) is equivalent to rotation by an angle $\phi$ about z-axis and can be realized as follows.
\begin{eqnarray}
\Big[\phi_z\Big] &=& \Big[(\pi/2)_{-x}\:.\phi_y\:.(\pi/2)_x\Big] \\ 
                 &=& e^{-i(\pi/2)I_x}\:e^{-i\phi I_y}\:e^{i(\pi/2)I_x} = e^{-i\phi/2}\begin{pmatrix} 1 & 0\\ 0 & e^{i\phi} \end{pmatrix} = U_{P}(\phi).
\end{eqnarray}
Construction of multiqubit gates (e.g. CNOT) is achieved mainly by the proper exploitation of evolution of coupling and qubit specific RF pulses. The Hamiltonian for a pair of weakly coupled system in an isotropic medium is given by 
\begin{equation}
\mathcal{H}_{weak}= \omega_1I_z^1 + \omega_2I_z^2 + 2\pi JI_z^1I_z^2,
\label{Hweak}
\end{equation}
where $J$ is the coupling constant and $\omega_1$ and $\omega_2$ are Larmor frequencies of two spins. Now these J-coupling constant and Larmor frequencies are time independent fixed quantities and can not be turned off. But little tricks with the refocusing scheme can make it possible to overcome the problem. Historically most of the NMR experiments rely on this same refocusing technique. Consider at time $t = 0$, the density matrix of the system is given by $I_x^1 + I_x^2$, then The density matrix after time $t = \tau$ is given by  
\begin{eqnarray}
I_x^1 + I_x^2 \stackrel{\omega_1I_z^1 + \omega_2I_z^2}{\xrightarrow{\hspace*{1.5cm}}} &&I_x^1\cos(\omega_1\tau) + I_y^1\sin(\omega_1\tau) + I_x^2\cos(\omega_2\tau) + I_y^2\sin(\omega_2\tau)\\ \nonumber
\stackrel{2\pi JI_z^1I_z^2}{\xrightarrow{\hspace*{1.5cm}}} &&I_x^1\cos(\omega_1\tau)\cos{\pi J\tau} + 2I_y^1I_z^2\sin(\omega_1\tau)\sin{\pi J\tau}\\ \nonumber
&& + I_y^1\sin(\omega_1\tau)\cos{\pi J\tau} - 2I_x^1I_z^2\sin(\omega_1\tau)\sin{\pi J\tau}	\\ \nonumber
&& + I_x^2\cos(\omega_2\tau)\cos{\pi J\tau} + 2I_z^1I_y^2\sin(\omega_2\tau)\sin{\pi J\tau}	\\
&& + I_y^2\sin(\omega_2\tau)\cos{\pi J\tau} - 2I_z^1I_x^2\sin(\omega_2\tau)\sin{\pi J\tau}
\end{eqnarray}
This generalized calculations can be simplified for many practical situations. For example, a $\pi$ pulse on spin 2 at the middle of $\tau$ period refocuses the J-coupling as well as chemical shift evolution of the spin 2 (see Fig \ref{nmrgates1}a),
\begin{equation}
I_x^1 + I_x^2 \stackrel{[\frac{\tau}{2}-\pi^{(2)}-\frac{\tau}{2}]}{\xrightarrow{\hspace*{1.5cm}}} I_x^1\cos(\omega_1\tau) + I_y^1\sin(\omega_2\tau) + I_x^2.
\end{equation}
Similarly, a $\pi$ pulse applied on both the spins in the middle of $\tau$ period refocuses the chemical shift while retaining the J-coupling evolution (see Fig. \ref{nmrgates1}b)
\begin{equation}
I_x^1 + I_x^2 \stackrel{[\frac{\tau}{2}-\pi^{(1,2)}-\frac{\tau}{2}]}{\xrightarrow{\hspace*{2cm}}} I_x^1\cos(\pi J\tau) + 2I_y^1I_z^2\sin(\pi J\tau) + I_x^2\cos(\pi J\tau) + 2I_z^1I_y^2\sin(\pi J\tau).
\end{equation} 
Both J-coupling and chemical shift can be refocused over a time $\tau$ by the pulse program shown in Fig. \ref{nmrgates1}c.
\begin{equation}
I_x^1 + I_x^2 \stackrel{[\frac{\tau}{4}-\pi^{(2)}-\frac{\tau}{4}-\pi^{(1)}-\frac{\tau}{4}-\pi^{(2)}-\frac{\tau}{4}-\pi^{(1)}]}{\xrightarrow{\hspace*{4cm}}} I_x^1 + I_x^2
\end{equation}
A general method for refocusing has been described by Linden et. al. \cite{freeman99, freeman992}.
A CNOT gate can be achieved by the pulse sequence shown in Fig. \ref{nmrgates1}d.
\begin{equation}
\Bigg[ \left(\frac{\pi}{2}\right)_{-y}^{2}\;\left(\frac{\pi}{2}\right)_{-z}^{1,2}\;\frac{\tau}{2}\;\left({\pi}\right)_{y}^{1,2}\;\frac{\tau}{2}\;\left({\pi}\right)_{y}^{1}\;\left(\frac{\pi}{2}\right)_{-y}^{2}\;\Bigg] = -(1+i)\begin{pmatrix} 
1&0&0&0\\0&1&0&0\\0&0&0&1\\0&0&1&0\\
\end{pmatrix},
\end{equation}
where $\tau = \frac{1}{2J}$. Various gates and corresponding unitary operators has been studied in this context \cite{cory98, moscajmr, madi}.

\subsection{Numerically optimized quantum gates}
Apart from the above described methods of preparing quantum gates using ordinary `hard' RF pulses, there are more robust ways to synthesize any desired gate with high fidelity. These kind of pulses are often called `strongly modulated pulses' (SMPs). SMPs are made up with suitable sequences of RF pulses whose amplitude, frequency, and phase are made time-dependent in such a way that it produces the best sequence of pulses with maximum robustness against RF inhomogeneity \cite{fortunato, MaheshNGE, dieterrev}. There are few techniques available to find suitable SMPs for a given target operation \cite{fortunato, MaheshNGE, khaneja05}. Most of the techniques rely on the numerical optimization of the overall transformation by searching the available parameter space. Fortunato et al used the stochastic search methods to construct SMPs \cite{fortunato}, while Khaneja et al described the gradient ascent pulse engineering (GRAPE) technique \cite{khaneja05}. Designing an SMP for a given target operator reduces the numerical search problem to set the control parameters that maximizes the fidelity \cite{dieterrev}. In this thesis we have used SMPs in many cases in later chapters. Our SMPs are synthesized by the stochastic search methods powered by genetic search algorithm \cite{MaheshNGE}.  
  
\subsection{Measurement}
NMR being an ensemble technique, the measurement of a quantity (say D) is done by measuring its expectation value \cite{ErnstBook},
\begin{eqnarray}
s(t) \;\propto &&\sum_k{p^k(t)\langle\psi^k(t)\vert D\vert\psi^k(t)\rangle} \nonumber\\
&&= \sum_k{p_k(t)}\sum_{r,s}{c_r^{k*}(t)c_s^{k}(t)\langle r\vert D\vert s\rangle} \nonumber\\
&&= tr\{D\rho(t)\},
\end{eqnarray}
where $\rho$ is the density matrix and $D$ is the detection operator. The free evolution of spin system under Zeeman and coupling Hamiltonian is detected over a time scale. The detection period is normally decided by the relaxation rate and it is recorded in the time domain. By Fourier transform, the NMR spectra can be transformed into frequency domain.\

The quantum algorithms are designed such that the final output state is an eigenstate in the computational basis \cite{chuangbook}. In NMR, the computational basis is generally same as Zeeman basis (product basis). The eigenstates of product basis (diagonal elements of density matrix) correspond to the population distribution of a pseudopure state. For a two spin system, the general population distribution can be written as,
\begin{eqnarray}
\rho = c_1I_z^1 + c_2I_z^2 + c_32I_z^1I_z^2,
\end{eqnarray}
where $c_1, c_2, c_3$ are real constants. However, the traceless density matrix $\rho$ characterized by $\{c_1, c_2, c_3 \}$ does not give rise to any signal, since it corresponds to longitudinal magnetization. Applying a $\pi/2$ pulse results,
\begin{eqnarray}
\rho' = c_1I_x^1 + c_2I_x^2 + c_32I_x^1I_x^2. 
\end{eqnarray} 
From the above equation one can get the values of $c_1$ and $c_2$ but not $c_3$ since the last term does not contain any single quantum coherences. The signal received for the above applied $\pi/2$ pulse will be,
\begin{eqnarray}
s(t) &=& tr\left\{e^{-i\mathcal{H}t}\rho'e^{i\mathcal{H}t}\left( I_+^1+I_-^2\right) \right\}\\
     &=& tr\left\{e^{-i\mathcal{H}t}\left( c_1I_x^1 + c_2I_x^2\right)e^{i\mathcal{H}t}\left( I_+^1+I_-^2\right)\right\}
\end{eqnarray}
where the $c_3$ term won't produce any signal since `multiple quantum' under such a evolution remains `multiple quantum'. Here $\left(I_+^1+I_-^2\right)$ represents the detection operator and $\mathcal{H}$ is the evolution Hamiltonian as given in Eq. \ref{Hweak}. 

The complete procedure of characterizing a quantum state is through density matrix tomography. By this method all the coherence orders can be measured by a series of experiments. A detailed analysis of density matrix tomography is given in Appendix A and B and its explicit application in practical situations is given in later part of Chapter 2. 
\subsection{Coherence order}
Decoherence time in NMR quantum computers is generally related to the spin-spin relaxation time ($T_2$), although this is a simplification \cite{jones2000}. $T_2$ generally represents the decoherence time of a single spin coherences and the decoherence times for a multiqubit systems can be quite different. However $T_2$ gives a rough estimation of the decoherence time of the system and in most of the cases it is of the order of multiple seconds in a liquid state system. The decoherence time in NMR is one of the best among all the available QC techniques. The coherence time can be prolonged by applying specific dynamical decoupling sequences. A thorough discussions given in Chapter 4.    
\subsection{Limitations of NMR-QIP}
There are multiple issues which posses serious challenge to NMR-QIP tasks.
\begin{enumerate}
	\item Scalability of liquid state NMR is a big challenge that we need to be overcome. Indeed, going beyond 10 qubits is a very difficult task. High resolution liquid state NMR-QIP relies on weakly coupled systems. For a large order qubit system (say 10), the J-coupling constants between two farthest spins are very small. Lower coupling constant means lack of well resolved spectra and more time consuming `quantum gates'. This problem can partly be resolved by taking partially orienting spin systems in liquid crystal solvents and thus introducing dipolar couplings along with J-couplings. Solid state NMR qubits has the potential for becoming the scalable qubits. However, at the moment solid state NMR system produces complicated spectra and allows lesser controlling technique.	
	\item Lack of creating a `pure' state is another limitation in NMR-QIP. In liquid state NMR, spins are in a highly mixed initial state and preparation of a `pure' state needs extreme experimental conditions. However, one can prepare a `pseudopure' state (PPS) which mimics a `pure' state and effectively able to perform as a `pure' state. PPS can be written as,
	\begin{eqnarray}
	\rho_{pps} = \frac{1}{2^N}\left( 1-\epsilon\right)\mathbbm{1} + \epsilon\vert\psi\rangle\langle\psi\vert,
	\end{eqnarray}
	where $\epsilon$ denoting the purity factor and at high temperature limit it can be written as,
	\begin{eqnarray}
		\epsilon\sim \frac{Nh\nu}{2^NkT},
	\end{eqnarray}
  $\epsilon$ is of the order of $10^{-5}$ under normal conditions. Hence it can be seen that the amount of magnetization (signal) decreases exponentially with the number of qubits. However for 15-20 qubits, the magnetization may not be enough to carry out QIP tasks. Many ideas come forward to tackle this issue. Carrying out quantum computation in mixed state is one of them \cite{vazirani98}. `One clean qubit' protocol needs only one `pure' qubit having rest of them in `mixed' state \cite{lafllamme98}.
	\item One of the major requirement to perform some algorithm is to create an entangled state. Many believe that the power of quantum computers is largely depend on its entangling phenomena. But it is proved by Braunstein et. al. that, small scale liquid state system at normal conditions lacks any kind of `entanglement'. But still NMR is the only technique which implements the Shor's factoring algorithm till date and this algorithm needs an entangling state. These two contradictory aspects, led many people to think that `entanglement' might not the necessary condition to be fulfilled to perform QIP tasks. A new measure of calculating non-classical correlations known as `discord' is introduced recently \cite{vedral, ollivier, katiyar}. 
	\item Crowding of frequency space is another challenge as the number of qubits increased. A weakly coupled spin system of N spin-1/2 nuclei gives rise to $N2^{N-1}$ resonance lines. Thus, a 10 qubit system would have 10 sets of 512 lines. Resolving spectra is a daunting task for this kind of situation. This can partly be resolved by synthesizing special molecules and using sophisticated spectrometers.	
	\item Decoherence can be a potential problem for large scale qubit and also for solid state qubits. However, the relevant parameter is not the decoherence time itself but the ratio of decoherence time to the gate-time.    
\end{enumerate}

\section{NMR - An ideal platform for studying quantum mechanical phenomena} 
Although `nature prefers quantum mechanics' and it is the most fundamental physics, capable of explaining almost everything starting from photosynthesis effect to blackhole formation, it is rather a challenging job to `feel' it in our daily life \cite{natphoto, sciphotolee, pnasphoto, natQbio, natdrum}. The lack of experimental proof of quantum mechanical phenomena in the early days of its introduction led many eminent scientists commenting on its `utility' and `completeness' \cite{einstein, qdebate}. Due to its very nature, even today it is highly challenging to perform quantum mechanical experiments in laboratory \cite{qdebate}.

The nuclear spins in an NMR provides an excellent test bed for performing various kind of quantum mechanical experiments in a highly controlled way \cite{ryanpra}. The principles of an NMR can only be fully understood by quantum theory \cite{Abragam}. In return, NMR can be used as a prototype quantum mechanical testbed. The outcome of the quantum mechanical probabilistic calculations mostly produces counter intuitive results which are difficult to `digest'. Nonetheless, NMR has proven to be one of the leading architecture in performing various quantum mechanical phenomena experimentally \cite{soumyaqdc, soumyalgi}. 

In this thesis we have shown some important experimental implementations of quantum mechanical phenomena which earlier thought to be intractable. There are various examples where a quantum mechanical phenomenon does not have a classical analogue and in this kind of situation it is rather difficult to `understand' it \cite{soumyaeit}. For example, quantum contextuality is a kind of quantum mechanical phenomenon which has been proved by various quantum platform including NMR recently \cite{moussa}. The experimental results clearly shows quantum mechanically expected values which are counter intuitive to our macrorealistic world \cite{moussa}.

Quantum objects behave differently than a macrorealistic object and there are certain inequalities (e.g. Bell's inequality) which can only be violated by quantum objects \cite{bell, leggettgarg}. To prove this kind of violation one needs to have an excellent quantum platform. In Chapter 5, we have shown the violation of Leggett-Garg inequality for nuclear spins as predicted by quantum theory \cite{soumyalgi}.

Bohr's complementary principle is another famous description regarding quantum mechanical objects. A consequence of the complementary principle is that one can not observe both `wave' and `particle' nature of a quantum object simultaneously \cite{bohr}. However, recently it was proposed that, by using certain special experimental setups, one can simultaneously observe wave and particle nature of a quantum system \cite{ionicioiu}. This requires a reinterpretation of Bohr's complementary principle. In Chapter 6, we have shown a detailed experimental study of this new experimental proposal and our results clearly suggest that there is indeed a necessity of revisiting Bohr's principle \cite{soumyaqdc}.

Many new fields related to experimental quantum mechanics are coming up due to the fact that now we have some excellent quantum platform which are capable of carrying out experimental work in a highly controlled way. Quantum chemistry and quantum biology are such two emerging fields which are making lot of progresses \cite{natQbio, Qchem}. Understanding all these phenomena experimentally is vital in pursuit of understanding quantum mechanics and its practical applications at large.

\thispagestyle{empty}
\chapter{Density Matrix Tomography of Long Lived Singlet States}

The lifetime of nuclear singlet states can be much longer than any other non-equilibrium states under suitable conditions. In section 2.2, we introduced long-lived singlet (LLS) states and it's preparation by standard methods. In section 2.3, we introduced a robust density matrix tomography scheme which is particularly suited to study homonuclear spin systems with small chemical shift differences. In section 2.4, we have applied the tomography scheme to characterize the singlet states under various experimental conditions, revealing interesting features of LLS. This chapter ends with a conclusion given in section 2.5.

\section{Introduction}
The long lifetimes of nuclear spin coherence enables NMR spectroscopists to carry out
a variety of spin choreography \cite{LevBook, ErnstBook}. Nuclear spin coherences decay over time
mainly due to spin-spin relaxation and magnetic field inhomogeneity.
Often, coherences are converted into longitudinal nuclear spin orders to study 
slow dynamical processes.  But even the longitudinal spin orders
decay toward equilibrium state due to spin-lattice relaxation.  Hence 
for a typical NMR experiment consisting of preparing and measuring certain correlated 
spin states, the ultimate time barrier was assumed to be defined by the spin-lattice 
relaxation time constant $T_1$ \cite{Abragam}.

It has recently been demonstrated that there exist certain `long-lived states'
which decay slower than the $T_1$ values of individual spins 
\cite{LevPRL04,LevittJACS04,LevittJCP05,BodenJMR06,LevJMR06,LevJMR07,GrantJMR07,LevittJACS08}.
The long lived singlet states (LLS) has found wide range of applications ever since it 
was discovered by Carravetta, Johannessen, and Levitt in 2004 \cite{LevPRL04}.
Overcoming the $T_1$ barrier has led to several exciting applications in
studying slow molecular dynamics and transport processes \cite{BodenJACS05,BodenJACS07}, 
precise measurements of NMR interactions \cite{LevPRL09}, and the 
transport and storage of hyperpolarized nuclear spin orders \cite{GrantJMR08,BargonJCP06,WeitekampJACS08,BodenPNAS09,AdamsScience09,WarrenScience09}.  

Bodenhausen and co-workers have demonstrated that the singlet spin-lock can also be achieved by  
RF modulations which are used in heteronuclear spin-decoupling \cite{BodenCPC07}.  Detailed theoretical
analysis of zero-field singlet states as well as singlet spin-lock have already been
provided by Levitt and co-workers \cite{LevittJCP05,LevJCP09} and by 
Karthik et al \cite{BodenJMR06}.  Recently, long-lived states in
multiple-spin systems are also being explored \cite{GrantJMR08,BodenCPC09}.

\section{Long-lived singlet states}
   Let us begin with a simplest model consisting of a pair of spin-1/2 nuclei. These two spins are labeled as $I^{1}$ and $I^{2}$. The free-precession Hamiltonian of this system at laboratory frame can be written as,  
\begin{equation}
   \mathcal{H} = \omega_{1}I_{z}^{1}+\omega_{2}I_{z}^{2}+2\pi J_{12}I^{1}\cdot I^{2},
\end{equation}
where $\omega_{1}$ and $\omega_{2}$ are denoting the resonant frequency of the two spins respectively, whereas $J_{12}$ denotes the spin-spin coupling (J-coupling) between the two spins.

The quantum states of the system can always be expressed with the combination of superposition of Zeeman states, namely $|00\rangle$, $|01\rangle$, $|10\rangle$, $|11\rangle$. Here $|0\rangle$ denotes the angular momentum of $\hbar/2$ along the magnetic field direction (`up' direction) and $|1\rangle$ denotes the angular momentum of $-\hbar/2$ along the exact opposite direction of the magnetic field (`down' direction). 
\ The four Zeeman product states together lead to one singlet state and three triplet states,
\begin{equation}
\begin{aligned}
&|S_{0}\rangle = \frac{|01\rangle - |10\rangle}{\sqrt{2}} ,\\
&|T_{+1}\rangle = |00\rangle ,\\
&|T_{0}\rangle = \frac{|01\rangle + |10\rangle}{\sqrt{2}} ,\\
&|T_{-1}\rangle = |11\rangle .\\
\end{aligned}
\end{equation}
  Singlet states have many different properties compared to its triplet counterparts. Two most important properties are:
  
(a) Singlet state is anti-symmetric with respect to spin-exchange, whereas triplet states are symmetric.

(b) Singlet state has a zero total nuclear spin angular momentum quantum number [$I^{2}|S_{0}\rangle = 0$], whereas triplet states have non-zero total nuclear spin angular momentum quantum number [$I^{2}|T_{M}\rangle = 2|T_{M}\rangle$].\ 

In the case of magnetically equivalent nuclear pair, the singlet state and the triplet states form an orthonormal eigenbasis of the internal Hamiltonian $\mathcal{H}_J = 2\pi J I^{1}\cdot I^{2}$. 
Singlet states can be prepared between two assymetric spins by imposing equivalence condition (either by lifting the sample out of Zeeman field or by applying suitable RF field acting as `spin-lock'). But, being a zero-quantum coherence, singlet state itself is inaccessible to macroscopic observable directly. The traditional methods by Caravetta et al \cite{LevPRL04}, described the way to access the singlet states by breaking its magnetic symmetry to convert into observable single quantum coherences. In this context we may note that, protons in Hydrogen molecule or in water is already in magnetic equivalence, but there is still no way to break the symmetry.

\subsection{Why singlet state is long lived ?}
     Any quantum state, deviating from its thermal equilibrium conditions, will return to its stable thermal equilibrium state through a mechanism known as relaxation. Hence it is needless to say that in NMR any observable quantum state is in non-equilibrium condition and that is the reason each state has its own lifetime. There are two major factors behind relaxation, (i) spin-lattice relaxation ($T_{1}$) and (ii) spin-spin relaxation ($T_{2}$). In majority of the cases $T_{2}$ relaxation is much faster than $T_{1}$. So the upper limit of the nuclear spin memory is bounded by the $T_{1}$ irrespective of any experimental safe guard. However, there are some specialized cases where exceptions can be found, such as in the case of `parahydrogen', where the spin state isomers lived much longer than $T_{1}$ \cite{mcdtpara}. Though the major reasons behind $T_{1}$, $T_{2}$ relaxation depend on individual molecular property, other controllable parameters such as magnetic field inhomogeneity, temperature fluctuations, sample concentration etc. also contributes to the relaxation mechanism.\

Levitt and co-workers have successfully demonstrated \cite{LevPRL04, LevittJACS04} that the singlet state lifetime can be made many orders of magnitude longer than $T_{1}$ for `ordinary' molecules in solution state of homonuclear system. Now we will discuss some physics behind this astonishing long-lifetime of singlet states  \cite{llsencyclo}. The Hamiltonian for a pair of spins in magnetically equivalent environment is written as bellow:
\begin{equation}
   \mathcal{H} = \omega_{0}\left(I_z^1 + I_z^2\right) + 2\pi J I^{1}\cdot I^{2},
\end{equation} 
where, $\omega_{0} = \gamma B_{0}$ denotes the Larmor frequency of both (equivalent) the spins and $B_{0}$ is the applied static magnetic field. The matrix representation of the Hamiltonian in singlet-triplet basis can be written \cite{llsencyclo} as follows: 
\begin{equation}
\mathcal{H} = \bordermatrix{~ & |S_{0}\rangle & |T_{+1}\rangle & |T_{0}\rangle & |T_{-1}\rangle \cr
                  \langle S_{0}| & -\frac{3}{2}\pi J & 0 & 0 & 0 \cr
                  \langle T_{+1}| & 0 & \omega_{0} + \frac{1}{2}\pi J & 0 & 0 \cr
                  \langle T_{0}| & 0 & 0 & \frac{1}{2}\pi J & 0 \cr
                  \langle T_{-1}| & 0 & 0 & 0 & -\omega_{0} + \frac{1}{2}\pi J \cr}.
\end{equation}

From the earlier equation it is seen that at zero field ($\omega_{0}$ = 0), the triplet states are degenerate with same energy eigenvalues ($\frac{1}{2}\pi J$). The energy difference between the singlet and the triplet sates is $2\pi J$ which is independent of the field. 
Since the Hamiltonian is diagonal, there will not be any mix-up of singlet state population with triplet states' populations \cite{llsencyclo}. However, triplet states among themselves equilibrate quickly. Eventually there will be a singlet-triplet transition resulting in the re-establishment of thermal equilibrium much slower than $T_{1}$ relaxation time scale. The time constant for singlet-triplet equilibration is loosely termed as `singlet lifetime' ($T_{S}$) \cite{llsencyclo}.

We already know that singlet states are `antisymmetric' with respect to spin exchange, whereas triplet states are `symmetric' with respect to the spin exchanges. The major relaxation processes, including intra-molecular dipolar relaxation mechanism, are `symmetry preserving' in nature. Hence these relaxation mechanisms will not affect the singlet-triplet conversion which requires symmetry transformations. These conditions make singlet states as a `special' state which is immune to intra-molecular dipolar relaxation, though it is the major reason behind $T_{1}$ relaxation \cite{llsencyclo}.

Previous discussion shows how necessary it is to get a magnetically equivalent pair of nuclear spins to realize the LLS. We need to create such a magnetically equivalent condition to create and persist in singlet states, but to get signal out of singlet states we need to break the symmetry. In the next paragraphs we will discuss about the techniques for magnetically inequivalent pair of nuclear spins.
\ The Hamiltonian for a pair of chemically inequivalent nuclei in present of Zeeman field can be written as follows:
\begin{eqnarray}
\mathcal{H} &=& \omega_{1} I_{z}^{1} + \omega_{2} I_{z}^{2} + 2 \pi J I^{1}\cdot I^{2}\nonumber\\
            &=& \omega_{0} (1+\delta_{1}) I_{z}^{1} + \omega_{0} (1+\delta_{2}) I_{z}^{2} + 2 \pi J I^{1}\cdot I^{2},
\end{eqnarray}
where $\delta_{1}$ and $\delta_{2}$ are the two chemical shifts of the two spins. The matrix representation of this Hamiltonian in singlet-triplet basis can be expressed as \cite{llsencyclo}:
\begin{equation}
\mathcal{H} = \bordermatrix{~ & |S_{0}\rangle & |T_{+1}\rangle & |T_{0}\rangle & |T_{-1}\rangle \cr
                  \langle S_{0}| & -\frac{3}{2}\pi J & 0 & \frac{1}{2}\omega_{0}\Delta\delta & 0 \cr
                  \langle T_{+1}| & 0 & \omega_{0}(1+\frac{1}{2}\sum{\delta}) + \frac{1}{2}\pi J & 0 & 0 \cr
                  \langle T_{0}| & \frac{1}{2}\omega_{0}\Delta\delta & 0 & \frac{1}{2}\pi J & 0 \cr
                  \langle T_{-1}| & 0 & 0 & 0 & -\omega_{0}(1+\frac{1}{2}\sum{\delta}) + \frac{1}{2}\pi J \cr},
\end{equation}

where, 
\begin{equation}
\begin{aligned}
& \sum{\delta} = \delta_{1} + \delta_{2} , 
& \Delta\delta = \delta_{1} - \delta_{2}.
\end{aligned}
\end{equation}

   In this case, the matrix is not a diagonal matrix, hence the singlet and triplet states are not the eigenstates of this Hamiltonian. The off-diagonal term in the matrix ($\frac{1}{2}\omega_{0}\Delta\delta$) represents the possible conversion of singlet-triplet transition. This transition is directly dependent on the chemical shift difference between the two spins. Hence, even if we are able to prepare singlet states in an inequivalent pair of nuclei, it will not be long lived till it has some dependency on the chemical shift differences. Still, it gives us a clue to experience long-lived singlet states if somehow the chemical shift difference ($\Delta\delta$) is suppressed \cite{llsencyclo}. In the next subsection we will discuss this method of chemical shift suppression in detail.

\subsection{Singlet Preparation in NMR}
   So far we have learn that singlet states can not be observed for magnetically equivalent pair of spins, as it does not give any observable NMR signal, and even for the magnetically inequivalent spin pairs because of the chemical shift difference barrier.
   
   The key to LLS revelation is to switch the magnetic equivalence `on' and `off' by some experimental manipulations \cite{llsencyclo}. There are at least two well established procedures to do so - (i) field cycling and (ii) radiofrequency spin-locking \cite{LevPRL04, LevittJACS04}. By field cycling method, we can switch-off the magnetic field manually so that magnetic equivalence is established and then once again switch-on the magnetic field to convert into single quantum coherences. The other method (radiofrequency spin-locking) is more practical with least manual work. We will discuss this method in detail. 
   
   Getting pure singlet states may be seen as a three step procedure:
\begin{enumerate}[(i)]
\itemsep-1.5em 
	\item Building singlet population.\\
  \item Applying spin-lock.\\
  \item Singlet detection.  
\end{enumerate}
 
\subsubsection{Building singlet population}
   With the application of suitable \textit{rf} pulses and delays it is possible to create a density matrix operator which represents a part of singlet states in it. The density matrix for a singlet population can be represented by the Cartesian product operator formalism as follows:   
\begin{equation}
\begin{aligned}
|S_0\rangle\langle S_0| 
&= \frac{1}{2}\left(|01\rangle-|10\rangle)(\langle01|-\langle10|\right)\\
&=\frac{1}{2}\left(|01\rangle\langle01|-|10\rangle\langle01|-|01\rangle\langle10|+|10\rangle\langle10|\right)\\
&= \frac{1}{2}\left(I_{\vert0\rangle}^{1}I_{\vert1\rangle}^{2}-I_+^{1}I_-^{2}-I_-^{1}I_+^{2}+I_{\vert1\rangle}^{1}I_{\vert0\rangle}^{2}\right)\\
&= -\frac{1}{2}\left(I_+^{1}I_-^{2}+I_-^{1}I_+^{2}\right)-I_z^{1}I_z^{2} + \frac{1}{4}\mathbbm{1}.
\end{aligned}
\end{equation}

  The earlier equation shows that singlet populations can be constructed from zero quantum coherences and longitudinal magnetization of both the spins. Hence a little trick with the excitation of zero quantum coherences with appropriate phase can leads us to the singlet populations \cite{llsencyclo}. The following pulse sequence is found \cite{LevPRL04} suitable to create singlet state populations starting from equilibrium condition. 
\begin{equation}
90_0 - \tau_1 - 180_0 - (\tau_1 +\tau_2) -90_{90} - \tau_3,
\end{equation}\\
where $\tau_{1} = 1/{4J}$, $\tau_{2} = 1/{2 \Delta\nu}$ and $\tau_{3} = 1/{4 \Delta\nu}$. J and $\nu$ are denoting the spin-spin coupling constant and chemical shift difference in Hz respectively. The `offset' should be placed at the middle of the two spins for simplification. The above written pulse sequence works as follows :

Initial $90_0$ pulse brings the longitudinal magnetization to transverse plane.
\begin{center}
\begin{math}
I_z^1 + I_z^2  \stackrel{90_0}{\xrightarrow{\hspace*{1cm}}} -I_y^1 - I_y^2,
\end{math}
\end{center} 
followed by the spin-echo with only J evolution for the duration of 1/{2J} :
\begin{center}
\begin{math}
-I_y^1 - I_y^2  \stackrel{\tau_{1}-180_{0}-\tau_{1}}{\xrightarrow{\hspace*{1.5cm}}} 2I_{x}^{1}I_{z}^{2} + 2I_{z}^{1}I_{x}^{2} .
\end{math}
\end{center}
During the subsequent $\tau_2$ interval, there will be evolution under the isotropic chemical shifts. This delay ($\tau_{2} = 1/{2 \Delta\nu}$) is relatively shorter and can be ignored for any significant $J$-evolution during this time. The product operator formalism goes as follows:
\begin{center}
\begin{math}
2I_{x}^{1}I_{z}^{2} + 2I_{z}^{1}I_{x}^{2}  \stackrel{\tau_2}{\xrightarrow{\hspace*{1cm}}} 2I_{y}^{1}I_{z}^{2} - 2I_{z}^{1}I_{y}^{2}.
\end{math}
\end{center}
Now a 90 degree y pulse will bring the density operator into zero quantum coherences.
\begin{center}
\begin{math}
2I_{y}^{1}I_{z}^{2} - 2I_{z}^{1}I_{y}^{2}  \stackrel{90_{90}}{\xrightarrow{\hspace*{1cm}}} 2I_{y}^{1}I_{x}^{2} - 2I_{x}^{1}I_{y}^{2} = -i(I_{+}^{1}I_{-}^{2} - I_{-}^{1}I_{+}^{2}).
\end{math}
\end{center}
A further chemical shift evolution required for a phase corrected zero quantum coherence.
\begin{center}
\begin{math}
2I_{y}^{1}I_{x}^{2} - 2I_{x}^{1}I_{y}^{2}  \stackrel{\tau_3}{\xrightarrow{\hspace*{1cm}}} -2I_{x}^{1}I_{x}^{2} - 2I_{y}^{1}I_{y}^{2} = -(I_{+}^{1}I_{-}^{2} + I_{-}^{1}I_{+}^{2}).
\end{math}
\end{center}
This may be rewritten as follows:
\begin{equation}
\begin{aligned}
-I_{+}^{1} I_{-}^{2} - I_{-}^{1}I_{+}^{2} 
& = -|01\rangle\langle10| - |10\rangle\langle01| \\
& = |S_{0}\rangle\langle S_{0}| - |T_{0}\rangle\langle T_{0}|.
\end{aligned}
\end{equation}

   Hence from the above calculations it is seen that the resulting density operator is in fact combination of the singlet state and one of the triplet states' population. Now our aim is to filter out the singlet state from the singlet-triplet population distribution. This can be done by radio frequency spin locking as discussed below in detail.
   
\subsubsection{Radio frequency spin-lock}
  A spin-lock is a low power on-resonant continuous radio frequency pulse along the spin magnetization in transverse plane. This low frequency \textit{rf} pulse keeps the magnetization from precessing in transverse plane. Hence this pulse can be used as a possible way to suppress the chemical shift differences. It is popularly known as a `spin-lock' as it arrests the spin precession. 

 The duration of the spin-lock may last for several minutes, triggering the possibility of severe probe damage. Hence one must be careful to select the \textit{rf} spin-lock power and duration. There are two basic kinds of spin-lock. (i) Unmodulated spin-lock, and (ii) modulated spin-lock.
 
     (i) Unmodulated \textit{rf} field is commonly known as `continuous wave' (CW) irradiation. CW irradiation has constant amplitude and has no phase modulation over time. CW has shorter bandwidth and hence not useful for large chemical shift differences. Theoretically it is possible to apply more power for higher chemical shift difference systems, but that may cause serious damage to \textit{rf} probes. 
     
     (ii) Modulated lock can be realized by using CPD (composite pulse decoupling) pulses. As the name suggests, it is a phase modulated composite pulse, routinely used as a decoupling pulse sequence. In many cases it can outperform CW pulses as a spin-lock sequence. The bandwidth of CPD pulses are much larger compared to CW pulses and hence useful for larger chemical shift difference singlets. Commonly used CPD pulses are WALTZ-16, GARP etc.
     
     During spin-lock the three triplet states' populations equilibrate rapidly under normal relaxation procedure, whereas singlet population being itself immune to \textit{rf} spin-lock, decays much slowly. After the fast decay of triplet states, singlet state achieve its maximum purity (singlet correlation may reach upto 0.997). Eventually singlet state also decays despite \textit{rf} spin-lock shielding, but with much slower rate than any other states.
     
     Now here we can recap the fact that singlet state itself is a zero quantum coherences and can not be directly accessible. Hence we must transfer the zero-quantum coherences to a observable single-quantum coherences to detect it. The following section describes the method in detail.
     
\subsubsection{Singlet detection}
     The simplest method to detect singlet is to evolve it for a $1/(4\Delta\nu)$ chemical shift evolution and followed by a strong 90$^{\circ}$ pulse. The transformation of density matrix operator are as follows:
\begin{eqnarray}
\centering
|S_{0}\rangle\langle S_{0}|  = &-\frac{1}{2}(I_{+}^{1} I_{-}^{2} + I_{-}^{1}I_{+}^{2}) - I_{z}^{1}I_{z}^{2} +\frac{1}{4}\mathbbm{1}& \nonumber\\
&{{\downarrow} \tau_3}& \nonumber \\
&-\frac{1}{2i}(I_{+}^{1} I_{-}^{2} - I_{-}^{1}I_{+}^{2}) - I_{z}^{1}I_{z}^{2} +\frac{1}{4}\mathbbm{1}.\nonumber
\end{eqnarray}

  This can also be written in terms of Cartesian product operator formalism:
\begin{center}
\begin{math}
-\frac{1}{2i}(I_{+}^{1} I_{-}^{2} - I_{-}^{1}I_{+}^{2}) - I_{z}^{1}I_{z}^{2} +\frac{1}{4} = \frac{1}{2}(2I_{x}^{1}I_{y}^{2} + I_{y}^{1}I_{x}^{2}) - I_{z}^{1}I_{z}^{2} + \frac{1}{4}.
\end{math}
\end{center}
  
  Now a simple $90_0$ pulse brings the magnetization into observable single quantum coherence.
  
\begin{equation}
\frac{1}{2}(2I_{x}^{1}I_{y}^{2} + I_{y}^{1}I_{x}^{2}) - I_{z}^{1}I_{z}^{2} + \frac{1}{4}\mathbbm{1} \stackrel{90_0}{\xrightarrow{\hspace*{1cm}}} \frac{1}{2}(2I_{x}^{1}I_{z}^{2} - I_{z}^{1}I_{x}^{2}) - I_{y}^{1}I_{y}^{2} + \frac{1}{4}\mathbbm{1}.
\end{equation}

These shows the antiphase transverse magnetization for the spin pair. The characteristic spectra for this kind of antiphase magnetization shows a typical ``up-down-down-up'' pattern in the NMR peaks.

   However one might notice that this way of detecting singlet states has less qualitative information. A better quantitative study can be carried out through tomographic method. In this context we have developed a robust density matrix tomographic technique which is particularly suitable for this problem. In the following section we will discuss the `density matrix tomography' scheme in detail. Later we will apply this tomography sequence on singlets for its characterization in various experimental conditions.
   
\section{Density Matrix Tomography}
    The delicate nature of `quantum states' makes it vulnerable to macroscopic-world. The inevitable last step for most of the quantum information processing and quantum simulation is the measurement of derived quantum states. In the case of an ensemble quantum system, the states are presented by density matrix. In order to measure these density matrices, many sophisticated schemes have been envisaged. `Density matrix tomography' technique has proven its utility for mapping any quantum states with high accuracy. It enables us to measure all the elements of a general density matrix at any time point. The knowledge of the full density matrix of any quantum state is important for many reason e.g. (i) one can find the error in the experiment, since we already have the knowledge about theoretical density matrix. (ii) Measuring density matrix at different time points of a dynamic quantum algorithm gives the pattern of population and coherence transfers. In the following section we have presented a robust tomographic scheme in the context of NMR \cite{maheshjmr10}.

Earlier schemes of tomography
were designed in the context of quantum information processing \cite{ChuangPRSL98,ChuangPRA99}.  
They required spin-selective rotations and transition selective integrations of spectra.  
In homonuclear spin systems, particularly in $^1$H spin systems, it is hard to design high
fidelity spin-selective rotations owing to the small differences in chemical shifts  
(on the other hand, the heteronuclear singlet state is predicted to be short-lived
\cite{LevJCP09}).
These spin selective pulses generally tend to be long in duration, still introduce significant
errors.  Integration of individual transitions is also problematic since the
transitions, particularly those with mixed line shapes corresponding to a general density matrix, 
may severely overlap.  Tomography based on two-dimensional NMR spectroscopy had also been 
proposed \cite{AnilPRA03}.  This is a general method in the sense only one 
2D experiment is needed to be carried out irrespective of the size of the spin system.
However, the 2D method is time consuming.  Also since it
relies on fitting the 2D cross-sections (along the indirect dimension) 
to mixed Lorentzian, the accuracy is limited by the quality of the fit that
is achieved.  In the following we present a robust density matrix tomography for a homonuclear 
weakly coupled two spins-1/2 system which needs only non-selective RF pulses and integrations over each spin
instead of individual transitions \cite{maheshjmr10}.\ 

The general traceless deviation density matrix
consists of 15 independent real numbers:
\begin{eqnarray}
\rho = 
\left(
\begin{array}{cccc}
  p_0 \;\;\;\; &  r_3 + i s_3  & r_1 + i s_1 &  r_5 + i s_5  \\
  & p_1 &  r_6 + i s_6  &  r_2 + i s_2  \\
 & & p_2 &  r_4 + i s_4  \\
 & & & -\sum_{i=0}^2p_i \\
\end{array}
\right).
\end{eqnarray}
Here real elements $p_k$ are populations and the complex elements $r_k+i s_k$
correspond to single ($k = 1 \mathrm{\;to\;} 4$), double ($k = 5$), and zero ($k = 6$)
quantum coherences.  
The elements below the diagonal are determined by the 
Hermitian condition $\rho_{jk} = \rho_{kj}^*$.
Since only single quantum coherences are directly
observable, four combinations $R_1 \coloneqq (r_1+r_2)$, $S_1 \coloneqq (s_1+s_2)$, 
$R_2 \coloneqq (r_3+r_4)$, and $S_2 \coloneqq (s_3+s_4)$ can be obtained from the integration of complex
line shapes of spins 1 and 2 respectively.
Now consider an RF sequence with propagator $U$, that 
transforms the original density matrix $\rho$ into $\rho' = U \rho U^\dagger$. 
Single quantum coherences of $\rho'$ will lead to different linear combinations 
of various elements in $\rho$.  Thus, by applying different propagators on $\rho$,
we can measure the values of different linear combinations of various elements of $\rho$.
The real and imaginary values of the integration of $j$th spin in $k$th experiment
will be labeled as $R_j^k$ and $S_j^k$ respectively.  Following six 
one-dimensional NMR experiments were found to be sufficient to tomograph a two-spin 
density matrix:
\begin{itemize}
\item[1.]{$\mathbbm{1}$
}
\item[2.]{$90_x$
}
\item[3.]{$\frac{1}{4J} \cdot 180_x \cdot \frac{1}{4J}$
}
\item[4.]{$45_x \frac{1}{4J} \cdot 180_x \cdot \frac{1}{4J}$
}
\item[5.]{$45_y \frac{1}{4J} \cdot 180_x \cdot \frac{1}{4J}$
}
\item[6.]{$\frac{1}{2\Delta \nu} \cdot 45_y \frac{1}{4J} \cdot 180_x \cdot \frac{1}{4J}$
}
\end{itemize}
Here $\mathbbm{1}$ is the identity i.e., direct observation without applying any extra pulses.
$\Delta \nu$ and $J$ are the chemical shift difference  and the scalar coupling 
respectively (both in Hz).  The offset is assumed to be at the center of the two doublets and the RF amplitudes are assumed to 
be much stronger than $\Delta \nu$.  By calculating the propagator for each of these 
experiments, 24 linear equations are achieved. Solving these equations, gives the all unknown parameters of the density matrix. A detailed analysis of the density matrix tomography scheme is given in Appendix A.\\

Now we will use this tomographic scheme on long-lived singlet states \cite{maheshjmr10}.
 
\section{Singlet State Characterization}
\begin{figure}[t]
\begin{center}
\includegraphics[width=5.5cm,angle=-90]{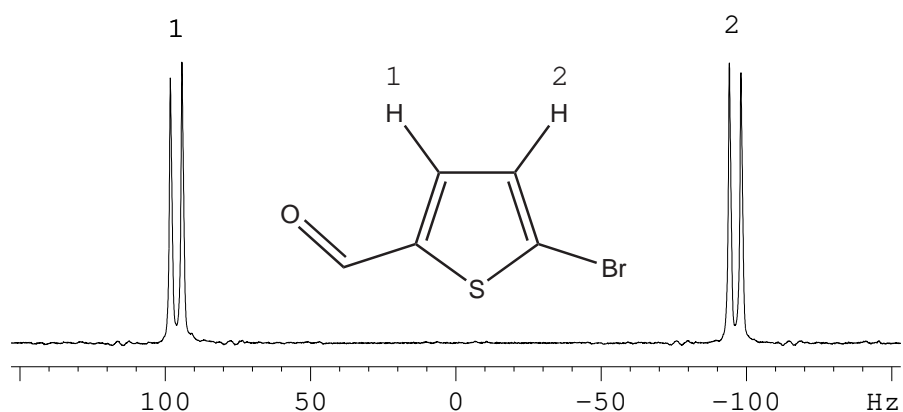}
\end{center}
\caption[$^1$H spectrum of 5-bromothiophene-2-carbaldehyde]{\label{btpspec} Part of the $^1$H spectrum of 5-bromothiophene-2-carbaldehyde (inset), displaying the doublets corresponding to the two $^1$H spins used to study the singlet state.  
The sample was dissolved in dimethyl sulphoxide-D6 and
all the experiments are carried out at 300 K.  The difference $\Delta \nu$ in chemical shifts
is 192.04 Hz and the scalar coupling $J$ is 4.02 Hz.  
Scalar coupling to aldehyde proton was too weak to be observed.  
The spin lattice relaxation time constants (T$_1$) for the two spins obtained from 
inversion recovery experiment are 5.2 s and 6.2 s respectively for the spins 1 and 2.}
\end{figure}

\subsection{Observing through antiphase magnetization}
\begin{figure}[t]
\begin{center}
\hspace{0.5cm}
\includegraphics[width=8cm,angle=-90]{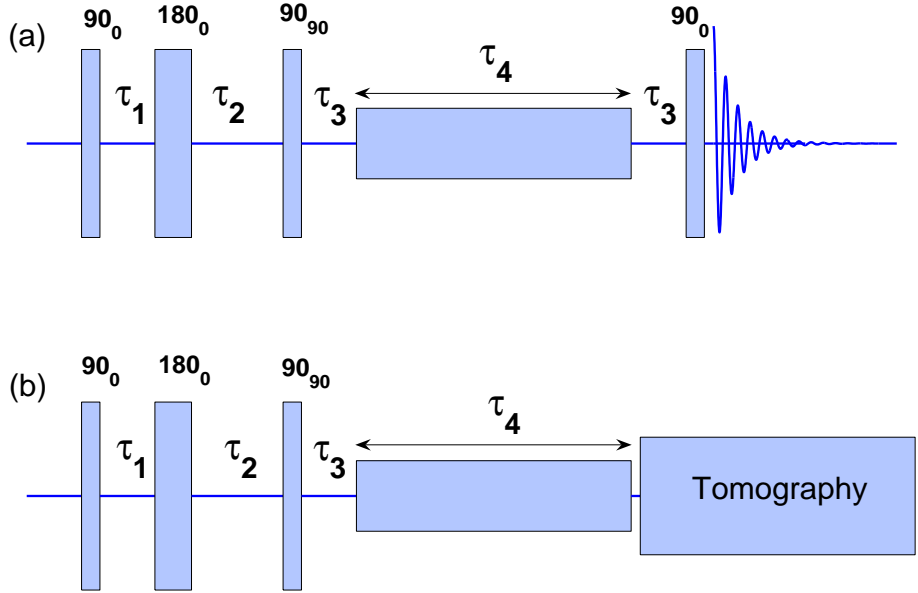}
\end{center}
\caption[The pulse sequences for the preparation of 
singlet states and detection via]{\label{sspul} The pulse sequences for the preparation of 
singlet states and detection via (a) converting to antiphase single quantum
magnetization and (b) tomography of singlet states.  Here $\tau_1 = 1/(4J)$,
$\tau_2 = 1/(4J)+1/(2 \Delta \nu)$, and $\tau_3 = 1/(4\Delta \nu)$,
with $\Delta \nu$ and $J$ being the chemical shift difference (in Hz) and
the scalar coupling respectively. $\tau_4$ is the duration of spin-lock.}
\end{figure}

The singlet state was prepared by the RF spin-lock method and converted into
antiphase magnetizations as described by Carravetta and Levitt
\cite{LevittJACS04} using the pulse sequence shown in Figure \ref{sspul}a.  The
RF spin-lock was achieved by either CW irradiation or by WALTZ-16 modulations.
The RF offset was set to the center of the two chemical shifts in these experiments.
The total magnitude of the antiphase 
magnetizations decays at different rates depending on the spin-lock conditions
(Figures \ref{sscompcw} and \ref{sscompcpd}).  The
decay constants with CW spin-lock are
16.6 s (Figure \ref{sscompcw}a) and 13.4 s (Figure \ref{sscompcw}h) respectively at RF amplitudes of 2 kHz  and 500 Hz.
Under WALTZ-16 spin-lock, the decay constants are slightly smaller, 16.2 s (Figure \ref{sscompcpd}a) 
and 12.8 s (Figure \ref{sscompcpd}h) respectively at 2 kHz and 500 Hz. 
Nevertheless, these values are about
2 to 3 times the $T_1$ values of the individual spins implying the preparation of long-lived singlet state.  

In this scheme, the integrated magnitude spectrum is usually monitored as
a function of spin-lock time.  The contributions from the spurious coherences
may not be eliminated in this process.  Further, the double quantum coherences,
if any, are not observed at all.  
Our interest is to quantify the singlet content in the 
instantaneous state $\rho(t)$ during the spin-lock.  
One might guess that the singlet content is maximum in the beginning and exponentially decays with the
spin-lock time.  Further, one may also guess that CW spin-lock is
superior to WALTZ-16 spin-lock at all timescales.
But the following tomography results provide a different picture \cite{maheshjmr10}.

\subsection{Tomography under varying spin-lock duration}
\begin{figure}
\begin{center}
\hspace{0.5cm}\vspace{-.5cm}
\includegraphics[width=13cm]{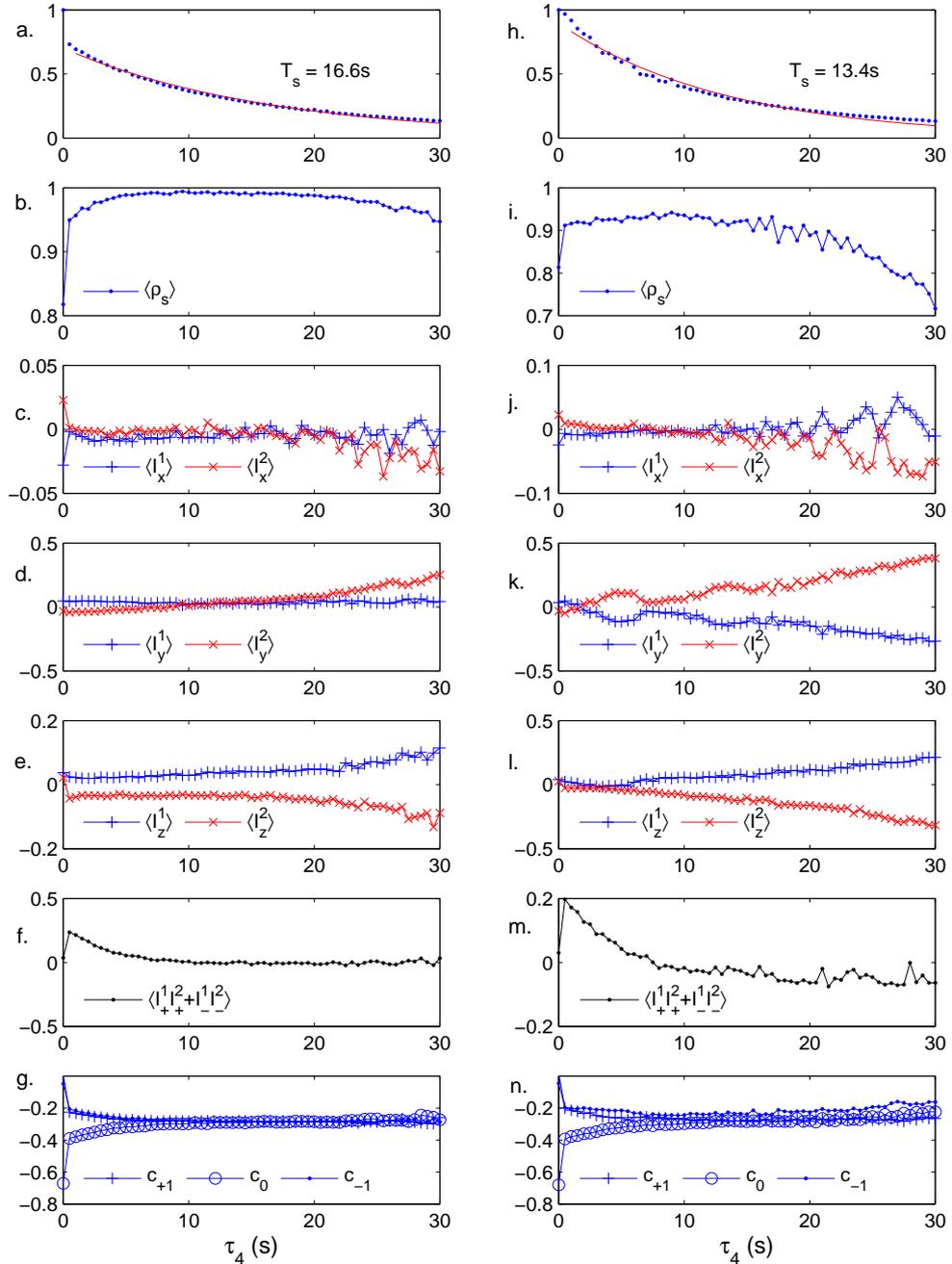}
\caption[Data characterizing the singlet state under CW spin-lock at an RF amplitude]{\label{sscompcw} 
Data characterizing the singlet state under CW spin-lock at an RF amplitude
of 2 kHz (a to g) and of 500 Hz (h to n).
The spin-lock duration $\tau_4$ was varied from  0 s to 30 s in steps of 0.5 s
in each case.
Dots in (a) and (h) correspond to the total magnitude of antiphase magnetization 
obtained from the pulse sequence in Figure \ref{sspul}a.
Singlet decay constant $T_s$ was obtained by using an exponential fit (smooth lines in (a) and (h)). 
During each fit, first two data points were omitted in view of strong spurious coherences 
created by the imperfections in the pulses.  Remaining graphs are the results obtained 
from tomography using the pulse sequence shown in Figure \ref{sspul}b.  They
correspond to the correlations:
$\langle \rho_s \rangle$ (b and i), 
$\langle I_x^{p} \rangle$ (c and j),
$\langle I_y^{p} \rangle$ (d and k),
$\langle I_z^{p} \rangle$ (e and l),
$\langle I_+^1I_+^2+I_-^1I_-^2 \rangle$ (f and m), and
$ c_q = \langle \vert T_q \rangle \langle T_q \vert \rangle$ (g and n),
with spin numbers $p = \{1,2\}$ and triplet subscripts $q=\{-1,0,+1\}$.
}
\end{center}
\end{figure}

\begin{figure}
\begin{center}
\includegraphics[width=13cm]{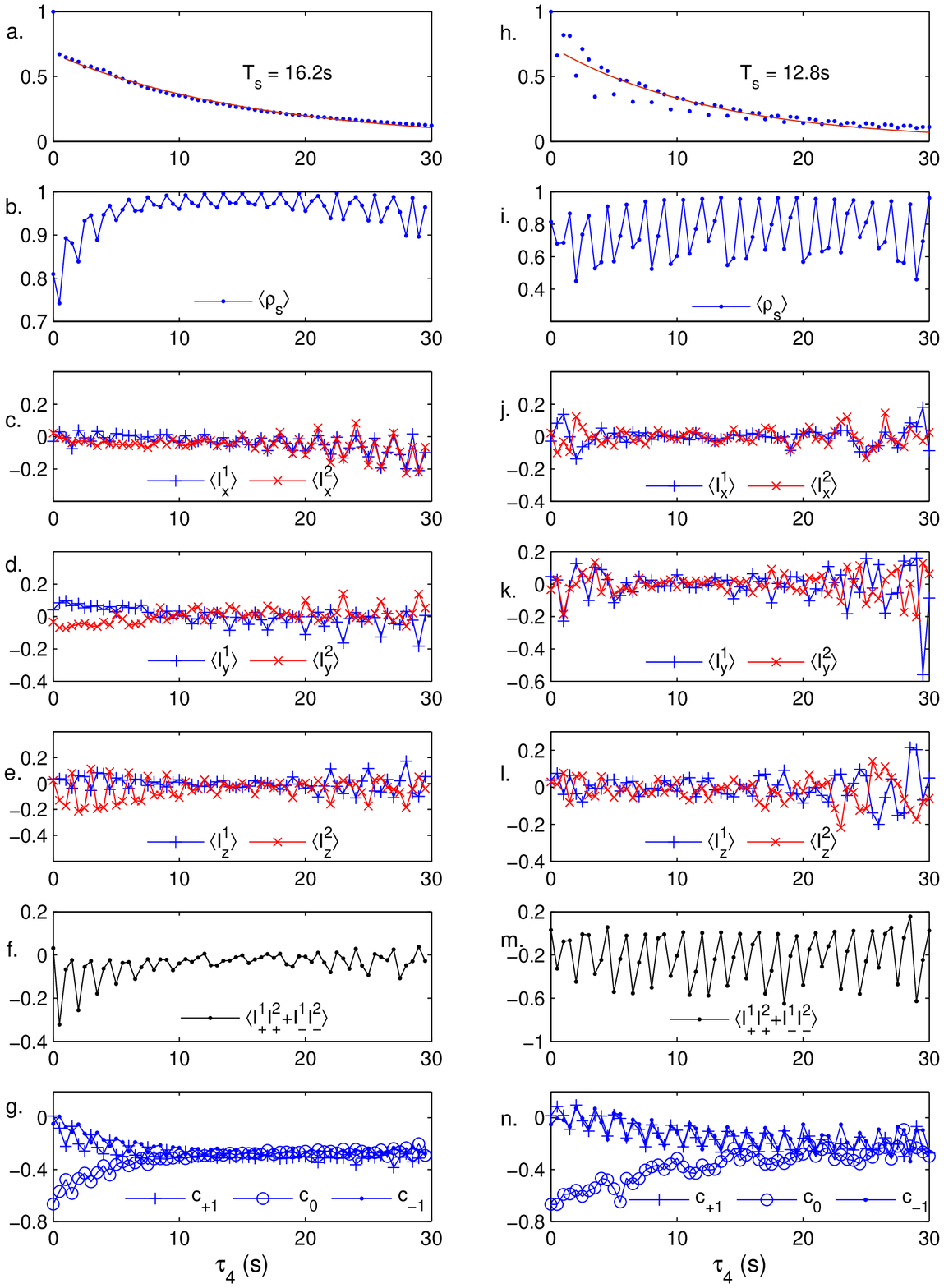}
\caption[Data characterizing the singlet state under WALTZ-16 (CPD) spin-lock at an RF amplitude]{\label{sscompcpd} 
Similar data as in Figure \ref{sscompcw}, but under 
WALTZ-16 spin-lock at an RF amplitude
of 2 kHz (a to g) and of 500 Hz (h to n).
The graphs correspond to 
total magnitude of antiphase magnetization (a and h),
$\langle \rho_s \rangle$ (b and i), 
$\langle I_x^{p} \rangle$ (c and j),
$\langle I_y^{p} \rangle$ (d and k),
$\langle I_z^{p} \rangle$ (e and l),
$\langle I_+^1I_+^2+I_-^1I_-^2 \rangle$ (f and m), and
$ c_q = \langle \vert T_q \rangle \langle T_q \vert \rangle$ (g and n),
with spin numbers $p = \{1,2\}$ and triplet subscripts $q=\{-1,0,+1\}$.
}
\end{center}
\end{figure}

\begin{figure}[t]
\begin{center}
\hspace{-1cm}
\includegraphics[width=9cm,angle=-90]{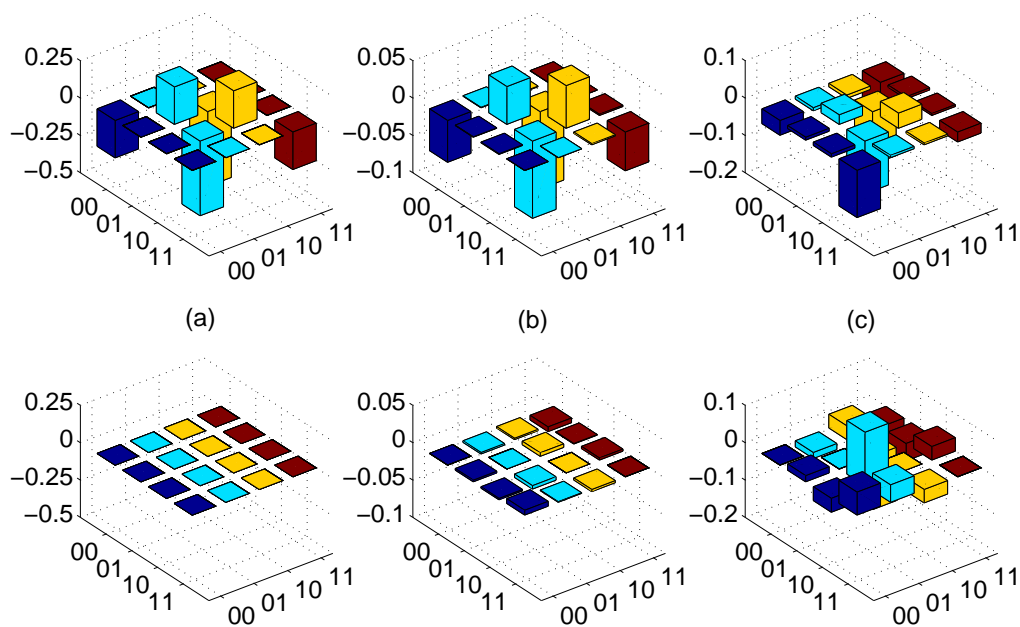}
\caption[Bar plots showing traceless part $\rho_s$ of the theoretical singlet state density matrix]{\label{sstomoeg} 
Bar plots showing (a) traceless part $\rho_s$ of the theoretical singlet state density matrix, 
(b) experimental state after 15 s of WALTZ-16 spin-lock at an RF amplitude of 2 kHz,
and (c) experimental state after 14 s of WALTZ-16 spin-lock at an RF amplitude of 500 Hz.
The upper and lower traces correspond to the real and imaginary parts respectively.
The singlet correlations in (b) and (c) are respectively 0.997 and 0.547.  The density
matrix in (b) shows significant decay, but still has high singlet content!  The real part of 
the density matrix in (c) shows significant double quantum artifact.
}
\end{center}
\end{figure}

The pulse sequence for the tomography of singlet states is shown in Figure \ref{sspul}b.
The density matrix of the singlet state is 
$\vert S_0 \rangle \langle S_0 \vert = \frac{1}{4} \mathbbm{1} + \rho_s$, with the
traceless part $\rho_s = -I^1 \cdot I^2$ being the
product of spin angular momentum operators of spins 1 and 2.  The correlation 
of the theoretical singlet state operator $\rho_s$ in the
instantaneous experimental density matrix $\rho(t)$ (obtained from tomography), 
\begin{eqnarray}
\langle \rho_s \rangle(t) = \frac{\mathrm{trace}\left[ \rho(t) \cdot \rho_s \right]}{\sqrt{ \mathrm{trace} \left[\rho(t)^2 \right] \cdot  \mathrm{trace} \left[\rho_s^2 \right]}},
\end{eqnarray}
gives a measure of singlet content in $\rho(t)$.  
The normalization used in the above expression disregards the attenuation 
of $\rho(t)$ itself.  
Similar definitions can be applied to
calculate the correlations $\langle I_x^1 \rangle$, $\langle \vert T_0 \rangle \langle T_0 \vert  \rangle$, etc.
We monitored the correlations as a function of spin-lock time $\tau_4$ from
0 s to 30 s in steps of 0.5 s under different spin-lock conditions using the 
sequence shown in Figure \ref{sspul}b.
The results are shown in Figures \ref{sscompcw} and \ref{sscompcpd}.  
3D bar plots of full density matrices at two particular spin-lock conditions  
are shown in Figure \ref{sstomoeg}.

The Figures \ref{sscompcw}b, \ref{sscompcw}i, \ref{sscompcpd}b, and \ref{sscompcpd}i
indicate correlation $\langle \rho_s \rangle$ 
as a function of spin-lock time under various spin-lock conditions.  In all the
cases, the initial correlation is about 0.8.  This is expected, since the initial
state prepared by the pulse sequences in Figure \ref{sspul} just before the
spin-lock is actually 
\begin{eqnarray}
\rho(0) = \vert S_0 \rangle \langle S_0 \vert - \vert T_0 \rangle \langle T_0 \vert.
\end{eqnarray}
With CW spin-lock at a high RF amplitude of 2 kHz (Figure \ref{sscompcw}b - \ref{sscompcw}g), 
the singlet correlation $\langle \rho_s \rangle$ quickly
reaches to 0.95 in 0.5 s of spin-lock time (Figure \ref{sscompcw}b).  
Most of the spurious coherences and the residual
longitudinal magnetizations created during the preparation are destroyed by the RF inhomogeneity during spin-lock.
Figures \ref{sscompcw}g and \ref{sscompcw}n reveal that the initial correlation  
$\langle \vert T_0 \rangle \langle T_0 \vert \rangle (0)$ is $-0.7 \sim 1/\sqrt{2}$
which is just expected .  Within 0.5 s, 
the $\vert T_0 \rangle \langle T_0 \vert$ content is rapidly reduced.
But complete equilibration of triplet levels takes about 5 s.
Interestingly, there is a sudden build-up and gradual fall of double quantum 
coherence as seen in Figures \ref{sscompcw}f and Figures \ref{sscompcw}m.
As the singlet state gets purified, $\langle \rho_s \rangle$ exceeds 0.99 in 6 seconds and reaches a 
maximum value of 0.994 at 9.5 s. 
After about 18 s, $\langle \rho_s \rangle$ starts decaying below 0.99, probably
due to the gradual conversion of singlet state
to other magnetization modes via the triplet states by relaxation mechanisms.  On the other hand, 
there is a gradual build up of y- and z-magnetizations (Figures \ref{sscompcw}d and \ref{sscompcw}e) 
in a similar way as that of a steady state experiment \cite{Abragam,LevittPRL92}.  
Nevertheless, the singlet correlation remained above 0.95 till 30 s.
The x-magnetization and the double quantum coherence (Figures \ref{sscompcw}c and \ref{sscompcw}f) 
remained small during the period of high correlation.  After the initial differences,
the triplet states equilibrate in about 6s, and remain steady then onwards
(Figures \ref{sscompcw}g and \ref{sscompcw}n).

With CW spin-lock at 500 Hz , 
the singlet correlation reaches only up to 0.94 again at about 9 s
and then steadily drops to 0.71 at 30 s (Figure \ref{sscompcw}i).  The increased 
buildup of x-, y-, and z- magnetizations with the reduction of the spin-lock
power can also be noticed (Figure \ref{sscompcw}j - \ref{sscompcw}l).

Under WALTZ-16 spin-lock (Figure \ref{sscompcpd}), 
all the graphs are characterized by oscillations that are either in-phase
or anti-phase.
The origin of oscillations probably lies in the
cyclic nature of WALTZ-16 modulation.  

At an RF amplitude of 2 kHz, the maximum singlet correlation of 0.997 was reached at 15 s(Figure \ref{sscompcpd}b).  
The 3D bar plot of the density matrix corresponding to
this case is shown in Figure \ref{sstomoeg}b. 
More interestingly, $\langle \rho_s \rangle$ peaks seem to maintain
above 0.99 till $\tau_4 = 28.5$ s, i.e., about 10 s longer than the CW case!
Thus, for certain values of spin-lock durations, WALTZ-16 provides 
purer singlet states than that of CW.

The singlet correlation under
WALTZ-16 spin-lock at 500 Hz displays stronger oscillations
(Figure \ref{sscompcpd}i).  
Despite the oscillations, the singlet correlation reaches as
high as 0.96 at 13.5 s.  Again it can be noticed that good singlet content is
held for longer periods by  WALTZ-16 than the CW of same amplitude.  
For example at 500 Hz RF amplitude, WALTZ-16 gives
a singlet correlation of 0.94 at $\tau_4 = $ 27 s, while that for CW it is only 0.79.

\begin{figure}
\begin{center}
\includegraphics[width=12cm,angle=-90]{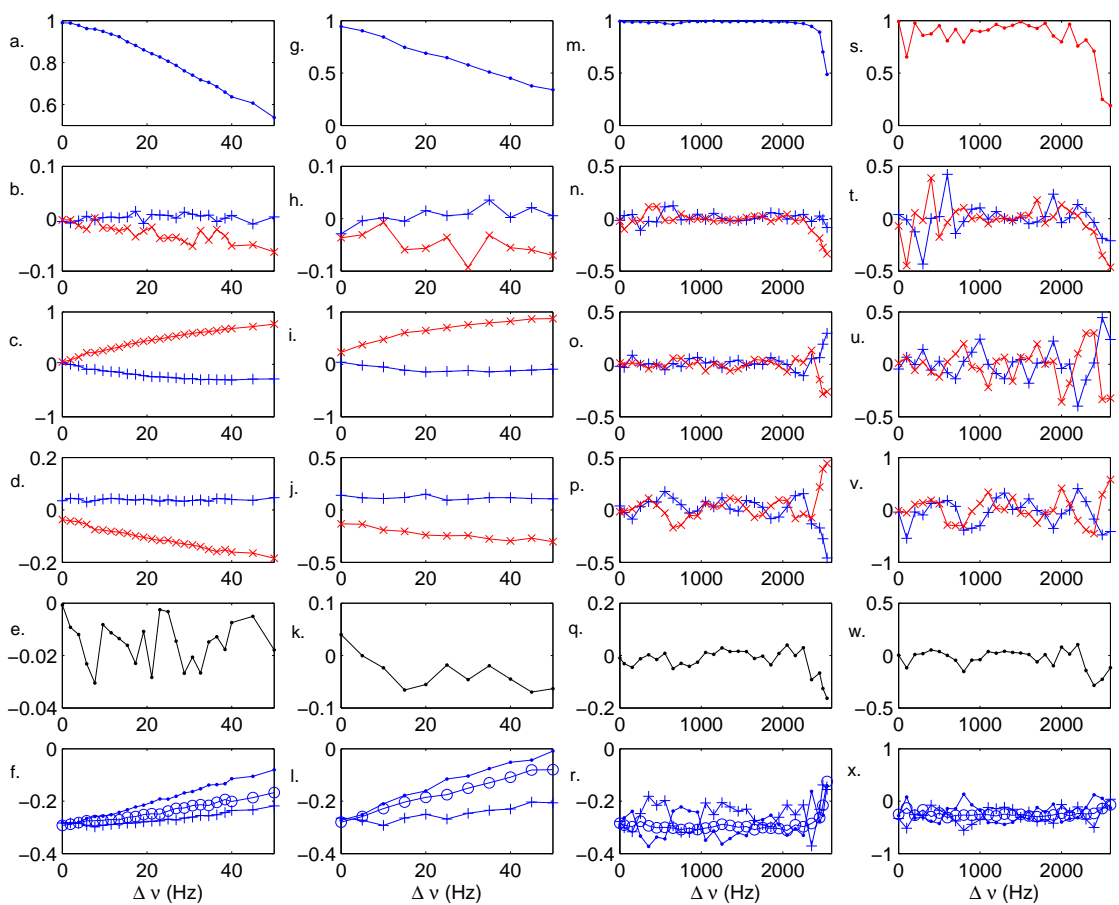}
\caption[Correlations calculated using the density matrix tomography of singlet states
prepared with different spin-lock conditions:]{\label{ssoff} 
Correlations calculated using the density matrix tomography of singlet states
prepared with different spin-lock conditions: 
(i) CW spin-lock at 2 kHz for 15s (a to f),
(ii) CW spin-lock at 2 kHz for 28.5s (g to l),
(iii) WALTZ-16 spin-lock at 2 kHz for 15s (m to r), and,
(iv) WALTZ-16 spin-lock at 2 kHz for 28.5s (s to x).  In each case, 
the horizontal axis indicates the RF offset $\Delta \nu$
during the spin-lock.  The offset is measured from the center of the two
chemical shifts.
The rows correspond to :
$\langle \rho_s \rangle$ (b and i), 
$\langle I_x^{p} \rangle$ (c and j),
$\langle I_y^{p} \rangle$ (d and k),
$\langle I_z^{p} \rangle$ (e and l),
$\langle I_+^1I_+^2+I_-^1I_-^2 \rangle$ (f and m), and
$ c_q = \langle \vert T_q \rangle \langle T_q \vert \rangle$ (g and n),
where spin numbers $p = \{1$ (pluses), 2 ($\times$'s)\} 
and triplet subscripts $q=\{-1$ (dots), 0 (circles), $+1$ (pluses)\
}.
\end{center}
\end{figure}
\subsection{Offset dependence}
Theoretical and numerical investigations on the offset dependence of singlet spin-lock
has been have been carried out by Karthik and Bodenhausen 
\cite{BodenJMR06} and by Pileio and Levitt \cite{LevJCP09}.  
Robustness of various modulation schemes with regard to
offset of singlet spin-lock  have been demonstrated 
by Bodenhausen and co-workers \cite{BodenCPC07}.  Here we
probe the offset dependence of singlet evolution using tomography \cite{maheshjmr10}.
Figure \ref{ssoff}
shows the experimental data obtained from a series of singlet state tomography
experiments, each time
varying the RF offset of the spin-lock.  The RF offset was measured from the
center of the two chemical shifts.  Again the experiments were carried out under the 
following spin-lock conditions:
\begin{enumerate}[(i)]
\itemsep-1.5em
	\item CW for 15s (Figure \ref{ssoff}a-\ref{ssoff}f),\\
  \item CW for 28.5s (Figure \ref{ssoff}g-\ref{ssoff}l),\\
  \item WALTZ-16 for 15s (Figure \ref{ssoff}m-\ref{ssoff}r), and\\
  \item WALTZ-16 for 28.5s (Figure \ref{ssoff}s-\ref{ssoff}x).    
\end{enumerate}
The graphs indicate
that the WALTZ-16 scheme is far superior compared to CW in preserving the singlet
correlation at high RF offsets.  The singlet correlations with 2 kHz CW drops
below 0.5 for an offset of 50 Hz.  
However, WALTZ-16 at 2 kHz amplitude maintains a high correlation 
of 0.97 at 28.5s, even with an offset of 2.1 kHz.  In the case of CW spin-lock, rapid
build up of y-magnetizations can be noticed with the increase of the RF offset \cite{maheshjmr10}.
\section{Long lived singlet states in multi-spin systems}

\begin{figure}
\centering
\includegraphics[width=13cm]{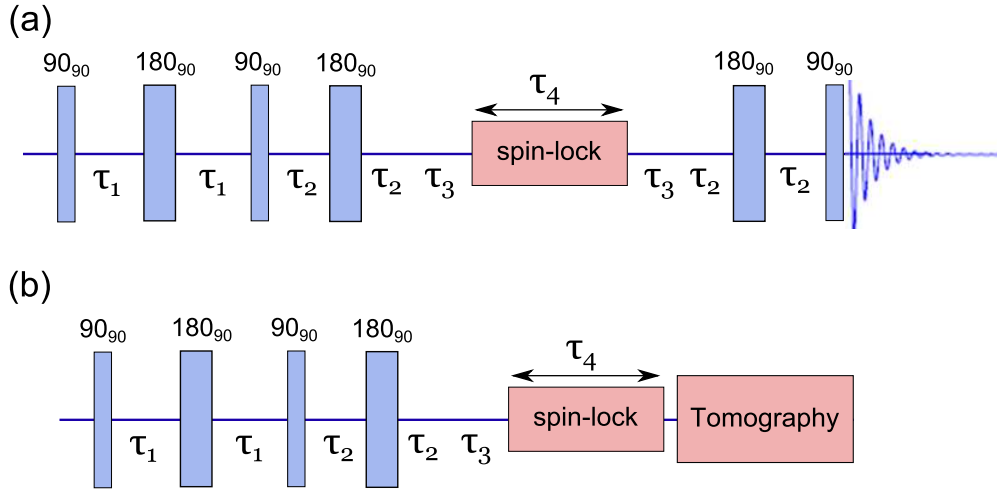}
\caption[Pulse sequence for the creation of long lived singlet states in a 3 spin system]{Pulse sequence for the creation of long lived singlet states in a 3 spin system ($AMX$). (a) Anti-phase singlet magnetization to be accessed via spin-3, (b) qualitative measure of singlet correlation is done by state tomography. $\tau_1$ and $\tau_2$ are optimized delays in a way that both $J_{13}$, $J_{23}$ get a $\pi/2$ $J$- evolution, $\tau_3=\frac{1}{4\Delta\nu_{12}}$, and $\tau_4$ is the spin-lock duration.}
\label{lls3pp}
\end{figure}

\begin{figure}
\centering
\includegraphics[width=14cm]{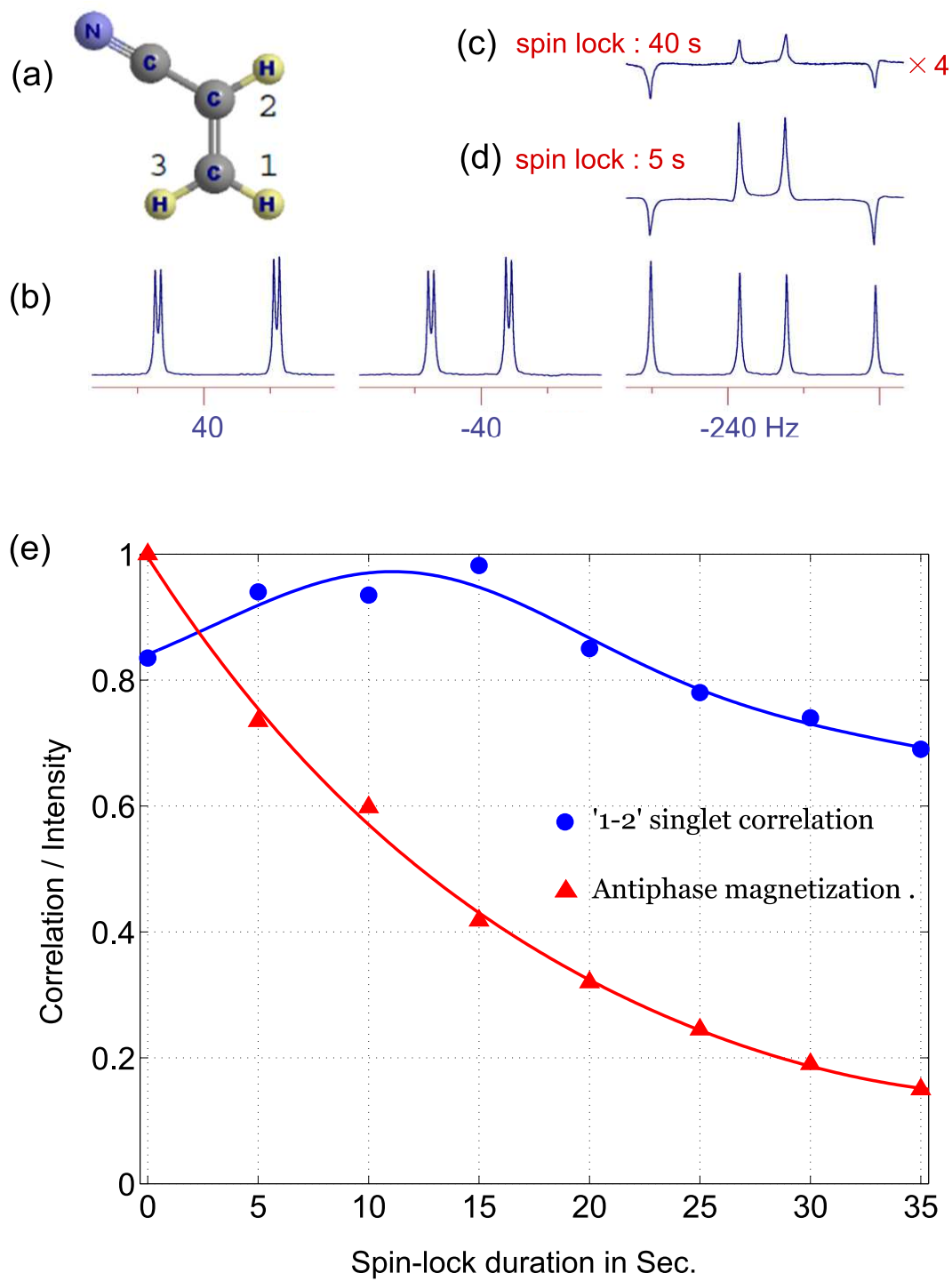}
\caption[Experimental results of 3-spin LLS]{Experimental results of 3-spin LLS. (a) Acrylonitrile dissolved in CDCl$_3$, where 3 protons acting as a three spin homonuclear system. (b) The $^1H$ reference spectra of Acrylonitrile in a 500 MHz spectrometer. (c) and (d) showing the antiphase spectra of spin-3 after a spin lock duration of 5 s and 40 s respectively. (e) The solid curve showing the antiphase magnetization decay and dotted curve showing the singlet correlations obtained from tomography over a duration of spin-lock ($\tau_4$) time.}
\label{lls3s}
\end{figure}

\subsection{Long lived singlet states in a 3-spin system}
In this subsection we will describe the methods for preparing long lived singlet states in a 3-spin system ($AMX$). We have extended the procedure of the 2-spin system as described in the previous sections. The singlet population distribution between any two spins can be prepared in presence of a third spin. The pulse sequence relies on the refocusing of the unnecessary couplings. The NMR pulse sequence is shown in Fig. \ref{lls3pp}. In this particular example we have prepared the singlet population between spin-1 and spin-2. Singlet population is accessed by transferring the magnetization into spin-3 . The quantitative measure of singlet magnetization is done by the pulse sequence shown in Fig. \ref{lls3pp}a. The extensive tomographic method of accessing singlet correlation has also been performed. The experimental results are shown in Fig. \ref{lls3s}. The decay of antiphase magnetization and tomographic correlation is shown in Fig.  \ref{lls3s}e. The $T_1$ time for all the three spins are roughly 6 sec, while the LLS time ($T_{LLS}$) is found to be 17.9 sec. Hence the ratio $T_{LLS}/T_1 \approx 3$. The nature of this plot is similar to the spin-2 system and the reason for this is given in previous section. The 3-spin density matrix tomography scheme is described in detail in Appendix B.

\begin{figure}
\centering
\includegraphics[width=13cm]{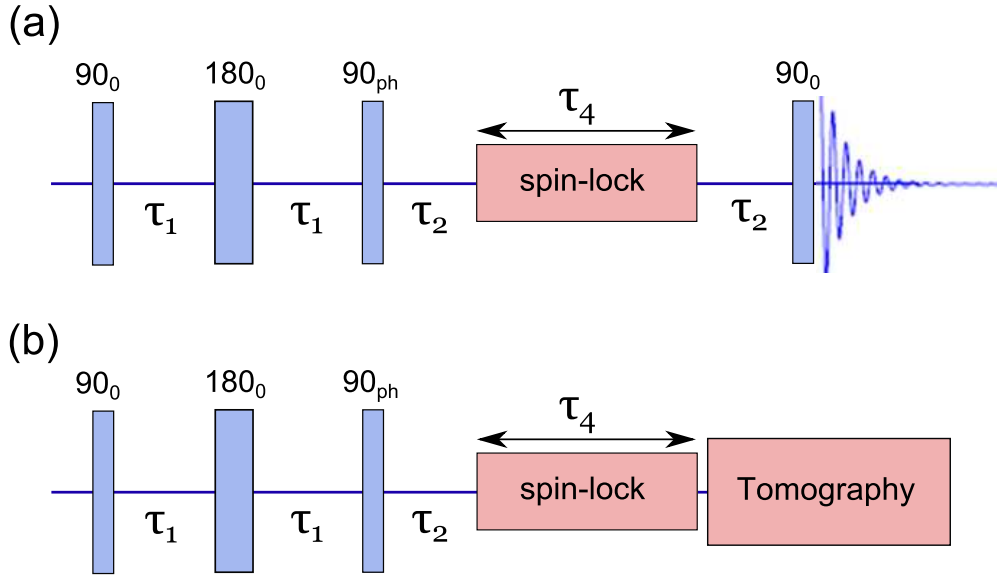}
\caption[Pulse sequence for the creation of long lived singlet states in a 4 spin system]{Pulse sequence for the creation of long lived singlet states in a 4 spin system ($AMXY$). (a) Singlet states accessed by transferring it into anti-phase magnetization, (b) qualitative measure of singlet correlation is done by state tomography. $\tau_1$, $\tau_2$ are optimized delays and $\tau_4$ is the spin-lock duration. $90_{ph}$ denoting a optimized phase $\pi/2$ non-selective pulse.}
\label{lls4pp}
\end{figure}
\begin{figure}
\centering
\includegraphics[width=14cm]{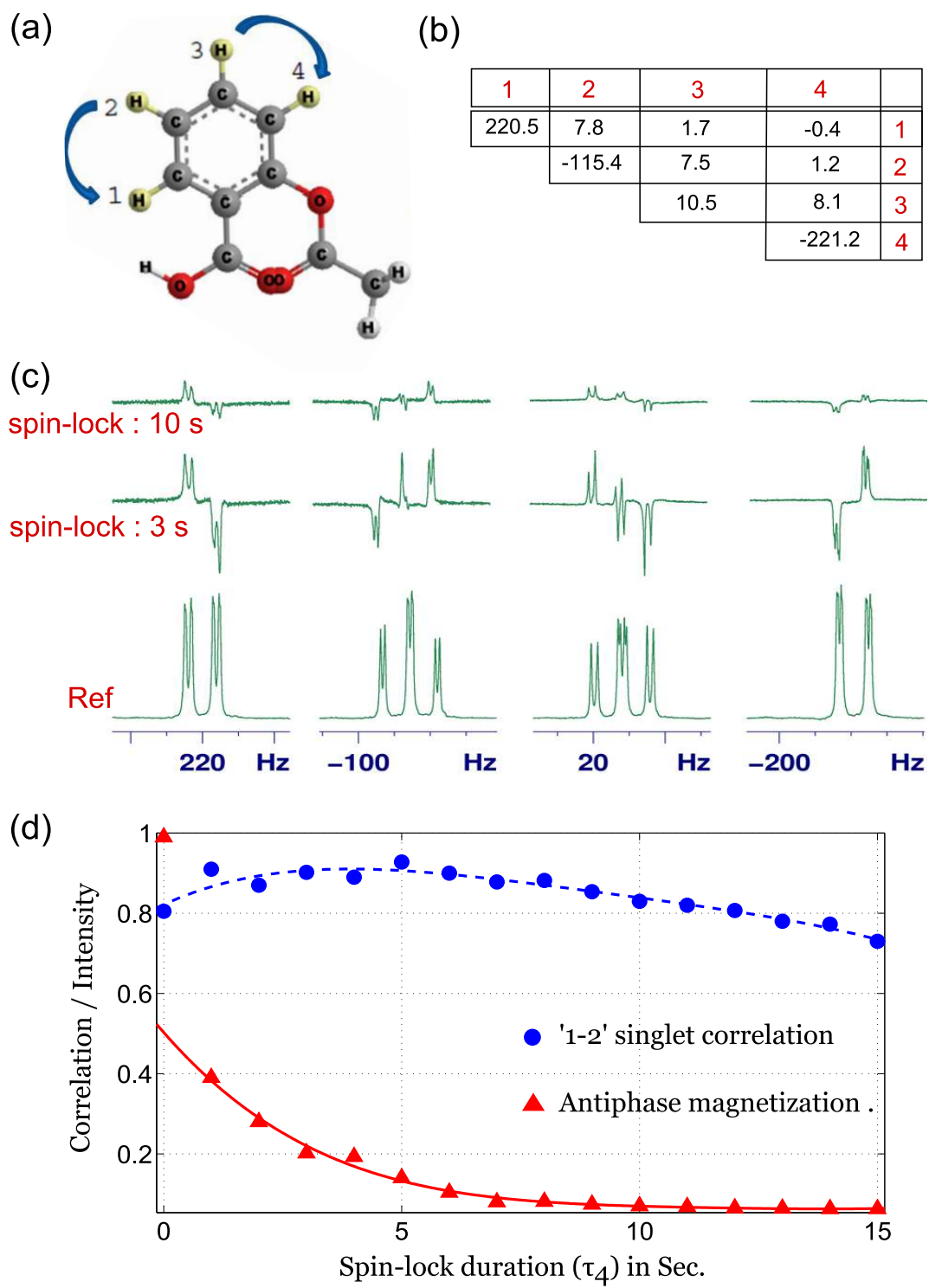}
\caption[Experimental results of 4-spin LLS]{Experimental results of 4-spin LLS. (a) Aspirin dissolved in CD$_3$OD, where 4 protons acting as a four spin homonuclear system. (b) The chemical shifts and J-coupling constants in Hz are shown in a table format. (c) The $^1H$ reference spectra of Aspirin is at the bottom trace. The antiphase spectra are shown in upper trace after a spin lock duration of 3 s and 10 s respectively. (d) The solid curve showing the antiphase magnetization decay and dotted curve showing the 1-2 singlet correlations obtained from tomography over a duration of spin-lock ($\tau_4$) time.}
\label{lls4s}
\end{figure}
\subsection{Long lived singlet states in a 4-spin system}
We have prepared two pair of singlet states in a 4 spin $AMXY$ system. The exact pulse sequence is shown in Fig. \ref{lls4pp}. We were able to prepare simultaneous singlet states in between spin-1 and spin-2 and also in between spin-3 and spin-4. The J-evolution delays ($\tau_1$ and $\tau_2$) are calculated in a optimized way. The traditional method of accessing singlet states is by converting it into single quantum coherences. The antiphase spectra of aspirin are shown in Fig. \ref{lls4s}c. We have done density matrix tomography to calculate the correlation at various spin-lock duration. The nature of antiphase decay and singlet correlation profile matches with the previous cases (2-spin and 3-spin) and hence got the similar explanation. We have found $T_1 \approx 3 s$ and $T_{LLS} \approx 6 s$, hence the ratio $T_{LLS}/T_1 \approx 2$. This also proves the long lived nature of the prepared singlet states. 

\section{Conclusions}
Study of singlet state is important not only because of the 
interesting Physics that makes it long-lived, but also  
because of its potential for a number of applications.
We have studied the singlet state directly and quantitatively using density matrix 
tomography.  A new set of tomography sequences have been introduced for this purpose.  
The density matrix tomography provides a tool not only for characterizing various spin-lock 
schemes but also for understanding the spin dynamics during the spin-lock period.

The singlet state is preserved with CW spin-lock as well as with WALTZ-16 spin-lock
at two different RF amplitudes: 2 kHz and 500 Hz.
The results indicate that at high RF amplitudes, both CW and WALTZ-16 achieve high
singlet content.  An important feature of singlet state
is that it gets purified by itself during the spin-lock, simply because of its longer 
life time compared to the spurious coherences.  There exist optimum 
spin-lock values at which the singlet correlations are maximum.
While WALTZ-16 shows significant oscillations in the singlet purity,
for certain intervals of spin-lock it gives better performance than CW and holds the 
singlet content for longer intervals of time.  
The dependence of correlations with the RF offset during the spin-lock are also studied
under both CW and WALTZ-16 schemes.  It is found that WALTZ-16 is far superior in preserving
the singlet state at large RF offsets.

\chapter{Preparation of Pseudopure States Using Long Lived Singlet States}

  An introduction to pseudopure states (PPS) is given in section \ref{ppsintro}. In section \ref{ppsmethods}, a general review of various methods proposed for preparation of pseudopure states in NMR have been described. In section \ref{ppslls}, a new method for creating pseudopure states from long-lived singlet states is presented. In this context, we have developed a robust, scalable quantum circuit for creating PPS for any number of qubits. Experimental results are shown and discussed for multi-qubit systems in the later part of the section \ref{ppslls}.

\section{Introduction}
\label{ppsintro}
\subsection{A pure state and a mixed state}
 A `pure state' is a quantum state which can be represented by a single ket $\vert i \rangle$. A quantum state which can not be represented by a single ket format in any of the basis is termed as `mixed state'. In a mixed state system, we are only able to say that the system has certain probabilities (say $p_1$, $p_2$... ) of being in different states (say $\vert 1 \rangle$, $\vert 2 \rangle$... ). For example, let us consider a two level system of energy difference $\Delta E$ of about $2eV$ (e.g. electronic levels of sodium). At room temperature of 300 K, the ratio between ground and excited states population is of the order of $1 : e^{-\Delta E/kT}$ $\sim 1 : 10^{-34}$. For all practical purposes, we can assume that such a state exists in a pure ground state at thermal equilibrium. In another case, if we consider a much shorter energy gap difference of the order of $2 \mu eV$ (such as the case of nuclear spin's energy gap for hydrogen at $10T$ external magnetic field), then the populations of the two states are roughly in the order of $1 : 0.9999$. The later case described a perfect mixed state condition. 
 
 While a pure classical state is represented by a single moving point in phase space having definite spatial ($q_i, ...  q_f$) and momentum coordinates ($p_i, ...  p_f$), a classical mixed state is described by a non-negative density function $\rho(q_i,... q_f, p_i,... p_f )$, such that the probability that a system is found in the interval $dq_i...dp_f$ at time $t$ is $\rho dq_i...dp_f$. The `quantum analog' of the classical pure state is represented by a single state vector, while the quantum analog of a classical mixed state is represented by the density matrix \cite{schiff}.

 An ensemble is in an idealized pure state if all of its' members are in state $\vert \psi (t)\rangle$. The density operator of a pure state $\vert \psi (t)\rangle = \sum_{i} c_i(t)\vert i\rangle$ is given by,
\begin{equation} 
\rho_{pure} = \vert \psi\rangle\langle\psi\vert = \sum_{i}\sum_{j} c_i(t)c_{j}^{*}(t)\vert i\rangle\langle j\vert. 
\end{equation}
 A pure state density matrix, $P_{\vert n\rangle}=\vert n\rangle\langle n\vert$ is a projection operator since,
\begin{equation}
P_{\vert n\rangle}\sum_{i}c_i\vert i\rangle = c_n\vert n\rangle, \quad \mathrm{and} \quad \sum_{n}P_{\vert n \rangle}=\mathbbm{1}.
\end{equation}
  Since all the subsystems behave identically, one has complete knowledge of all parts of the system. Pure states are also known as the `states of maximum information'. A pure state has two basic properties : 
\begin{equation}
\rho_{pure}^{2} = \rho, \quad \mathrm{and} \quad tr(\rho_{pure}^{2}) = 1.
\end{equation} 
The density matrix of an ensemble in a mixed state is obtained by ensemble average : 
\begin{equation} 
\rho_{mix} = \sum_{i}\sum_{j} \overline{c_i(t)c_{j}^{*}(t)}\vert i\rangle\langle j\vert = \sum_{k} p_{k} \vert \psi_{k}\rangle\langle\psi_{k}\vert , 
\end{equation}
where $p_{k}$ denotes the probability of the system of being in state $\psi_{k}$ with $\sum_{k} p_k = 1$ and the bar denotes the ensemble average. $\rho_{mix}$ is not a projection operator and is also known as state with `less than maximum information'. It holds the property of :
\begin{equation}
\rho_{mix}^{2} \neq \rho, \quad \mathrm{and} \quad tr(\rho_{mix}^{2}) < 1.
\end{equation}  

\begin{figure}
	\centering
		\includegraphics[width=6cm]{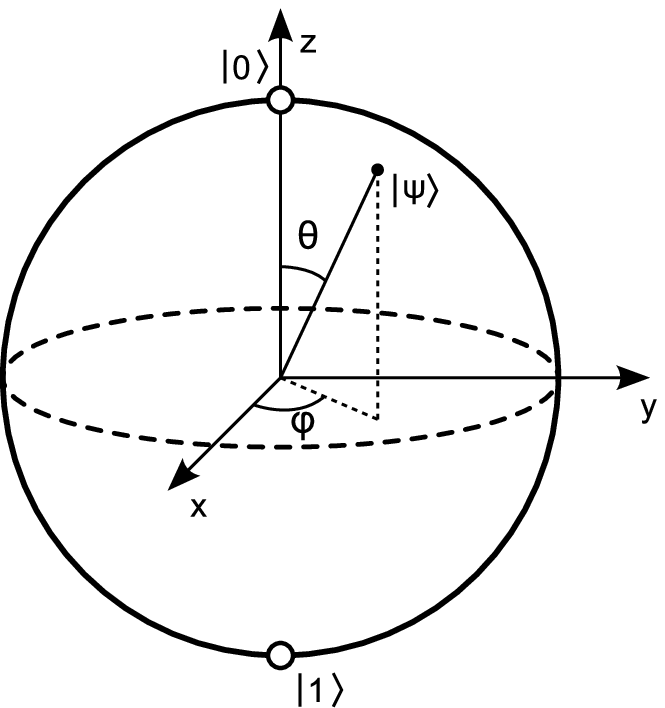}
		\caption[Bloch sphere representation of pure states, mixed states]{Bloch sphere: All the surface points represent a pure states, any non-surface points are mixed states. The origin corresponds the maximally mixed state.}
	\label{fig:bloch}
\end{figure}

 Another way to represent a wave function and hence a density matrix is through the `Bloch sphere'. A pure state $\vert\psi\rangle$ is a geometrical point on the Bloch surface.  For a single qubit, $\vert\psi\rangle$ can be written as :
\begin{equation}
\vert\psi\rangle = \cos(\theta/2)\vert 0\rangle + e^{i\phi}\sin(\theta/2)\vert 1\rangle,
\end{equation}
with $0\leq\theta\leq\pi$ and $0\leq\phi < 2\pi$. Here $\vert 0\rangle$ and $\vert 1\rangle$ are two orthonormal basis vectors of the two-level system (single qubit). Any arbitrary density matrix $\rho$ can be expanded in terms of Pauli operators $\sigma_{i}$:
\begin{equation}
\rho = \frac{1}{2}(\mathbbm{1}+\vec{r}.\vec{\sigma}),
\label{rhoeq}
\end{equation}
where $\vec{r}$ is called as Bloch vector, the radius vector of the state from the origin. The eigenvalues of $\rho$ are $\frac{1}{2}(1 \pm \vert\vec{r}\vert)$. As density operator must be positive, we get $\vert\vec{r}\vert\leq 1$. The purity of the density matrix can be measured by squaring it,

\begin{equation}
\begin{split}
tr(\rho^{2}) =& tr\left(\frac{1}{4}(\mathbbm{1} + 2\vec{r}.\vec{\sigma} + (\vec{r}.\vec{\sigma})(\vec{r}.\vec{\sigma}))\right) \\
             =& \frac{1}{2}(1 + \vert\vec{r}\vert^2).
\end{split}
\end{equation}
 For a pure state $tr(\rho^{2}) = 1$, i.e., $\vert\vec{r}\vert = 1$ and for a mixed state $\vert\vec{r}\vert < 1$. Hence a pure state represents a point on the surface of a Bloch sphere, whereas any point other than on the surface represents a mixed state. The origin of the Bloch sphere ($\vert\vec{r}\vert$ = 0) represents the maximally mixed state with $tr(\rho^{2}) = \frac{1}{2}$.

\subsection{Necessity of Pure states in QIP}
 In order to carrying out information processing, a quantum register
must satisfy a set of criteria laid out by DiVincenzo \cite{divincenzo}.
An important criterion is the ability to precisely initialize the
register to a desired ket of the computational basis. It has been shown that 
highly mixed state may not be used to create an entanglement \cite{braunstein99} in QIP. Again, a mixed input state leads to a mixed 
output state which are generally difficult for analysis \cite{preskill}. In one particular case it can be show that a mixed state input gives wrong output results. Grover search algorithm for a two-qubit system starting from a pure state gives the desired answer with full probability. While starting from a mixed state, instead of getting a definite state, one gets a result with probabilities of different states. On the otherhand, we can find some algorithms which have no requirement of starting from pure initial state, e.g. Deutsch-Josza algorithm \cite{deutsch92}. Deutsch-Josza algorithm decides whether the function is `balanced' or `constant' in one iteration. If the initial state is a pure state, then the in-phase output signal denotes it as a balanced and the anti-phase output signal denotes it as a constant function \cite{jones98}. In other case, when the initial state is a mixed state, the output signal consists of all possible transitions for a constant function, whereas for a balanced function, at least one of the transition will be missing \cite{kavita00, mahesh01, freeman98, deutsch92}. Since the Deutsch-Josza algorithm is used only to distinguish between a balanced function and a constant function, mixed initial state should be sufficient for this purpose. Again there are some algorithms that take advantage of the mixed nature of an ensemble systems \cite{collins, brusch, miao}. These algorithms utilize the power of parallel processing of the identical quantum registers. In a broad overview, however, initialization of quantum registers remain an important requisite for a large scale quantum computer.

\begin{figure}
	\centering
		\includegraphics[width=12cm]{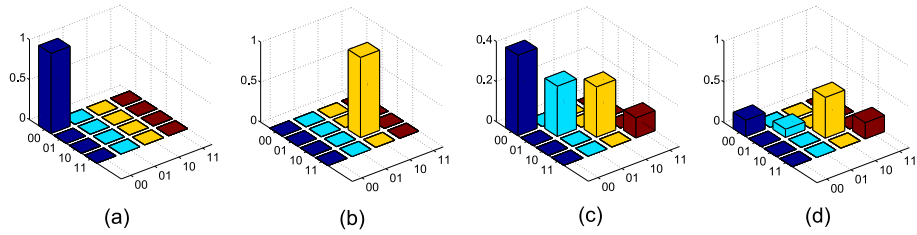}
	\label{fig:grover}
	\caption[Simulations of Grover's search algorithm]{Simulations of Grover's search algorithm for a two qubit system starting from pure-state and mixed-state condition. The state being selectively flipped is $\vert 01 \rangle$. (a) Pure input state $\vert 00 \rangle\langle 00\vert$, (b) output state $\vert 10 \rangle\langle 10\vert$ with full probability, (c) a complete random mixed state population input, (d) the answer of the Grover's search algorithm with no definite result.}
\end{figure}
   
\subsection{Pure states in NMR}
 The density operator of an ensemble system characterized by the Hamiltonian $\mathcal{H}$ at thermal equilibrium is given by :
\begin{equation}
\rho = \frac{e^{-\mathcal{H}/kT}}{Z},
\end{equation}
where $k$ is the Boltzmann constant and $T$ is the temperature of the system. Z represents the partition function and is given by $Z = tr \left[e^{-\mathcal{H}/kT}\right]$. For a N-spin homonuclear system at room temperature, the eigenvalues of $\mathcal{H}$ are much smaller than $kT$ value, hence the equilibrium density operator can be approximated to,
\begin{equation}
\begin{split}
\rho_{eq} =& \frac{1}{2^N} \left(\mathbbm{1}+\frac{\hbar\gamma B_0}{kT}\sum_{j=1}^{N}I_{z}^j\right)\\
          =& \frac{1}{2^N} (\mathbbm{1}+\epsilon\rho_{dev}),
\end{split}   
\label{rhodev}       
\end{equation}

where $\gamma$ is the gyromagnetic ratio, $B_0$ is the strength of the Zeeman field, $I_{z}^j$ is the longitudinal component
of spin operator for $j^{th}$ spin, and $\rho_{dev}$ is the trace-less `deviation density matrix'. The coefficient $\epsilon = \hbar\gamma B_0/kT$ is normally very small; $\epsilon \sim 10^{-5}$
for hydrogen nuclei in a Zeeman field of 10 T, at room
temperature (300 K). Thus a normal NMR system represents a highly mixed state \cite{Abragam,LevBook,goldman}. However, only the deviation
density matrix contributes to the NMR signal, which is smaller by a factor of $\epsilon$ compared to
the system which is in a pure state, i.e., if all the spins were in one state. It seems from the Eq. \ref{rhodev}
that the only way of preparing a pure state in NMR is by using extremely high magnetic fields at
extremely low temperatures. While achieving higher magnetic fields than a few tens of Tesla, is still
a technological challenge, carrying out experiments at low temperatures is associated with many
technical problems including among others, liquid to solid phase transition. However, in Eq.
\ref{rhodev}, we should notice that we have taken only the nuclear spin part. Efforts have been made to
bring-in other interactions which will enable preparing pure states without the requirement of using unrealistically
high magnetic fields or undesirable low temperatures. One technique is Optical pumping,
wherein one tries to transfer electron polarization into the nuclear spin system. This includes for
example, a polarization transfer from a laser polarized noble gas atoms like
$^{129}$Xe to molecule of
interest. While this technique is widely used in magnetic resonance imaging (MRI), it is expected to be
relatively inefficient with regard to preparing a pure state \cite{hubler}.

The spin temperature can nevertheless be reduced by 
using parahydrogens \cite{Anwar} or by using
Dynamic Nuclear Polarization (DNP)\cite{Morley}.  In future, either or both of
these techniques may be available for preparing NMR quantum registers 
into almost pure states \cite{laflammerev}.  
The existing approach for initializing NMR registers is
however based on specially prepared mixed states known as pseudopure states (PPS).

\subsection{Pseudopure states}
\begin{figure}
	\centering
		\includegraphics[width=12cm]{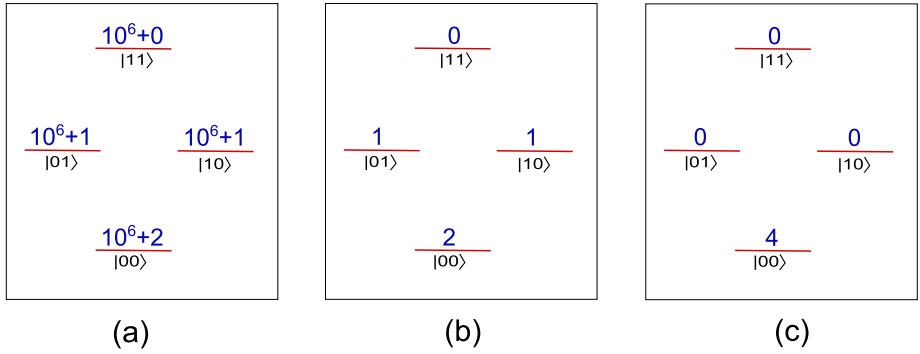}
	\caption[Population distribution of a two-spin system at room temperature]{Representative population distribution of a two-spin system at room temperature when
the system is in equilibrium (a), the deviation distribution (b), and the deviation distribution corresponding
to $\vert 00\rangle$ pseudopure state.}
	\label{fig:ppsdev}
\end{figure}

 In 1997 it was suggested independently by Cory et al \cite{corypps} and Chuang et al \cite{chuangpps} that a specially 
prepared mixed state can be prepared, known as pseudopure state (PPS), that can simulate a pure state in NMR. PPS are isomorphic to pure states for several computational problems \cite{corypps,chuangpps}. From Eq. \ref{rhodev}, it is seen that the equilibrium density operator ($\rho$) can be split into two parts, the identity part ($\mathbbm{1}$) and the deviation part ($\rho_{dev}$). The identity part comes as a uniform background and does not give any kind of NMR signals. The NMR signal solely depends on the deviational part of the density operator $(\rho_{dev})$. Although preparation of pure states in NMR is a very difficult and technically challenging task, preparing a pure deviational density operator is rather a easy one. Hence a pseudopure state (PPS) is an ensemble with a pure deviational density operator. It is also known as `effective pure state' since it mimics a `pure state'. Fig. \ref{fig:ppsdev} shows representative population distributions of a two-spin system in equilibrium (Fig. \ref{fig:ppsdev}a), the deviation part distribution (Fig. \ref{fig:ppsdev}b), and the pure deviational distribution (Fig. \ref{fig:ppsdev}c). Fig. \ref{fig:ppsdev}c represents the `pseudopure state'. Equation \ref{rhodev} can be rewritten as follows :
\begin{equation}
\begin{split}
\rho_{eq} =& \frac{1}{2^N} (\mathbbm{1}+\epsilon\rho_{dev})\\
          =& \frac{1}{2^N} (1-\epsilon)\mathbbm{1} + \epsilon\vert\psi\rangle\langle\psi\vert = \rho_{pps},
\end{split}   
\label{pps}       
\end{equation} 
where $\epsilon$ is a measure of the magnetization retained in
the pseudopure state and it usually gets halved with every additional 
qubit \cite{MaheshBEN}.
The unit background is invariant under the Hamiltonian
evolution, does not lead to NMR signal and is ignored \cite{corypps} :
\begin{equation}
U\rho_{pps}U^{\dagger} = \frac{1}{2^N} (1-\epsilon)\mathbbm{1} + \epsilon U\vert\psi\rangle\langle\psi\vert U^{\dagger}.  
\label{ppsuni}       
\end{equation}

 Thus the equilibrium density matrix of a single spin-1/2 nucleus
is always in a pseudopure state.  Initializing a multi-spin system into a pseudopure state
however is essentially a non unitary process \cite{dieterrev}.  
 So far, PPS remains as the leading technique to simulate a pure state in NMR QIP.

\section{Methods for preparing pseudopure state}
\label{ppsmethods}
 Several methods have earlier
been proposed for the preparation of pseudopure states. These methods involved in averaging 
the magnetization modes over the sample space 
(called `spatial averaging' \cite{corysp}), or 
over spin space (called `logical labeling' \cite{chuangpps,kavita00}), or over
several transients (called `temporal averaging' \cite{knillpps}).
In some cases subsystem pseudopure states are easier to prepare 
either by transition selective pulses \cite{maheshpps}
or by coherence selection using pulsed field gradients \cite{MaheshBEN},
but these methods invariably result in loss of a qubit for further computation.

 Implementation of a bulk quantum computation can be described in a general way as the transformation of an initial density matrix $\rho_0$
into a final density matrix $\rho_{out}$ according to \cite{knillpps},
\begin{equation}
\rho_{out} = \sum_{i,j}R_i C P_j \rho_0 P_j^{\dagger} C^{\dagger} R_i^{\dagger},
\end{equation}
where $P_j$ is the preparation operator, $C$ is the unitary transformation corresponding to the computation
to be performed, and $R_i$ is the post-processing unitary operator, usually the identity.

  It can be noticed that for a single spin-1/2 system in thermal equilibrium, only two energy levels are possible. Hence, with an excess population in the lower energy level than higher energy level, the system is in a pseudopure state by default. However, for higher order spin systems (more than one) one has to suitably manipulate the spin order to achieve the desired pseudopure state. In the following, we will describe a few of the most successful methods for preparing pseudopure states in NMR. In section \ref{ppslls} we will introduce a new method for preparing pseudopure states by exploiting long-lived nature of singlet states.

\subsection{Temporal averaging}

\begin{figure}
	\centering
		\includegraphics[width=13cm]{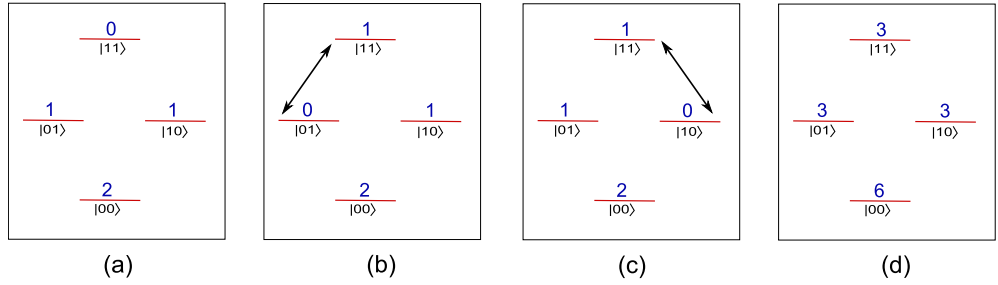}
	\caption[Preparation of $\vert 00\rangle$ pseudopure state by using temporal averaging method]
	{Preparation of $\vert 00\rangle$ pseudopure state by using temporal averaging method. Representative deviation population
	distribution at thermal equilibrium (a). Modified deviation populations after inverting populations between $\vert 01\rangle$ , $\vert 11\rangle$ (b) and $\vert 10\rangle$, $\vert 11\rangle$ (c). The population distribution of $\vert 00\rangle$ PPS is obtained by temporal averaging technique i.e. by adding (a), (b), and (c).}
	\label{fig:temporal}
\end{figure}
 Temporal averaging method is one of the earlier techniques that was proposed by Knill et. al. \cite{knillpps}. This method is based on the randomization of the probabilities of other states by adding multiple experiments. 
By averaging out all this experiments, we can retain the desired component whereas other undesirable terms cancel out.
The basic idea of this method is quite similar to phase-cycling technique, routinely used in NMR. 
For a N-qubit system, the exhaustive averaging involves cyclically permuting the non-ground states
in $2^N-1$ different ways. This permutation achieved by CNOT gates which can be implemented by using 
spin-selective pulses and evolution of couplings or by transition selective pulses.\\
 
 The method for preparing $\vert 00\rangle$ PPS for a 2 qubit system is described here (Fig. \ref{fig:temporal}).
Figure \ref{fig:temporal}a shows the deviational part of the equilibrium population distribution. After permuting the
populations of $\vert 01\rangle$ and $\vert 11\rangle$ by applying a transition selective $\pi$ pulse, we obtain the population distribution similar to Figure \ref{fig:temporal}b. Similarly, a permutation of populations of $\vert 10\rangle$ and $\vert 11\rangle$ can also be obtained as shown in Figure \ref{fig:temporal}c. Now adding these three population distributions, we can obtain the $\vert 00\rangle$ PPS as shown in Figure \ref{fig:temporal}d.\\
 
 Though this method is useful for smaller number of qubits, it becomes quite laborious for higher number of spin systems. The number of experiment (cyclic permutations) increases exponentially with the number of qubits. For example, preparing a 4-qubit PPS, would require 15 cyclic permutations. Because of this limitations various improvements on this method have been proposed such as by using non-cyclic permutations and applying unequal weightings to different permutations. Utilizing this improvements, one can find out the 4-qubit PPS by only 5 experiments \cite{mori05}. Choosing random permutations may turn out effective for very large spin systems \cite{knillpps}.

\subsection{Logical labeling}

\begin{figure}
	\centering
		\includegraphics[width=12cm]{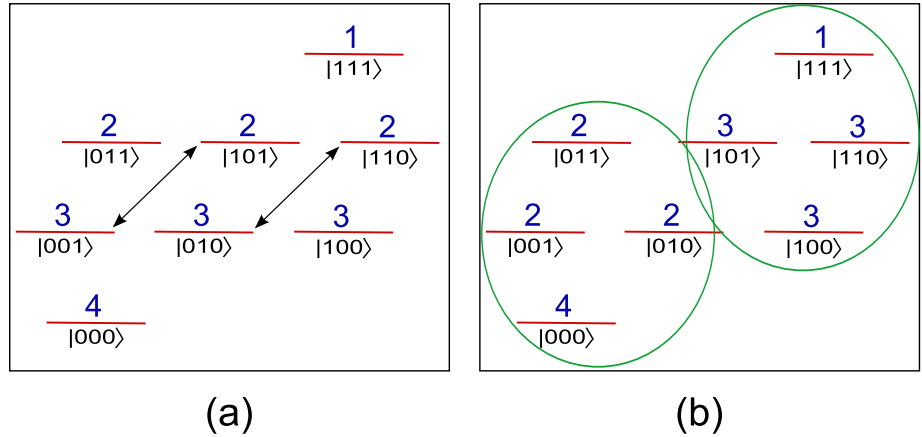}
	\caption[Preparation of pseudopure state by using logical labeling technique]
	{Preparation of pseudopure state by using logical labeling technique in a three qubit system. (a) Representative deviation population at thermal equilibrium (b) The three qubit system can be thought of consisting two subsystems corresponding to 0 and 1 state of the first qubit. After inverting the transitions $001 \leftrightarrow 101$ and $010 \leftrightarrow 110$, 0 subsystem is in $\vert 00\rangle$ pseudopure state and the 1 subsystem is in -$\vert 11\rangle$ pseudopure state.}
	\label{fig:logical}
	\end{figure}

 The logical labeling technique \cite{chuangpps, ChuangPRSL98} can be used to prepare conditional (or subsystem) pseudopure states. It exploits the fact that in an N-spin homonuclear system , there are $^NC_{N/2}$ levels having equal population distributions. The basic idea of this approach is to find the subset having similar kind of pattern of a pseudopure state and then `relabel' these states.
 
 Let us consider a case for a three-spin homonuclear system with the equilibrium deviation populations as shown in Figure  \ref{fig:logical}a). By permuting the states $\vert 001\rangle$ and $\vert 010\rangle$ with $\vert 101\rangle$ and $\vert 110\rangle$ respectively, we obtain the population distribution similar to the Figure \ref{fig:logical}b. Now these states correspond to the deviation density matrix,

\begin{equation}
\rho = \vert 0\rangle\langle 0 \vert \otimes (2\vert 00\rangle\langle 00\vert + 2I) + \vert 1\rangle\langle 1\vert \otimes (3I - 2\vert 11\rangle\langle 11\vert).
\end{equation}
 From the above equation it is clearly visible that depending on the state of the first qubit, we can have $\vert 00\rangle$ and $\vert 11\rangle$ pseudopure state of the remaining qubit. Again, the permutations can be carried out by spin-selective pulses and evolution of couplings or by transition selective pulses. Here the first qubit is acting as a ancilla qubit and this should not be disturbed during the computation. The two subsystems (two circles in Fig. \ref{fig:logical}b) undergo independent and parallel evolution. \\
 
 For higher number of qubits, more labeling qubits may be required. However, it is not necessary that all the subsystems are in pseudopure states. One clear disadvantage of this technique is the requirement of one `extra' qubit as ancilla. 

\subsection{Spatial averaging}

\begin{figure}
	\centering
		\includegraphics[width=12cm]{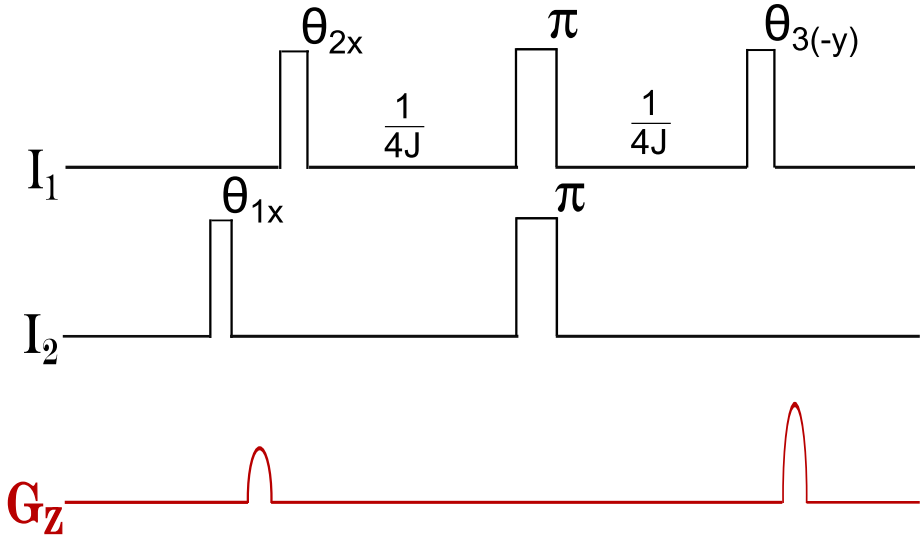}
	\label{fig:spatial}
	\caption[Pulse sequence for preparing $\vert 00\rangle$ pseudopure state by using spatial averaging method]
	{Pulse sequence for preparing $\vert 00\rangle$ pseudopure state by using spatial averaging method in a two qubit system ($I_1$, $I_2$). $\theta$ representing particular pulse angles and subscript denoting phase of the pulses. $J$ is the spin-spin coupling constant between two spins and $G_Z$ denoting pulse field gradient pulses.
	   }
	\end{figure}

Spatial averaging technique relies on the application of suitable selective pulses and magnetic field gradients \cite{corypps}. This method was proposed by Cory et. al. and since then it is proven to be one of the popular techniques to prepare pseudopure states. Our aim is to prepare the state $I_z^1 + I_z^2 + 2I_z^1I_z^2$ from the state $\gamma_1 I_z^1 + \gamma_2 I_z^1$. Depending on the situation whether $\gamma_1$ is greater, less, or equal to $\gamma_2$, we will have exact values for $\theta$s. For a homonuclear system ($\gamma_1 = \gamma_2$) or $\vert\gamma_1/2\gamma_2\vert \leq 1$, we will have: $(\theta_1, \theta_2, \theta_3) = (\pi/3, \pi/4, \pi/4)$. On the other hand, suppose $\vert\gamma_1/2\gamma_2\vert \geq 1$ e.g., $^1H - ^{13}C$ system, we will have the $\theta$ values as $(\theta_1, \theta_2, \theta_3) = (0, \pi/12, 5\pi/12)$.\\

 A particular case for a homonuclear spin system for the preparation of $\vert 00\rangle$ pseudopure state is described below by product operator formalism.

\begin{eqnarray}
&I_z^1 + I_z^2& \nonumber \\
&{\color{blue}{\downarrow} (\frac{\pi}{3})_{x}^{2}}& \nonumber \\
&I_z^1 + \frac{1}{2}I_z^2 - \frac{\sqrt 3}{2}I_y^2& \nonumber \\
&{\color{blue}{\downarrow} G_1}& \nonumber \\
&I_z^1 + \frac{1}{2}I_z^2& \nonumber \\
&{\color{blue}{\downarrow} (\frac{\pi}{4})_{x}^{1}}& \nonumber \\
&\frac{1}{\sqrt2}I_z^1 - \frac{1}{\sqrt2}I_y^1 + \frac{1}{2}I_z^2& \nonumber \\
&\hspace{2.5cm}{\color{blue}{\downarrow}\left[ \frac{1}{4J}-\pi_{x}-\frac{1}{4J} \right]} \nonumber \\
&\frac{1}{\sqrt2}I_z^1 + \frac{1}{\sqrt2}2I_x^1I_z^2 + \frac{1}{2}I_z^2& \nonumber \\
&{\color{blue}{\downarrow} (\frac{\pi}{4})_{-y}^{1}}& \nonumber \\
&\frac{1}{2}I_z^1 - \frac{1}{2}I_x^1 + \frac{1}{2}2I_x^1I_z^2 + \frac{1}{2}I_z^2 + \frac{1}{2}2I_z^1I_z^2& \nonumber \\
&{\color{blue}{\downarrow} G_2}& \nonumber \\
&\frac{1}{2}I_z^1 + \frac{1}{2}I_z^2 + \frac{1}{2}2I_z^1I_z^2&
\end{eqnarray}

 The pulse sequence for three spin homonuclear weakly coupled system is also described \cite{cory98, aviktsm, avik}. Generalization of spatial averaging method for N-qubits is given by sakaguchi et. al. \cite {sakaguchi}.
 
  In addition to these three main approaches , there are few other methods also available to prepare pseudopure states such as spatially averaged logical labeling technique (SALLT) \cite{maheshpps}, using `cat states', etc \cite{knill00}.

 In the next section we propose a different approach
that exploits long life-times of certain special states called
`singlet states' \cite{LevPRL04, LevittJACS04}. We don't need an extra qubit in this method as it was 
required in logical labeling technique \cite{maheshpra10}. 
The following section gives a detail description of the theory and pulse sequence required
for the preparation of pseudopure state by this new approach. Later,
experimental demonstrations on model systems consisting of two, three 
and four-qubit NMR registers are also described.

\section{Preparation of pseudopure states using Long-Lived Singlet States}
\label{ppslls}
 Consider an ensemble of identical molecules each having $n$ spin-1/2 nuclei in a magnetic
field.  The Zeeman Hamiltonian ${\cal H}_z = h \sum_j \nu_0^j I_z^j$,
is characterized by the frequency $\nu_0^j$ of Larmor precession,
and the z-component of the spin angular momentum operator $I_z^j$ of spins $j = 1 \cdots n$
\cite{LevBook}.
The eigenstates $\vert \pm 1/2 \rangle$ of ${\cal H}_z^j$
are labeled as $\vert 0 \rangle$ and $\vert 1 \rangle$ states
of a qubit, and the  multi-spin eigenbasis 
$\{\vert 00 \cdots 00 \rangle, \vert 00 \cdots 01 \rangle, \cdots \}$
is treated as the computational basis.
\subsection{preparation of singlet states}
\label{secsingstates}
 The Hamiltonian for an ensemble of spin-1/2 nuclear pairs of same isotope, in 
the RF interaction frame, can be expressed as
\begin{eqnarray}
{\cal H^{\mathrm{eff}}} =  h \left[ 
         \frac{\Delta\nu}{2}  I_z^1 
         - \frac{\Delta\nu}{2}I_z^2 
         +  J I^1 \cdot I^2
         +  \nu_{12} I_x^{1,2}
          \right].
\label{heff1}         
\end{eqnarray}
Here the RF frequency is assumed to be at the
mean of the two Larmor frequencies, and
$\Delta \nu$, $J$ and $\nu_{12}$ correspond  respectively to the
difference in Larmor frequencies (chemical shift difference), the
scalar coupling constant and the RF amplitude (all in Hz).

\begin{figure}
	\centering
		\includegraphics[width=3.5cm,angle=-90]{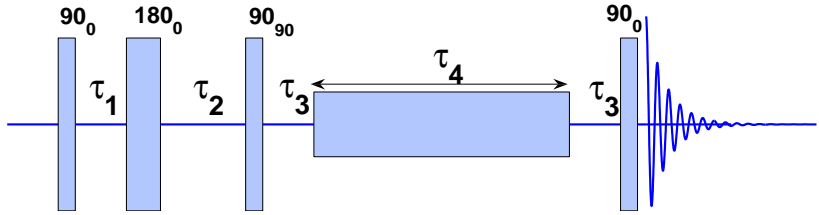}
	\caption[Pulse sequence for the preparation and detection of singlet states]{Pulse sequence for the preparation and detection of singlet states. 
	All the pulses are non-selective pulses.
	Here $\tau_1 = 1/(4J)$, $\tau_2 = 1/(4J) + 1/(2\Delta\nu)$, and $\tau_3 = 1/(4\Delta\nu)$, with $\Delta\nu$ and $J$
being the chemical shift difference (in Hz) and the scalar coupling respectively. $\tau_4$ is
the duration of spin-lock.}
	\label{ppsllsf1}
\end{figure}

 The detail version of singlet state preparation is shown in Chapter 2.2. The standard pulse sequence for the preparation of singlet state is shown in Fig. \ref{ppsllsf1}. 

The propagator form of this pulse sequence upto spin-lock can be written as follow :
\begin{eqnarray}
U_1^{1,2} = 
      \mathrm{e}^{-i \frac{\pi}{4} (I_z^1 - I_z^2)}
 \mathrm{e}^{-i \frac{\pi}{2} I_y^{1,2} }
 \mathrm{e}^{-i \frac{\pi}{2} (I_z^1 - I_z^2 )}
 \mathrm{e}^{-i \pi I_z^1 I_z^2}
 \mathrm{e}^{-i \frac{\pi}{2} I_x^{1,2}}.
\label{singprep}
\end{eqnarray}

 The singlet states by themselves are inaccessible to
macroscopic observables, but
can be indirectly detected by removing the equivalence and
transforming to observable single quantum coherence using the following propagator (see Fig. \ref{ppsllsf1}) \cite{LevPRL04,LevittJACS04}.
\begin{eqnarray}
U_D^{1,2} = 
\mathrm{e}^{-i \frac{\pi}{2} I_x^{1,2}}
\cdot \mathrm{e}^{i \frac{\pi}{4} (I_z^1 - I_z^2)}.
\label{singdet}
\end{eqnarray}   

 A more detailed and quantitative analysis of singlet states 
was carried out  using density matrix tomography in Chapter 2.3. 
In the next section we described the preparation of pseudopure states starting from this pure singlet states.   

\subsection{Initializing NMR Registers}
\subsubsection{2-qubit register}
\label{secsingini}
\begin{figure}
	\centering
		\includegraphics[width=12cm]{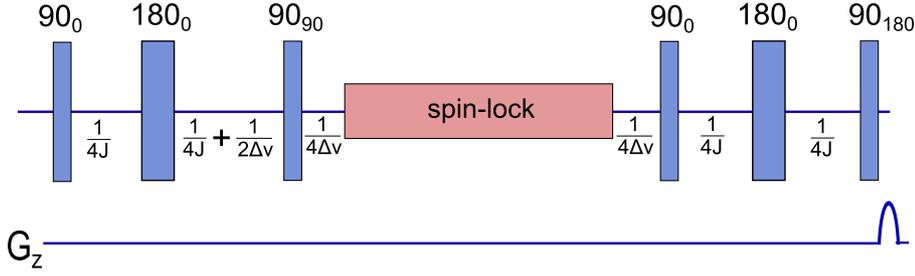}
	\caption[Pulse sequence for the creation of $\vert 01\rangle$ pseudopure state]{Pulse sequence for the creation of $\vert 01\rangle$ pseudopure state. $\Delta\nu$ and $J$
being the chemical shift difference (in Hz) and the scalar coupling respectively. A pulse field gradient ($G_z$)
used to destroy the unwanted coherences at the end of pulse sequence. }
	\label{pps2qPP}
\end{figure}

 The pulse sequence for initializing a 2-qubit NMR register 
via singlet states is shown in Fig.\ref{pps2qPP}.
An initially imperfect singlet density matrix gets purified during the spin-lock period 
as a result of the long life time, while the 
artifact coherences are destroyed by relaxation process as well as 
the inhomogeneities in the spin-lock itself. 
There exist
optimal spin-lock conditions at which one obtains singlet states
with high fidelity \cite{maheshjmr10, maheshpra10}.

Once the singlet state is prepared with high fidelity, the conversion 
$\vert S_0^{1,2} \rangle \rightarrow \vert 01 \rangle$ 
can be easily achieved by the propagator $U_2^{1,2}$ described by
\begin{eqnarray}
U_2^{1,2} = 
      \mathrm{e}^{i \frac{\pi}{2} I_x^{1,2}}
\cdot \mathrm{e}^{-i \pi I_z^1 I_z^2}
\cdot \mathrm{e}^{-i \frac{\pi}{2} I_x^{1,2}}
\cdot \mathrm{e}^{i \frac{\pi}{4} (I_z^1 - I_z^2)}.
\label{singtopps}
\end{eqnarray}

This propagator work as follows:

\begin{eqnarray}
&\vert S_0\rangle\langle S_0\vert = \frac{1}{4}{\mathbbm 1} - 2I_x^1I_x^2 - 2I_y^1I_y^2 - 2I_z^1I_z^2& \nonumber \\
&{\color{blue}{\downarrow} \frac{1}{4\Delta\nu}}& \nonumber \\
&2I_x^1I_y^2 - 2I_y^1I_x^2 - 2I_z^1I_z^2& \nonumber \\
&{\color{blue}{\downarrow} 90_{0}}& \nonumber \\
&2I_x^1I_z^2 - 2I_z^1I_x^2 - 2I_y^1I_y^2& \nonumber \\
&\hspace{2.5cm}{\color{blue}{\downarrow}\left[ \frac{1}{4J}-180_{0}-\frac{1}{4J} \right]} \nonumber \\
&-I_y^1 + I_y^2 - 2I_y^1I_y^2& \nonumber \\
&{\color{blue}{\downarrow} 90_{180}}& \nonumber \\
&I_z^1 - I_z^2 - 2I_z^1I_z^2 = \vert 01\rangle\langle 01\vert&
\end{eqnarray}

Finally a pulsed field gradient $G_z$ can be used to destroy the
residual single and multiple quantum coherences
generated due to pulse imperfections.  
If necessary, other pseudopure states can be obtained simply by 
applying NOT gates.

\subsubsection{3-qubit register}  
\begin{figure}
\begin{center}
\includegraphics[width=14cm]{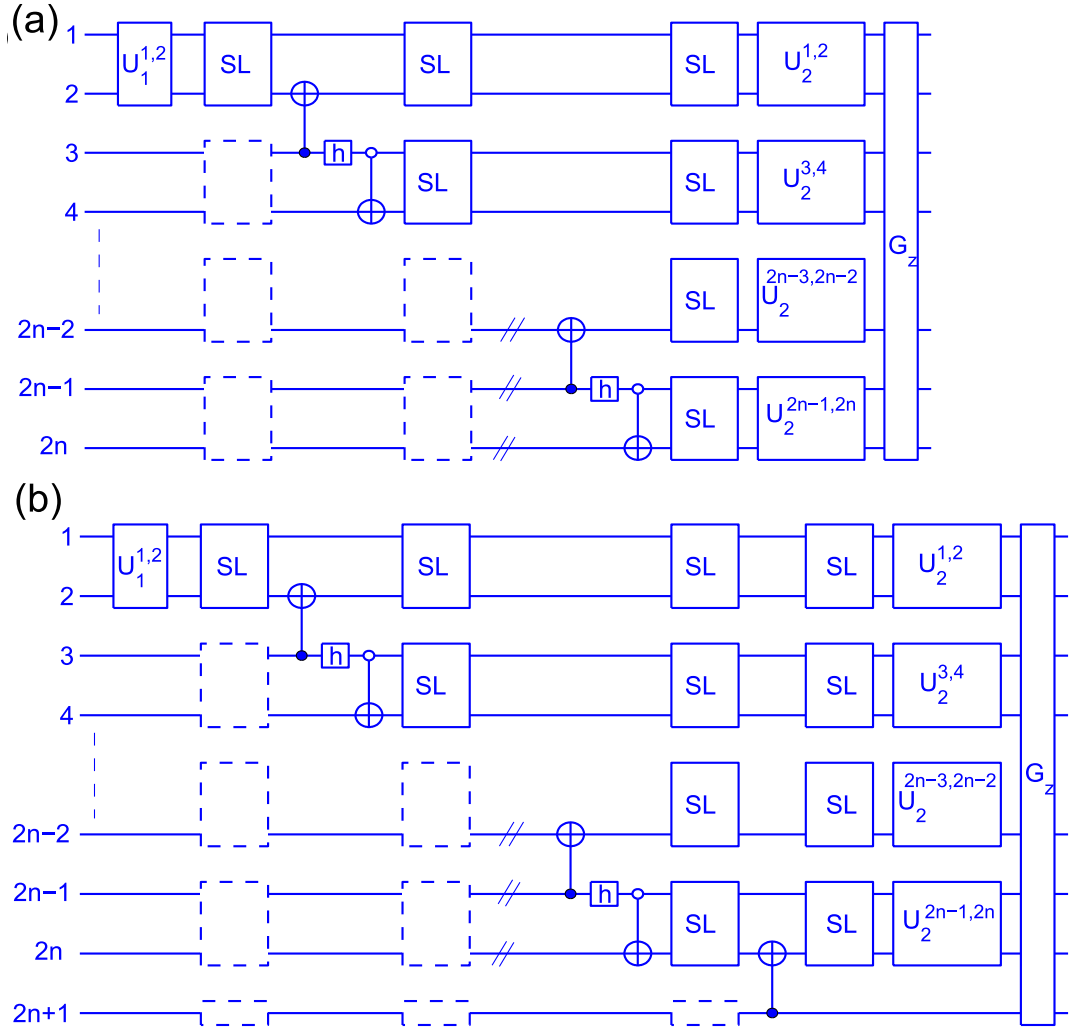} 
\caption[The circuit diagrams for initializing n-qubit register]{
The circuit diagrams for initializing (a) even-qubit register, 
and (b) odd-qubit register. Here SL and G$_\mathrm{z}$ denote spin-lock and the pulsed
field gradient.  $U_1$ and $U_2$ are gates as described the text.
In (a) and (b) the C'NOT gates with open circles correspond to NOT operation if the
control is 0 and identity if the control is 1. The h-gate corresponds to 
pseudo-Hadamard: 
$\vert 0 \rangle \stackrel{\mathrm{h}}{\rightarrow} (\vert 0 \rangle - \vert 1 \rangle)/\sqrt{2}$
and 
$\vert 1 \rangle \stackrel{\mathrm{h}}{\rightarrow} (\vert 0 \rangle + \vert 1 \rangle)/\sqrt{2}$.
The dashed boxes indicate optional spin-locks.
}
\label{q123pps} 
\end{center}
\end{figure}

 For a 3-qubit system (see Fig.\ref{q123pps}c),
we first prepare a two-qubit
singlet and apply CNOT(2,3), i.e., a NOT gate on qubit-2 controlled 
by qubit-3.  
Subsequent spin-lock and $U_2^{1,2}$
gate on qubits 1 and 2 initializes a three qubit system 
into $\vert 010 \rangle$ state \cite{maheshpra10}.

\begin{figure}
	\centering
		\includegraphics[width=12cm]{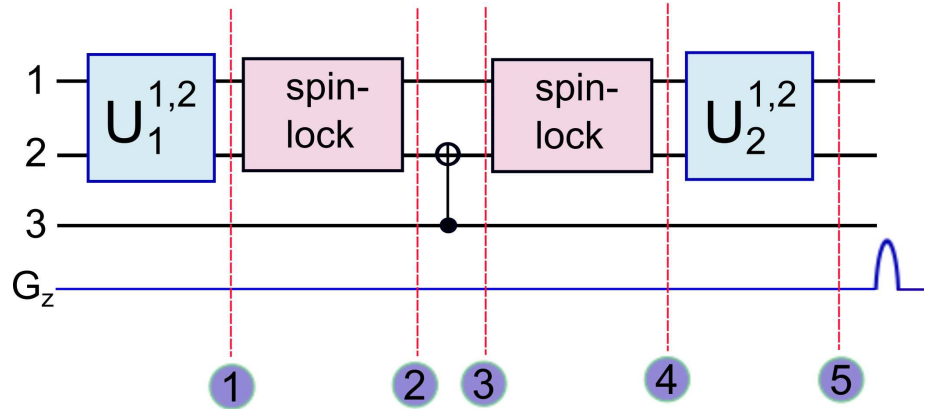}
	\caption[Circuit diagram for the preparation of a 3 qubit pseudopure state]{Circuit diagram for the preparation of a 3 qubit pseudopure state.
	$U_{1}^{1,2}$ (see Eq. \ref{singprep}) and $U_{2}^{1,2}$ (see Eq. \ref{singtopps}) are unitary propagator as described previously. the C'NOT gates with open circles correspond to NOT operation if the
control is 0 and identity if the control is 1 $G_z$ denoting a pulse field gradient. }
	\label{pps3qPP}
\end{figure} 

The circuit can be understood as follows.  After 
preparing the singlet on qubits 1 and 2 (at time point 2 in Fig. \ref{pps3qPP}), the third qubit remains in a mixed
state with a probability $p_0$ of being in state $\vert 0 \rangle$ and
a probability $p_1$ of being in state $\vert 1 \rangle$.
Since,
\begin{eqnarray}
\frac{\vert 01 \rangle - \vert 10 \rangle}{\sqrt{2}} \otimes \vert 0 \rangle
&\xrightarrow{\mathrm{CNOT(2,3)}}&
\frac{\vert 01 \rangle - \vert 10 \rangle}{\sqrt{2}} \otimes \vert 0 \rangle, 
\mathrm{\; and} \nonumber \\
\frac{\vert 01 \rangle - \vert 10 \rangle}{\sqrt{2}} \otimes \vert 1 \rangle
&\xrightarrow{\mathrm{CNOT(2,3)}}&
\frac{\vert 00 \rangle - \vert 11 \rangle}{\sqrt{2}} \otimes \vert 1 \rangle,
\label{s0topps}
\end{eqnarray}
the CNOT(2,3) gate transforms the mixed state according to (at time point 3 in Fig. \ref{pps3qPP}):
\begin{eqnarray}
\vert S_0^{1,2} \rangle \langle S_0^{1,2} \vert \otimes 
\left( p_0 \vert 0 \rangle \langle 0 \vert +  
 p_1 \vert 1 \rangle \langle 1 \vert \right)
\xrightarrow{\mathrm{CNOT(2,3)}}  
\nonumber \\
p_0 \vert S_0^{1,2} \rangle \langle S_0^{1,2} \vert \otimes 
\vert 0 \rangle \langle 0 \vert +  
p_1 \vert \phi_-^{1,2} \rangle \langle \phi_-^{1,2} \vert \otimes 
\vert 1 \rangle \langle 1 \vert.
\end{eqnarray}

During the second spin-lock applied on qubits 1 and 2, the
singlet part survives, where as the second term consisting of 
the Bell state $\vert \phi_-^{1,2} \rangle = (\vert 00 \rangle - \vert 11 \rangle)/\sqrt{2}$
decays fast (at time point 4 in Fig. \ref{pps3qPP}).  The singlet $\vert S_0^{1,2} \rangle$
is ultimately transformed into $\vert 01 \rangle$
by $U_2^{1,2}$ and thus we obtain $\vert 010 \rangle$ pseudopure state with 
a good approximation (at time point 5 in Fig. \ref{pps3qPP}).  
As mentioned in the introduction, at ordinary NMR conditions $p_0 \approx p_1$, and  
therefore discarding the second term means a loss of magnetization by a factor of 2.
Thus the magnetization is halved with every additional qubit in the register.
This scaling behavior is similar to that of other traditional methods \cite{mosca}.

\subsubsection{4-qubit register}
\begin{figure}
	\centering
		\includegraphics[width=12cm]{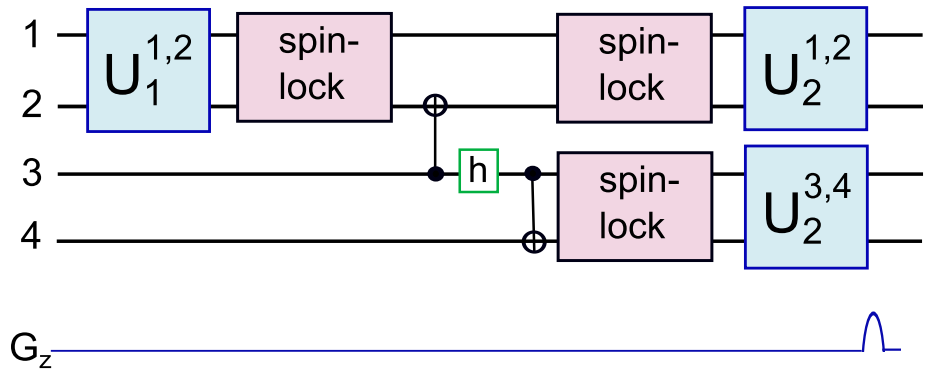}
	\caption[Circuit diagram for the preparation of a 4 qubit pseudopure state]{Circuit diagram for the preparation of a 4 qubit pseudopure state.
	$U_{1}^{1,2}$ (see Eq. \ref{singprep}), $U_{2}^{1,2}$ (see Eq. \ref{singtopps}), and $U_{2}^{3,4}$ are unitary propagators. $U_{2}^{3,4}$ is a similar propagator as $U_{2}^{1,2}$, acting on qubit 3 and 4. $G_z$ denoting a pulse field gradient to destroy unwanted coherences. the C'NOT gates with open circles correspond to NOT operation if the
control is 0 and identity if the control is 1. The h-gate corresponds to 
pseudo-Hadamard: 
$\vert 0 \rangle \stackrel{\mathrm{h}}{\rightarrow} (\vert 0 \rangle - \vert 1 \rangle)/\sqrt{2}$
and 
$\vert 1 \rangle \stackrel{\mathrm{h}}{\rightarrow} (\vert 0 \rangle + \vert 1 \rangle)/\sqrt{2}$.}
	\label{pps4qPP}
\end{figure}

 Let us now analyze a 4-qubit register (see Fig. \ref{pps4qPP}).  
After the first spin-lock, we have the singlet
state $\vert S_0^{1,2} \rangle$ on 1 and 2.  The third and fourth qubits 
are still in mixed states.  As seen before, if the third qubit is $\vert 1 \rangle$,
the CNOT(3,2) converts $\vert S_0^{1,2} \rangle$ into $\vert \phi_-^{1,2} \rangle$,
which will be eventually dephased out during the second spin-lock.  Therefore we
shall consider the third spin to be in state $\vert 0 \rangle$.  
Now if the fourth qubit is in state $\vert 0 \rangle$,
the pseudo-Hadamard h(3) gate followed by C'NOT(4,3)
(which applies NOT on qubit-4 only if qubit-3 is 
in state $\vert 0 \rangle$) will lead to a correlated state of singlet pairs:
\begin{eqnarray}
\vert S_0^{1,2} \rangle \otimes \vert 0 \rangle \otimes \vert 0 \rangle
\xrightarrow{\mathrm{h(3)}}
&\vert S_0^{1,2} \rangle \otimes \frac{\vert 0 \rangle - \vert 1 \rangle}{\sqrt{2}}  \otimes \vert 0 \rangle&
\nonumber \\
&\hspace{1cm}\downarrow \mathrm{C'NOT(4,3)} &\nonumber \\
&\vert S_0^{1,2} \rangle \otimes \vert S_0^{3,4} \rangle.&
\end{eqnarray}
By similar analysis one obtains 
$\vert S_0^{1,2} \rangle \otimes \vert \phi_-^{3,4} \rangle $ 
if the 4th qubit is originally in state $\vert 1 \rangle$.
This latter spin-order decays fast due to the shorter life-time
of $\vert \phi_-^{3,4} \rangle$.  Finally the long-lived spin-order
$\vert S_0^{1,2} \rangle \otimes \vert S_0^{3,4} \rangle$ can be
converted into $\vert 0101 \rangle$ pseudopure state by the
propagators $U_2^{1,2}$ and $U_2^{3,4}$.

The fact that only nearest-neighbor 
interactions are used is highly advantageous in practice \cite{maheshpra10}.  
Experimentally, the spin-lock of multiple singlet pairs can be achieved 
using sophisticated modulated RF sequences as described in the next section.

\subsubsection{Initialization for any number of qubits}
 Initialization of NMR qubits through long-lived singlet states
 can be extended to any number of qubit in principle \cite{maheshpra10}.
The circuits for initialization of registers with odd and even
number of qubits are shown in Fig.\ref{q123pps}b and \ref{q123pps}c 
respectively.  The basic idea is to divide the register into qubit-pairs 
and prepare a correlated state of singlets.  Each pair must
consist of two qubits of same nuclear species (homonuclear), 
but different pairs may be made up of different species (heteronuclear).
The correlated singlet states can be prepared by using CNOT gates
and pseudo-Hadamard gates as shown in Fig.\ref{q123pps}(b-c).
As each pair is converted into a singlet, an RF spin-lock is applied
on all the singlet pairs.  Under the RF spin-lock all the states except
the singlet states decay rapidly.  The circuits in Fig.\ref{q123pps}b and 
\ref{q123pps}c differ only at the last qubit which is unpaired in
the odd register.
If two or more qubit-pairs are made up of same nuclear species, then
it might be difficult to selectively spin-lock some of them leaving out
others.  
However, as described by the optional spin-locks shown by the dashed boxes 
in Fig.\ref{q123pps}, applying spin-locks on pairs which are not yet singlets 
has little effect on the overall scheme.  
Only exception is for the final spin-lock on an odd-qubit register,

which must be applied such that the
last unpaired qubit is left undisturbed.  It may be possible in some
cases to overcome even this requirement (see 3-qubit experiment).  
Nevertheless it may be desirable to have the unpaired qubit in
an odd-qubit register to be of different nuclear species than all
others.

\begin{figure}
\begin{center}
\includegraphics[width=10cm]{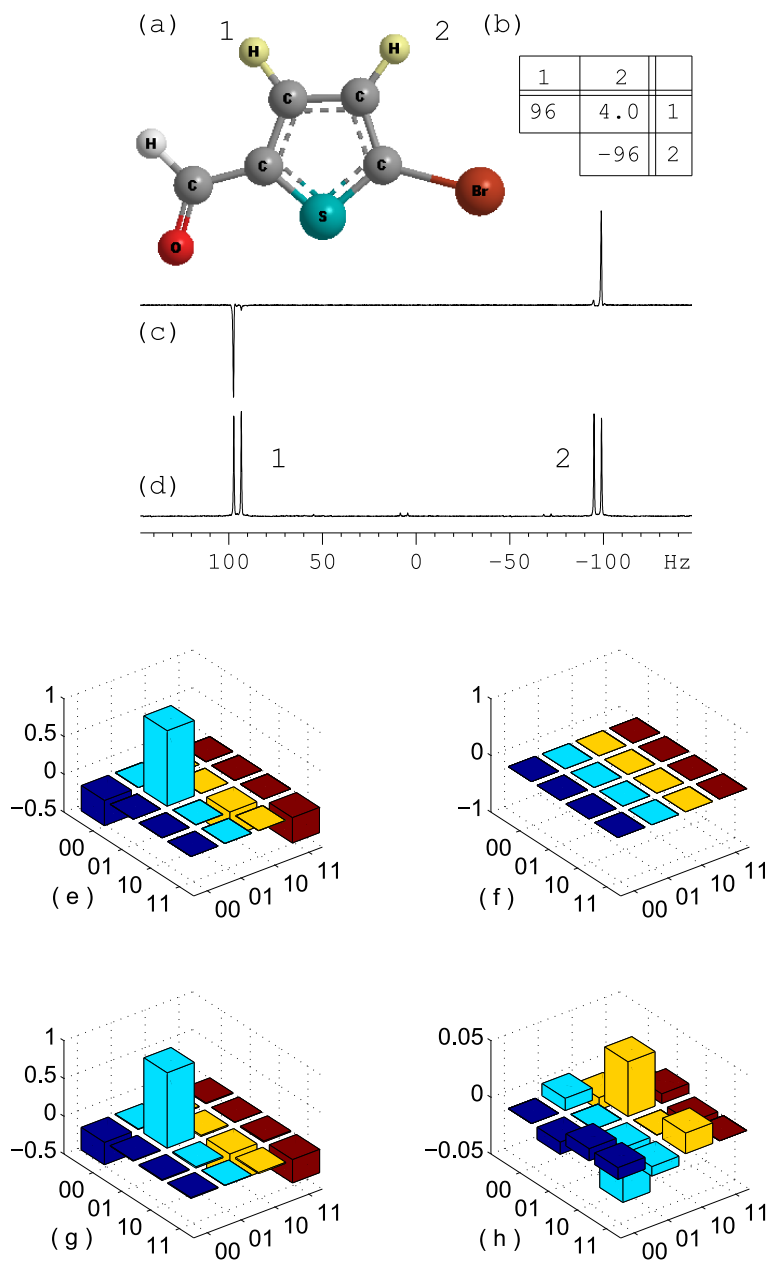}
\caption[Experimental results for a 2 qubit register]{
The molecular structure (a) and Hamiltonian parameters (b) of 
5 -bromothiophene-2-carbaldehyde (5 mg dissolved in 0.5 ml of dimethyl sulphoxide-d6), 
forming a homonuclear two-qubit register.
In (b) diagonal and off-diagonal elements correspond to the chemical shifts and
the scalar coupling constant respectively (in Hz).
The $^1$H spectra correspond to
the pseudopure state (c) and the equilibrium mixed state (d).
The barplots (e-h) correspond to the real (e,g) and imaginary (f,h)
parts of theoretical (e,f) and experimental (g,h) pseudopure
$\vert 01 \rangle$ state.
}
\label{pps2qexp}
\end{center}
\end{figure}

\section{Experiments}
 Strongly modulated pulses are used for designing high fidelity
local gates as well as CNOT gates  \cite{fortunato, MaheshNGE}.
The spin-lock was achieved by WALTZ-16 - a phase modulated RF sequence,
which is routinely used in broadband spin decoupling \cite{maheshjmr10}.  
Spectra corresponding to pseudopure states are obtained by linear detection
scheme using small flip angle RF pulses.  Since
the diagonal pseudopure states have one energy level more populated than
the equal distribution in all others, 
the spectrum should consist ideally of only one transition 
per qubit in each case. Quantitative analysis of the pseudopure states
are carried out using extended versions of density matrix tomography 
described in chapter 1 (Appendix A) \cite{maheshjmr10}.  Finally, the success of 
the experimental state $\rho$ in achieving a target pseudopure state
$\rho_\mathrm{pps}$ is measured by calculating the correlation \cite{fortunato},
\begin{eqnarray}
\langle \rho_\mathrm{pps} \rangle = \frac{\mathrm{trace}\left[ \rho \cdot \rho_\mathrm{pps} \right]}{\sqrt{ \mathrm{trace} \left[\rho^2 \right] \cdot  \mathrm{trace} \left[\rho_\mathrm{pps}^2 \right]}}.
\end{eqnarray}
Often only the diagonal elements of the density matrices are 
relevant and in such cases, the `diagonal correlation' can be expressed
by replacing all the operators in the above expression by their
diagonal parts  \cite{MaheshNGE}.
In the following we describe the individual cases of two-, three- and four-qubit
registers \cite{maheshpra10}.

\subsection{2-qubit register}
 The two-qubit system, Hamiltonian parameters, and the corresponding
pseudopure and the reference spectra are shown in Fig.\ref{pps2qexp}(a-d).
As shown in Fig.\ref{pps2qPP}, the experiment involved preparing singlet
using $U_1^{1,2}$, followed by RF spin-lock with amplitude 2 kHz and
duration 12.4 s, which are optimized for high singlet content \cite{maheshjmr10}.
The decay constant for singlet state was 16.2 s approximately three
times the $T_1$ values of the two spins.
The singlet is then converted into $\vert 01 \rangle$ 
pseudopure state using $U_2^{1,2}$.  A final gradient pulse served
to destroy the artifact coherences.  
The bar plots showing the real and imaginary parts of the theoretical
and experimental density matrix are shown in Fig.\ref{pps2qexp}(e-h).  
A very high correlation of 0.995 is obtained with $\vert 01 \rangle$
pseudopure stat \cite{maheshpra10}.


\begin{figure}
\begin{center}
\includegraphics[width=10cm]{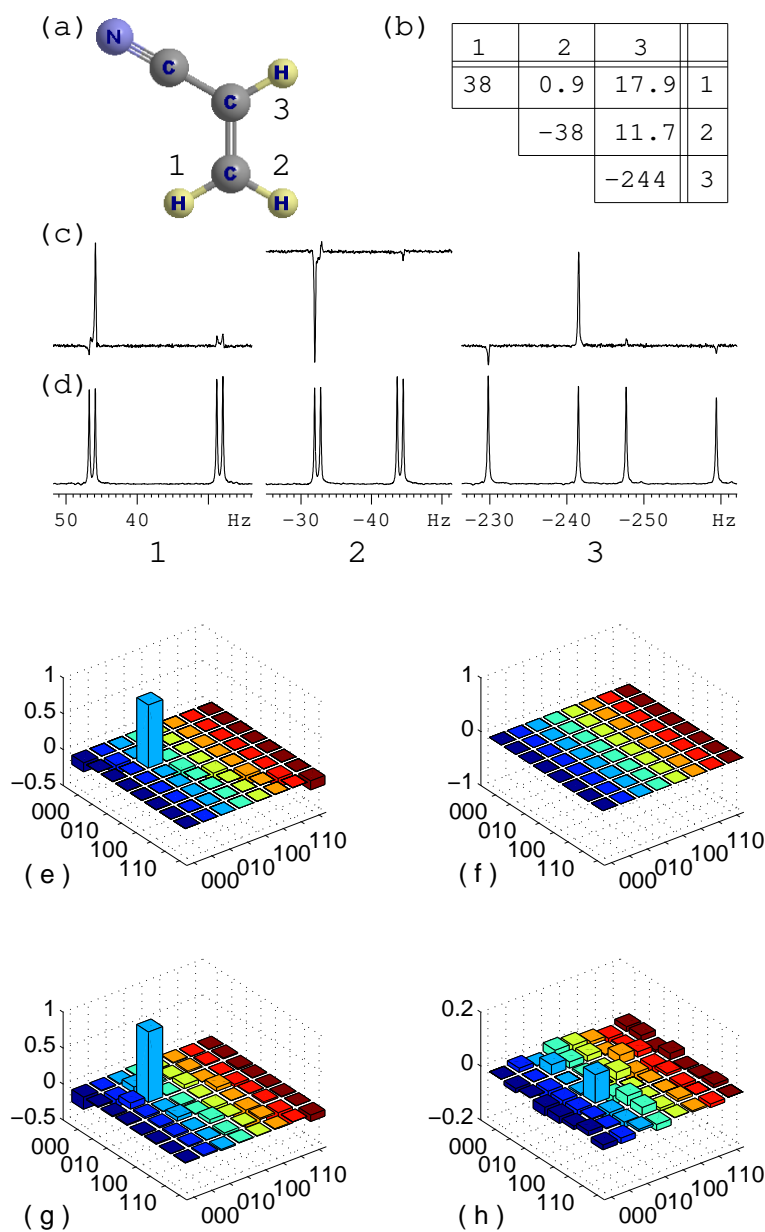}
\caption[Experimental results for a 3 qubit register]{
The molecular structure (a) and Hamiltonian parameters (b) of 
acrylonitrile (5 mg dissolved in 0.5 ml of CDCl$_3$), forming a homonuclear
3-qubit register.
In (b) diagonal elements are chemical shifts (in Hz) and the off-diagonal
elements are the scalar coupling constants (in Hz).  
The $^1$H spectra correspond to the pseudopure state (c) 
and the equilibrium mixed state (d). 
The bar plots are showing the real (e,g) and imaginary (f,h)
parts of theoretical (e,f) and experimental (g,h) pseudopure
$\vert 010 \rangle \langle 010 \vert$ state.  
}
\label{pps3qexp}
\end{center}
\end{figure}

\subsection{3-qubit register}
 The three-qubit system, Hamiltonian parameters and the corresponding
pseudopure and the reference spectra are shown in Fig.\ref{pps3qexp}(a-d).
The decay constant for singlet state of spins 1 and 2 was about 18 s, 
approximately three times of their $T_1$ values.
The pseudopure state was prepared using the circuit shown in 
Fig.\ref{pps3qPP}.
The off-set of the spin-lock was at the center of spins 1 and 2.
As described in section II-C, the second spin-lock is to be 
applied ideally on qubits 1 and 2 only (without disturbing 3rd qubit), 
which was harder to achieve in this homonuclear system.
However, we found that by carefully tuning the durations of the spin-locks,
we can obtain identical result to that of a no spin-lock on the 3rd spin.
This was possible due to 
(i) the cyclic nature of the WALTZ-16 spin-lock,
and 
(ii) the slower decay of $\vert 0^3 \rangle$ state during the 
off-resonant spin-lock compared to that of artifact coherences.
The two spin-locks consisted of 500 Hz WALTZ-16 
modulations whose durations were optimized to about 6.3 s.  
The CNOT gate was implemented using a 14 segment
strongly modulated RF pulse of duration approximately 60 ms and
of fidelity 0.96.
The bar plots showing the real and imaginary parts of the theoretical
and experimental density matrix are shown in Fig.\ref{pps3qexp}(e-h).  
The 3-spin tomography is an extension of the technique described in
reference \cite{maheshjmr10} and is described in appendix B.
The correlation of the experimental density matrix with the
theoretical pseudopure state is $\vert 010 \rangle$ is 0.952.  
The correlation is smaller compared to the two-qubit case, mainly
due to the errors in the CNOT gate and the non-selectivity of
the second spin-lock.  Nevertheless, the diagonal
correlation is achieved as high as 0.983 \cite{maheshpra10}.

\begin{figure}
\begin{center}
\includegraphics[width=10cm,angle=-90]{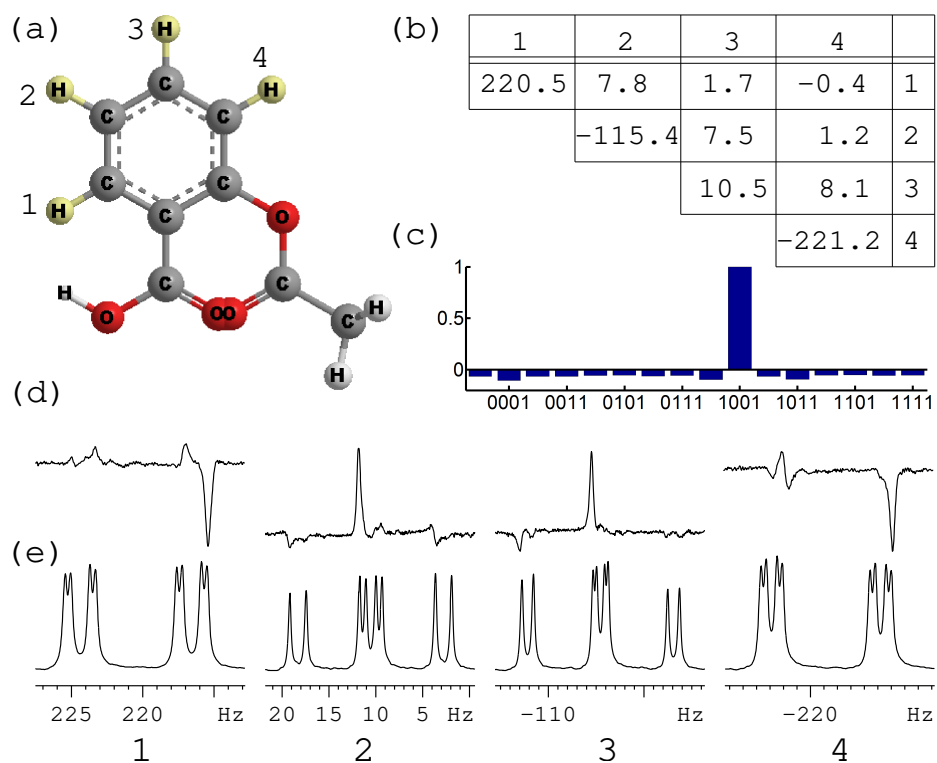}
\caption[Experimental results for a 4 qubit register]{
The molecular structure (a) and Hamiltonian parameters (b) of 
aspirin (5 mg dissolved in 0.5 ml of CD$_3$OD), forming a homonuclear 4-qubit register.
In (b) diagonal elements are chemical shifts (in Hz) and the off-diagonal
elements are the scalar coupling constants (in Hz).  
The barplot in (c) displays the diagonal elements of the density matrix
obtained by tomography of the pseudopure $\vert 1001 \rangle$ state.
The $^1$H spectra correspond to the pseudopure  state (d) 
and the equilibrium mixed state (e). 
}
\label{pps4q}
\end{center}
\end{figure}

\subsection{4-qubit register}
 The four-qubit system, Hamiltonian parameters and the corresponding
pseudopure and the reference spectra are shown in Fig.\ref{pps4q}.
The pseudopure state was prepared using the circuit shown in 
Fig.\ref{pps4qPP}.  
The singlet decay constants were about 6 s, approximately twice
the $T_1$ values of the individual spins.
We were able to carry out simultaneous 
spin-lock of two singlet pairs and initialize
a four-qubit register.  
The two spin-locks were achieved by 2 kHz WALTZ-16 modulations of
durations 2 s and 4.5 s each.
The two CNOT gates were made of 20 segments, approximately 61 ms duration
and of fidelities about 0.94.
The 10 segment h-gate was about 8.2 s long and of fidelity 0.98.
Complete tomography of a 4-qubit density matrix
is a laborious task.  After the preparation of the pseudopure state,
the non-zero quantum off-diagonal elements are efficiently
destroyed by the final gradient pulse.
Since only the diagonal elements are of main interest, we have 
carried out the four-qubit diagonal tomography \cite{MaheshNGE}.  
The bar plot showing the diagonal part of the 
experimental density matrix is shown in Fig.\ref{pps4q}.  
The diagonal correlation is estimated to be approximately
$0.97 \pm 0.01$ with $\vert 1001 \rangle$ pseudopure state.
The first pair has collapsed to $\vert 10 \rangle$ state instead 
of $\vert 01 \rangle$, due to an additional 180 degree pulse
that was applied on qubits 1 and 2 
for refocusing purposes during $U_2^{3,4}$ \cite{maheshpra10}.

\section{Conclusions}
\begin{figure}
\centering
\includegraphics[width=10cm]{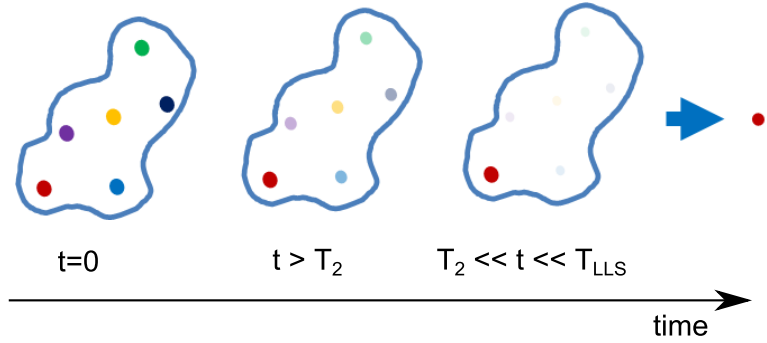}
\caption[Pictorial description of the evolution of PPS over time]{
Pictorial description of the evolution of PPS over time. At beginning (t=0), all the states are possible. After some time ($t\geq T_2$), most of the states other than singlet-PPS decay considerably. After a time period when $t\gg T_2$, only singlet-PPS remains while all other states decay close to zero.  }
\label{ppsfaint}
\end{figure}
 An ensemble of nuclear spin-pairs under certain conditions is
known to exhibit singlet state life-times much longer than other 
non-equilibrium states. This property
of singlet state can be exploited in quantum information processing
for efficient initialization of quantum registers.  Here we have described a general 
method of initialization and experimentally demonstrate it with two-, 
three-, and four-qubit nuclear spin registers. 
The basic idea is to prepare correlated state of singlets which
lives longer than other coherences and then convert this
long-lived state into a desired diagonal pseudopure state (see Fig. \ref{ppsfaint})
It is hard to initialize proton-based NMR registers using
traditional methods.  As a result many
popular NMR registers were based on carbon spins
or a combination of protons and carbons 
which permitted initialization by traditional methods.
The proposed method \cite{maheshpra10} of using long-lived is useful 
since only the nearest neighbor interactions are used, the present
method may facilitate
initialization of larger registers with weak long range
interactions.
Molecules are of interest where in the
inter pair dipolar couplings are sufficiently weak to keep the
singlet states long-lived, while the covalent bond mediated scalar 
interactions among the nearest neighbor spins are sufficiently strong.  
The method may also be applicable to registers based on parahydrogens
which naturally exist in singlet states.
Similar techniques may be used for multi-qubit initialization
in non-NMR systems exhibiting long-lived states. 
        
\thispagestyle{empty}
\chapter{Storing Entanglement Via Dynamical Decoupling}
  In this chapter, we have described the
experimental study of dynamical decouplings in preserving 
two-qubit entangled states using an ensemble of spin-1/2 nuclear pairs
in solution state. A brief introduction of decoherence 
and dynamical decouplings are given in section 4.1. In section 4.2, we have described Uhrig's dynamical decoupling 
and its usefulness in preserving coherence orders. In section 4.3, we have shown the creation of 
Bell states from long-lived singlet state. In section 4.4, experimental results
are shown for storing coherence orders in Bell states including singlet state.
We found that the performance of odd-order Uhrig sequences
in preserving entanglement is superior to both even-order 
Uhrig sequences and periodic spin-flip sequences.  We also found that there exists an optimal order of 
the Uhrig sequence using which the singlet state can be stored at high correlation for about 30 seconds.  

\section{Introduction}
Harnessing the quantum properties of physical systems have several potential applications, particularly in information processing, secure data communications, and quantum simulators \cite{chuangbook}. It is believed that such quantum devices may play an important role in
future technology \cite{milburn}.  But their physical realization is challenging mainly because of decoherence - the decay of the coherent states due to interaction with the surrounding environment \cite{joos, shlosshauer}. Therefore it is important to minimize the effects of decoherence using suitable perturbation on the quantum system \cite{viola99}. A technique, known as `dynamical decoupling' involves protecting the quantum states from decoherence by driving the system in a systematic manner such that the effective interactions with the environment at different instants of time cancel one another.
 
 In the following subsections we will describe the effect of decoherence and saving coherence through dynamical decouplings.
\subsection{Decoherence}
 As the name suggests, decoherence means loss of coherence order in a system which comprises with more than one entities. The idea of decoherence in quantum mechanics came much earlier in 1952 \cite{bohm1952} in the context of hidden variable model. It was first developed due to the possible explanation of the appearance of wave function collapse \cite{bohm1952, bohm19522}. Decoherence based explanation started getting acceptance in early 1980 \cite{zurek1981,zurek1982}. However, Decoherence became inevitable tool of understanding quantum mechanical interactions in the context of quantum information processing later. Soon it became clear that, decoherence is one of the biggest challenges to be overcome in order to realize a practical quantum computer. \\    
 
  Let us consider an isolated 2-level quantum system (a single spin-1/2 system). The wave function can be represented by following : 
\begin{eqnarray}
\vert\psi\rangle = c_0\vert 0\rangle + c_1\vert 1\rangle,\\
 \text{with}, \quad\quad \vert c_0\vert^2 + \vert c_1\vert^2 = 1. \nonumber
\end{eqnarray}
The system can be best represented by the density matrix formalism :
\begin{eqnarray}
\rho_s &=& \vert\psi\rangle\langle\psi\vert \nonumber\\
       &=& c_0c_0^* \vert 0\rangle\langle 0\vert + c_1c_1^* \vert 1\rangle\langle 1\vert +
           c_0c_1^* \vert 0\rangle\langle 1\vert + c_1c_0^* \vert 1\rangle\langle 0\vert,\nonumber\\
     \nonumber\\  &=& \bordermatrix{~ & \vert 0\rangle & \vert 1\rangle \cr
                  \langle 0\vert & c_0c_0^* & c_0c_1^* \cr
                  \langle 1\vert & c_1c_0^* & c_1c_1^* \cr}   
\end{eqnarray}

The diagonal elements are representing the population distribution of the system in two states. The off-diagonal elements are the coherence terms. Here, we are interested in the evolution of the coherence terms once the system is no more an isolated quantum system and is interacting with the environment. The interaction of system-environment is a non-unitary process and hence irreversible. Below we see the effect of environment on the 2-level super-positioned state.
\begin{equation}
\vert\psi\rangle\vert E\rangle =
(c_0\vert 0\rangle + c_1\vert 1\rangle)\vert E\rangle \stackrel{U(\tau)} {\longrightarrow}
(c_0\vert 0\rangle\vert E_0\rangle + c_1\vert 1\rangle\vert E_1\rangle).
\end{equation}
Now it can be noticed that the output state is an entangled state and can not be written as system and environment separately (unless $\vert E_0\rangle = e^{i\phi}\vert E_1\rangle$).
In terms of density matrix, the situation can be represented as below :
\begin{eqnarray}
\rho_s &=& \vert\psi\rangle\vert E\rangle\langle\psi\vert\langle E\vert \nonumber\\
       &=& c_0c_0^* \vert 0\rangle\langle 0\vert\otimes\vert E_0\rangle\langle E_0\vert
         + c_1c_1^* \vert 1\rangle\langle 1\vert\otimes\vert E_1\rangle\langle E_1\vert\nonumber\\
         &&+ c_0c_1^* \vert 0\rangle\langle 1\vert\otimes\vert E_0\rangle\langle E_1\vert
         + c_1c_0^* \vert 1\rangle\langle 0\vert\otimes\vert E_1\rangle\langle E_0\vert.
\end{eqnarray}
Now tracing out the environment from the system gives the necessary information about the system and can be written as :
\begin{eqnarray}
\rho_s &=& \text{trace}_E[\rho_{SE}]\nonumber\\
       &=& c_0c_0^* \vert 0\rangle\langle 0\vert + c_1c_1^* \vert 1\rangle\langle 1\vert
         + \langle E_1\vert E_0\rangle c_0c_1^* \vert 0\rangle\langle 1\vert
         + \langle E_0\vert E_1\rangle c_1c_0^* \vert 1\rangle\langle 0\vert \nonumber\\
\nonumber\\ &=& \bordermatrix{~ & \vert 0\rangle & \vert 1\rangle \cr
               \langle 0\vert & \vert c_0\vert^2 & \langle E_1\vert E_0\rangle c_0c_1^* \cr
               \langle 1\vert & \langle E_0\vert E_1\rangle c_1c_0^* & \vert c_1\vert^2 \cr} .
\end{eqnarray}

The above equation shows that the coherence terms obtains extra coefficients. Usually, when the environment has a large degree of freedom, these coefficients decay exponentially with time:
\begin{equation}
\vert \langle E_1(t)\vert E_0(t) \rangle \vert = e^{-\Lambda(t)}.
\end{equation} 
 Hence, after a certain time duration, the coherence terms decay to zero. 

\subsection{Dynamical decoupling}
Dynamical decoupling is a technique by which it is possible to suppress, at least to some extent, the environmental effect on a open quantum system under study. The idea
of dynamical decoupling has connections to the routinely used NMR decoupling sequences
where unwanted couplings are averaged out with the applications of suitable
modulated or unmodulated RF pulse sequences. The dynamical decoupling scheme relies on the
application of $\pi$ pulses at certain intervals.  
Preserving nuclear spin coherences
by spin flips at regular intervals was long been known in NMR as the famous Carr-Purcell-Meiboom-Gill (CPMG) sequence
\cite{carr,meiboom}.  
The CPMG sequence is widely used in NMR to 
measure the transverse relaxation time constants in the presence of 
spatial inhomogeneity of the static magnetic field and temporal fluctuations 
in the local fields arising due to the molecular motion \cite{LevBook}. 
The sequence involves a set of N number of $\pi$ pulses uniformly distributed
in a duration $[0,T]$ at time instants $\{t_1,t_2,\cdots,t_N\}$.
Assuming instantaneous $\pi$ pulses, $j^\mathrm{th}$ time instant
is linear in $j$,
\begin{eqnarray}
t_j^\mathrm{CPMG} = T \left( \frac{2j-1}{2N} \right).  
\end{eqnarray}
Of course, in practice the $\pi$ pulses do have finite duration 
owing to the limited power of electromagnetic irradiation 
generated by a given hardware. 
Further, the constant time period between these spin flips should ideally
be shorter than the correlation time of the spin-bath interaction.
Even, this delay is limited by the maximum duty-cycle that is
allowed for the hardware.  Dynamical decoupling with such bounded
controls have also been suggested \cite{viola03,gordon,jacob10,pasini10}.
For instance Hao et al. have been able to calculate, 
using a particular type of atomic systems, 
the maximum delay between spin-flips in order
to efficiently suppress decoherence due to a bath with a finite 
cut-off frequency \cite{liang}.
By studying the efficiency of the decoupling as a function of 
the CPMG period often it is possible to extract valuable informations 
about molecular dynamics and such studies are broadly categorized under
`CPMG dispersion' experiments \cite{palmer}.

Recently in 2007, Uhrig generalized the CPMG sequence by considering an
optimal distribution $\{t_1,t_2,\cdots,t_N\}$ of $N$ spin flips
in a given duration $[0,T]$ of time
that provides most efficient dynamical decoupling \cite{uhrig}.
Using a simple dephasing model, Uhrig proved that the 
time instants should vary as a squared sine bell:
\begin{eqnarray}
t_j = T \sin^2\left(\frac{\pi j}{2N+2} \right).
\label{tj}
\end{eqnarray}

UDD works well in systems having a high-frequency dominated
bath with a sharp cutoff \cite{Pasini,Biercuk,BiercukPRA}.
On the other hand, when the spectral density of the bath has
a soft cutoff (such as a broad Gaussian or Lorentzian), the CPMG sequence was found to 
outperform the UDD sequence 
\cite{duuhrig,dieteruhrig1,Cywinsky,Lange,Barthel,Ryan,sagi}. 
Suter and co-workers have studied these different regimes and
arrived at optimal conditions for the dynamical
decoupling \cite{dieteruhrig2}. 

Recently Agarwal has shown using theoretical and
numerical calculations that even 
entangled states of two-spin systems can be stored more efficiently
using UDD \cite{agarwal}.  
Since entangled states play a central role in QIP, 
teleportation, data encryption, and so on,
saving entanglement is crucial for the efficient physical
realization of quantum devices
\cite{chuangbook}.  More recently dynamical decoupling on an 
electron-nuclear spin-pair in a solid state system 
has been shown to prolong the pseudoentanglement lifetime by
two orders of magnitude \cite{duprl}.

While much of the experimental efforts have been on testing the loss of coherence due to $T_2$ processes, 
here in this chapter, we presented the first experiments where we study not only the loss of coherences, but also the
loss of entanglement due to both $T_1$ and $T_2$ processes. 
Though newer sequences have been suggested to decouple both of these
processes, these are yet to be studied experimentally \cite{uhrigprl09,lidarprl10}.
We have developed experimental techniques where we can prepare Bell states with high fidelity 
and characterize these states with high precision \cite{maheshjmr10,maheshpra10}. 
we explore the utility of different dynamical decoupling sequences on
systems wherein both T$_1$ and T$_2$ relaxations are significant.

\section{Uhrig dynamical decoupling}
Uhrig dynamical decoupling (UDD) claims to be more efficient than the CPMG sequence (which serves as the best known decoupling sequence for more than 50 years!) in preserving coherence orders. The efficiency of UDD over CPMG can be understood by various mathematical approaches. Uhrig \cite{uhrig} explained the efficiency of UDD by considering the standard spin-boson model in ohmic bath. This model predicts the noise-spectrum with a sharp cut-off. Here we describe the `filter function analysis' in brief for the study of UDD's efficiency \cite{Biercuk}.   
\subsection{Efficiency of UDD over CPMG}
 We can write the Hamiltonian of a system interacting with an environment as\cite{Biercuk, hao},
\begin{eqnarray}
\mathcal{H} = \frac{\hbar}{2}\left[\Omega + \beta(t) \right]\hat{\sigma}_z
\end{eqnarray}
 where $\Omega$ is the unperturbed part representing the system and $\beta(t)$ is the time dependent fluctuating part due to environmental interaction. As in ref \cite{hao}, the time evolution of a superposition state initially oriented along $\hat{Y}$ under the affect of this Hamiltonian can be written as
\begin{eqnarray}
\vert\psi(t) = \frac{1}{\sqrt 2}\left( e^{-i\Omega t/2}e^{-\frac{1}{2}\int_0^t{\beta(t')dt'}}\vert 0\rangle + 
 e^{i\Omega t/2}e^{\frac{1}{2}\int_0^t{\beta(t')dt'}}\vert 1\rangle \right)
\end{eqnarray}
where $\vert 0\rangle$ and $\vert 1\rangle$ representing the basis states and $\beta(t')$ adding the random phase errors. Accumulation of such phases lead towards decoherence. A fundamental technique for preserving coherence in NMR is `Spin-echo' given by Hahn \cite{hahn}. Spin-echo works as a refocusing technique by applying a $\pi$ pulse in between two exact delays.  Hahn echo became indispensable tool for coherence reorder and soon it was realized that the application of series of $\pi$ pulses at regular interval would be most effective in order to reduce dephasing \cite{carr, meiboom}. Hahn echo acts as a high pass filter for an arbitrary noise spectrum $S_\beta (\omega)$ and it neutralize the phase errors by slowly Fourier components of $\beta$. Now this one $\pi$ pulse logic can be extended to multiple $\pi$ pulses technique as well. Application of multiple pulses on a qubit system, leads to coherence state as,
\begin{eqnarray}
W(\tau) &=& \left| \overline{\langle\sigma_Y \rangle(\tau ')}\right| = e^{-\chi(\tau)},\nonumber\\
\mathrm{where,}\quad\quad \chi &=& \frac{2}{\pi}\int_0^{\infty}{\frac{S_{\beta}(\omega)}{\omega^2}F(\omega t)d\omega}.
\end{eqnarray}
Here, the filter function $F(\omega\tau)$ contains all the necessary information regarding the efficiency of pulse sequence for preserving coherence against the environment influence $S_\beta(\tau)$. Now, $F(\omega\tau)$ can be calculated from
\begin{eqnarray}
F(\omega\tau) = \left|\tilde{y}_n(\omega\tau)\right|^2,
\end{eqnarray}
where $\left|\tilde{y}_n(\omega\tau)\right|$ is the Fourier transform of time domain filter function $y_n(t)$. 
Any modification of filter function will give different efficiency power of that particular pulse sequence. CPMG sequence 
having $\pi$ pulses at regular interval was modified by Uhrig by repositioning the $\pi$ pulses at irregular intervals. Noise reduction is shown to be much more efficient for Uhrig sequence than CPMG \cite{uhrig, Biercuk}.

  Later, Agarwal has shown that this results of efficient UDD can be generalized for an entangled system as well \cite{agarwal}. Here our work mainly focuses on the experimental studies of UDD and CPMG on such an entangled states as well as on non-entangled states.


\section{Preparation of Entanglement}
\begin{figure}
\centering
\includegraphics[trim = 20mm 40mm 20mm 40mm, clip, width=12cm]{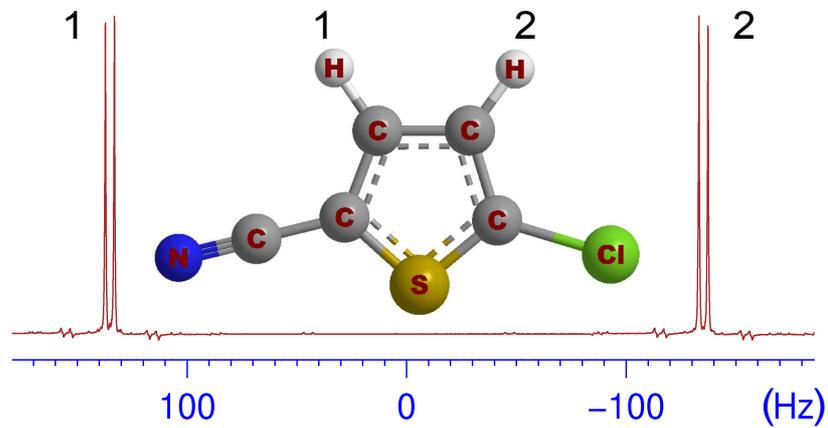} 
\caption[$^1$H NMR spectrum and the molecular structure
of 5-chlorothiophene-2-carbonitrile]{The $^1$H NMR spectrum and the molecular structure
of 5-chlorothiophene-2-carbonitrile.
}
\label{uhfig1} 
\end{figure}

We study storage of entanglement by dynamic decoupling
on a pair of spin-1/2 nuclei using liquid state NMR techniques.
The sample consisted of 5 mg of 5-chlorothiophene-2-carbonitrile 
dissolved in 0.75 ml of dimethyl sulphoxide (see Figure \ref{uhfig1}).
The two protons of
the solute molecule differ in the Larmor frequency by 
$\Delta \nu = 270.4$ Hz and have an indirect spin-spin coupling 
constant of $J = 4.1$ Hz.
The $T_2$ relaxation time constants for the two protons are about 2.3 s and
the $T_1$ relaxation time constants are about 6.3 s.

\subsection{Preparation of singlet states}
High fidelity entangled states are prepared via long
lived singlet states in a procedure described in chapter 2. 
%
%
\begin{figure}
\centering
\includegraphics[width=14cm]{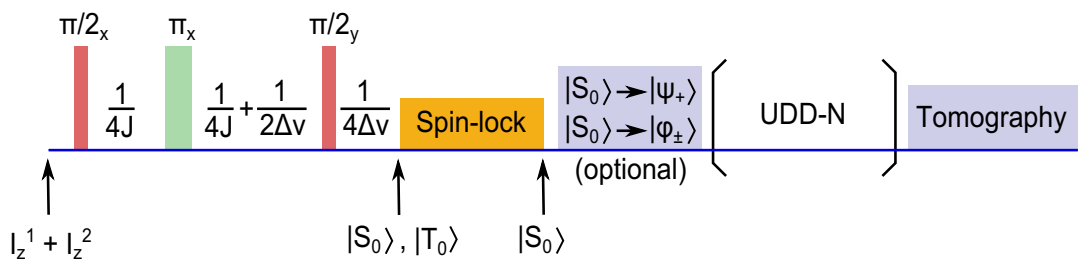} 
\caption[NMR pulse sequence to study dynamical decoupling on Bell states]
{NMR pulse sequence to study dynamical decoupling on
Bell states.  An incoherent mixture of singlet and triplet states
is prepared which under spin-lock purifies to singlet state.
The resulting singlet state can be converted to other Bell states.
Then dynamical decoupling sequence can be applied and the
performance of the sequence can be studied by characterizing the
residual state using density matrix tomography.}
\label{fullpulseq} 
\end{figure}

The long-lived nature of singlet states under the equivalence
Hamiltonian can be used to  prepare high-fidelity Bell states.
The experiment involves preparing an incoherent mixture of singlet
and triplet states,
\begin{eqnarray}
\rho(0) = - {\bm I}^1 \cdot {\bm I}^2 \equiv
\vert S_0 \rangle \langle S_0 \vert -\vert T_0 \rangle \langle T_0 \vert
\end{eqnarray}
from the equilibrium state $I_z^1 + I_z^2$ 
by using the pulse sequence shown in Figure \ref{fullpulseq}
\cite{LevJMR06}.
During the spin-lock $\vert T_0 \rangle$ state rapidly equilibrates
with the other triplet states. 
On the other hand, the decay constant 
of singlet state $\vert S_0 \rangle$ during the spin-lock
is much longer than the spin-lattice relaxation time constant ($T_1$)
(and hence the singlet state is known as a long-lived state) 
\cite{LevPRL04,LevittJACS04}. Hence at the end of suitable spin-lock we left out with 
high fidelity singlet states. The goodness of the prepared singlet state is measured 
by the tomographic method as described in Chapter 2.3. 
%
The correlation of the singlet state is given by,
\begin{eqnarray}
\langle \rho_s \rangle(t) = \frac{\mathrm{trace}\left[ \rho(t) \cdot \rho_s \right]}{\sqrt{ \mathrm{trace} \left[\rho(t)^2 \right] \cdot  \mathrm{trace} \left[\rho_s^2 \right]}},
\label{corr}
\end{eqnarray}
In the following we describe preparation of other Bell states from the singlet state
in a two-qubit NMR system.

\subsection{Preparation of other Bell states from singlet states}

Other Bell states can be obtained easily from the singlet
state:
\begin{eqnarray}
\vert S_0 \rangle & \xrightarrow{\mathrm{e}^{i\pi I_z^1}}    &  \vert \psi_+ \rangle = \frac{1}{\sqrt{2}}(\vert 01 \rangle + \vert 10 \rangle), \nonumber \\
\vert S_0 \rangle & \xrightarrow{\mathrm{e}^{i\pi I_x^1}}               & \vert \phi_- \rangle = \frac{1}{\sqrt{2}}(\vert 00 \rangle - \vert 11 \rangle), \nonumber \\
\vert S_0 \rangle & \xrightarrow{\mathrm{e}^{i\pi I_x^1} \cdot \mathrm{e}^{i\pi I_z^1}} & \vert \phi_+ \rangle = \frac{1}{\sqrt{2}}(\vert 00 \rangle + \vert 11 \rangle).
\end{eqnarray}
\begin{figure}
\centering
\includegraphics[width=12cm]{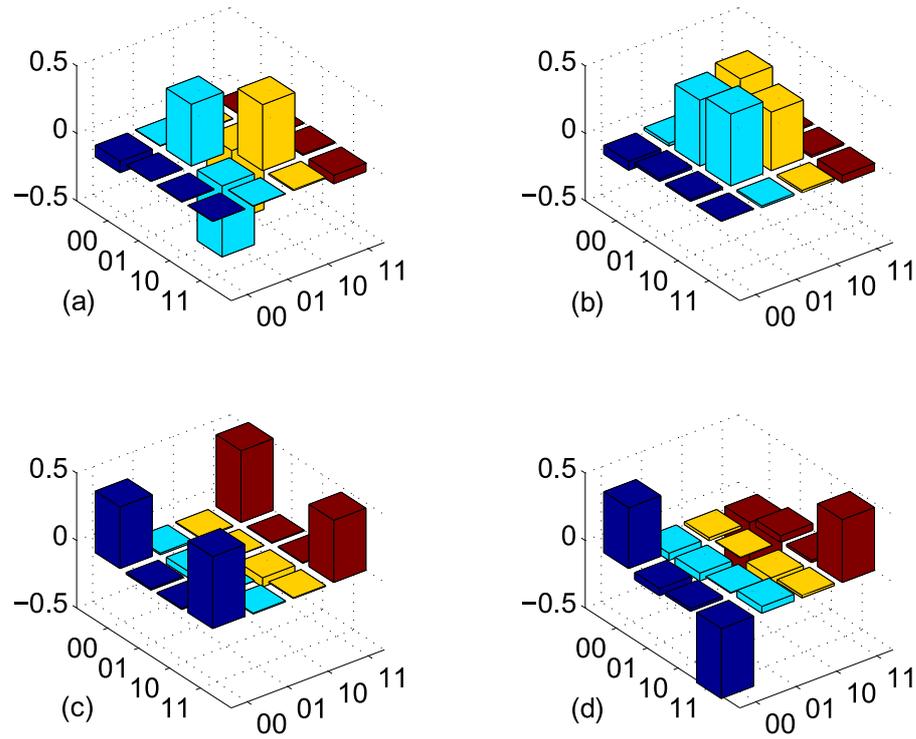} 
\caption[Density matrix tomography of Bell states]
{Density matrix tomography of Bell states : Real part of
(a) singlet state $\vert S_0\rangle = \frac{1}{\sqrt{2}}(\vert 01\rangle - \vert 10\rangle$ with correlation 0.99,
(b) $\vert \psi_+ \rangle = \frac{1}{\sqrt{2}}(\vert 01 \rangle + \vert 10 \rangle)$ with correlation 0.99,
(c) $\vert \phi_- \rangle = \frac{1}{\sqrt{2}}(\vert 00 \rangle - \vert 11 \rangle)$ with correlation 0.98, and
(d) $ \vert \phi_+ \rangle = \frac{1}{\sqrt{2}}(\vert 00 \rangle + \vert 11 \rangle)$ with correlation 0.97
}
\label{bell4} 
\end{figure}

The $z$-rotation in the above propagators can be implemented by 
using chemical shift evolution for a period of $1/(2\Delta \nu)$, and
qubit selective $x$-rotation can be
implemented by using radio frequency pulses\cite{maheshpra10}.
Details of dynamical decoupling on the Bell states
will be described in the next sections. In order to
investigate the decoupling performance, it is necessary to 
quantify the decay of Bell states with decoupling duration.
The Bell states by themselves are inaccessible to
macroscopic observables, but
can indirectly be detected 
transforming to observable single quantum coherences
\cite{LevPRL04,LevittJACS04}.
Alternatively, a more detailed and quantitative analysis of Bell states 
may be carried out  using density matrix tomography as described in Chapter2.3.
\cite{maheshjmr10}. 
We have utilized the density matrix formalism for the characterization
of the Bell states. The goodness of prepared Bell states can be evaluated from 
the definition of correlation using expressions similar to
(\ref{corr}). The density matrices for all the four Bell-state have been
shown in figure \ref{bell4}. We achieved high fidelity Bell states with
correlation around 0.99.
In the following we have shown the experimental implementations of
dynamical decoupling on such entangled states.

\section{Storage of entanglement by UDD}
\begin{figure}
\centering
\includegraphics[width=14cm]{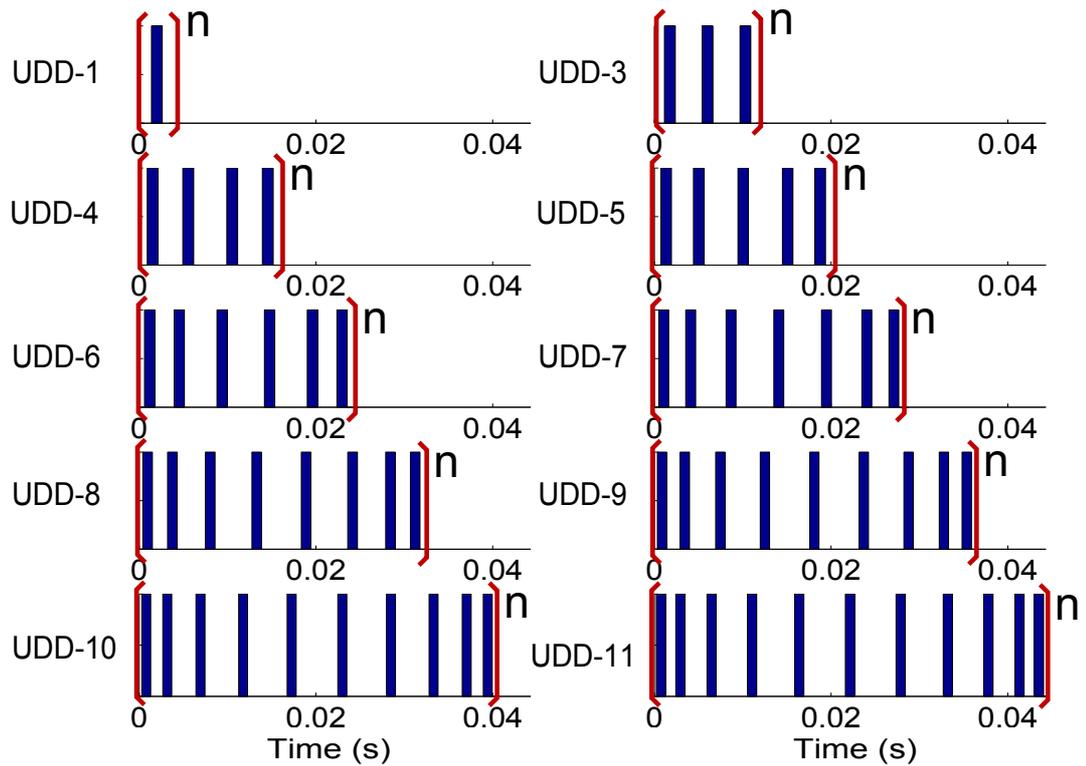} 
\caption[Pulse sequences for various orders of Uhrig Dynamical Decoupling]
{Pulse sequences for various orders of Uhrig Dynamical
Decoupling.  Note that both UDD-1 and UDD-2 are equivalent to CPMG.
The time instants are calculated according to the expression
(\ref{tj}), with $N$ being the order of UDD and the
total period $T = N \times 4.0272$ ms.
}
\label{uhfig2} 
\end{figure}

\subsection{Different orders of UDD}
As described earlier, the UDD scheme consists of a sequence of
spin flips placed at time instants given by the expression
(\ref{tj}).
Instead of applying the Uhrig's formula for the
entire duration of decoupling, we have applied the formula for a short
time interval ($T$) consisting of a small number ($N$) of pulses
and then repeating the sequence. 
Figure \ref{uhfig2} shows pulse sequences for various orders
of Uhrig Dynamical Decoupling (we refer to an N-pulse 
UDD sequence as UDD-N).
Note that UDD-1 (and UDD-2) are equivalent to CPMG sequences,
in which repeating segment consists of $[\tau_\mathrm{CPMG} - \pi - \tau_\mathrm{CPMG}]$.
In our experiments,  $\tau_\mathrm{CPMG}$ was set to 2 ms and
the duration $\tau_{\pi}$ of the $\pi$ pulse was 27.2 $\mu$s.
The total duration of UDD-N was set to $T = N(2\tau_\mathrm{CPMG}+\tau_{\pi})$,
such that for an extended period of time, the total number of 
$\pi$ pulses remain same irrespective of the order of UDD.  Only
the distribution of $\pi$ pulses varies according to the order 
of UDD.  For example, in one second of decoupling, there will be about 
250 $\pi$ pulses in all UDD-N.  Our investigation thus helps in studying
the efficiency of decoupling over a fixed duration of time
for a given number of $\pi$ pulses 
dispersed according to different orders of UDD.

\begin{figure}
\begin{center}
\hspace*{-1.5cm}
\includegraphics[width=14.5cm]{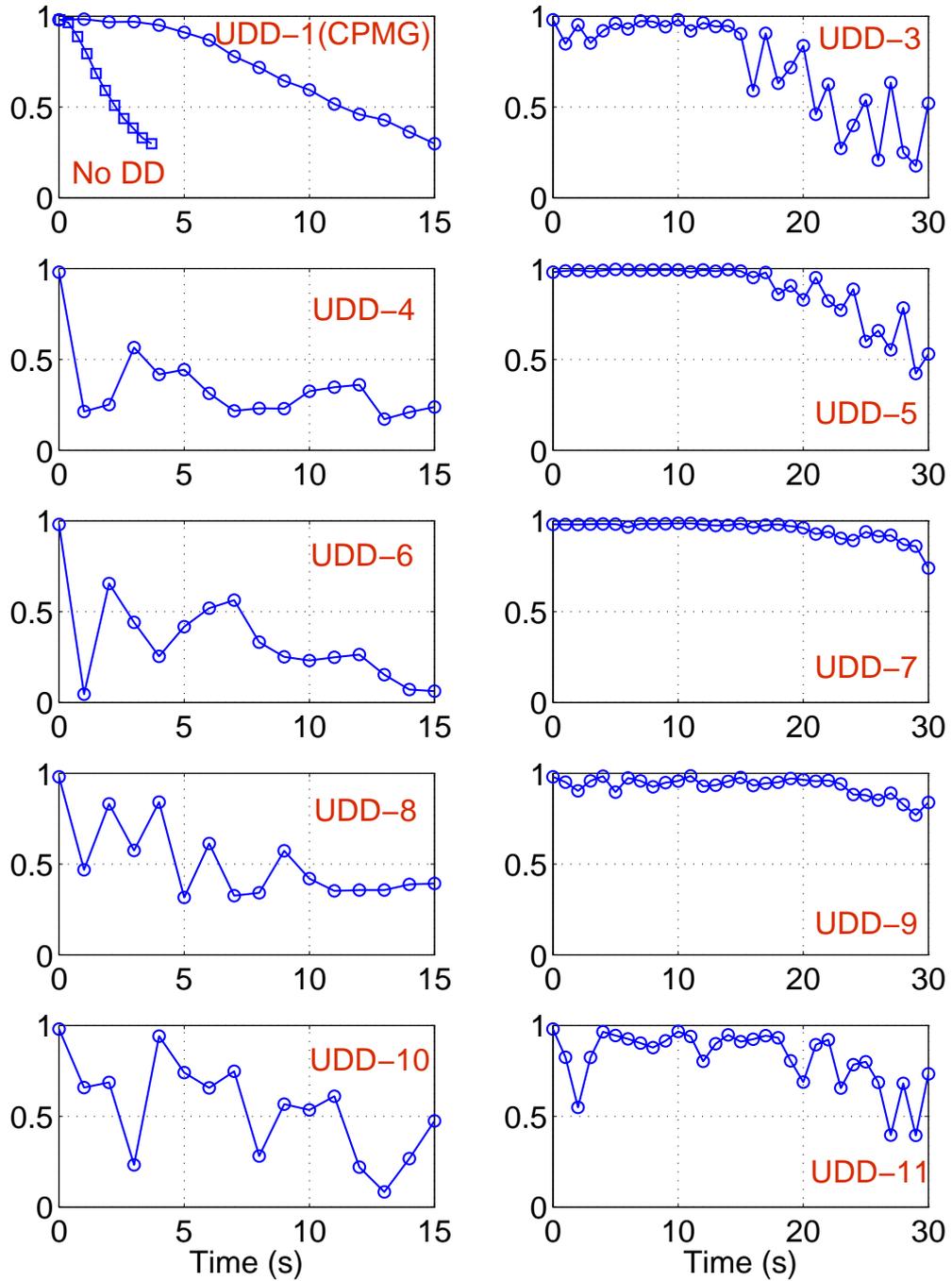} 
\caption[Experimental correlations (circles) of singlet state as a]
{Experimental correlations (circles) of singlet state as a function of 
decoupling duration of various orders of UDD.  Also shown in the
top-left figure is the correlation decay under no dynamical decoupling (squares).
}
\label{uhrigall} 
\end{center}
\end{figure}

\subsection{Performance of UDD over CPMG sequence}
Now we describe the performances of UDD-N on 
the singlet state which was prepared as explained before
(see Figure \ref{fullpulseq}).
After applying UDD-N for a fixed duration of time, we carried out
density matrix tomography and evaluated the correlation of the
preserved state with theoretical singlet density matrix.  The correlations
for various orders of UDD are displayed in Figure \ref{uhrigall}.
As can be seen from the figure, the singlet state can be
preserved for longer durations by UDD-1 (CPMG) than no-decoupling.
It is also clear that all even-order UDD sequences result in significant
fluctuations in the correlation of the singlet state.
However, the odd order UDD preserve the singlet state for tens of seconds.  
For example, the correlation of the singlet state under UDD-7 at all the
sampled time points till 20 seconds is above 0.96.
This rather surprising even-odd behavior is likely due to the differences
in the performances of the even and odd ordered sequences against
the spatial inhmongeneity of the RF pulses.

\begin{figure}
\centering
\includegraphics[width=8.5cm,angle=-90]{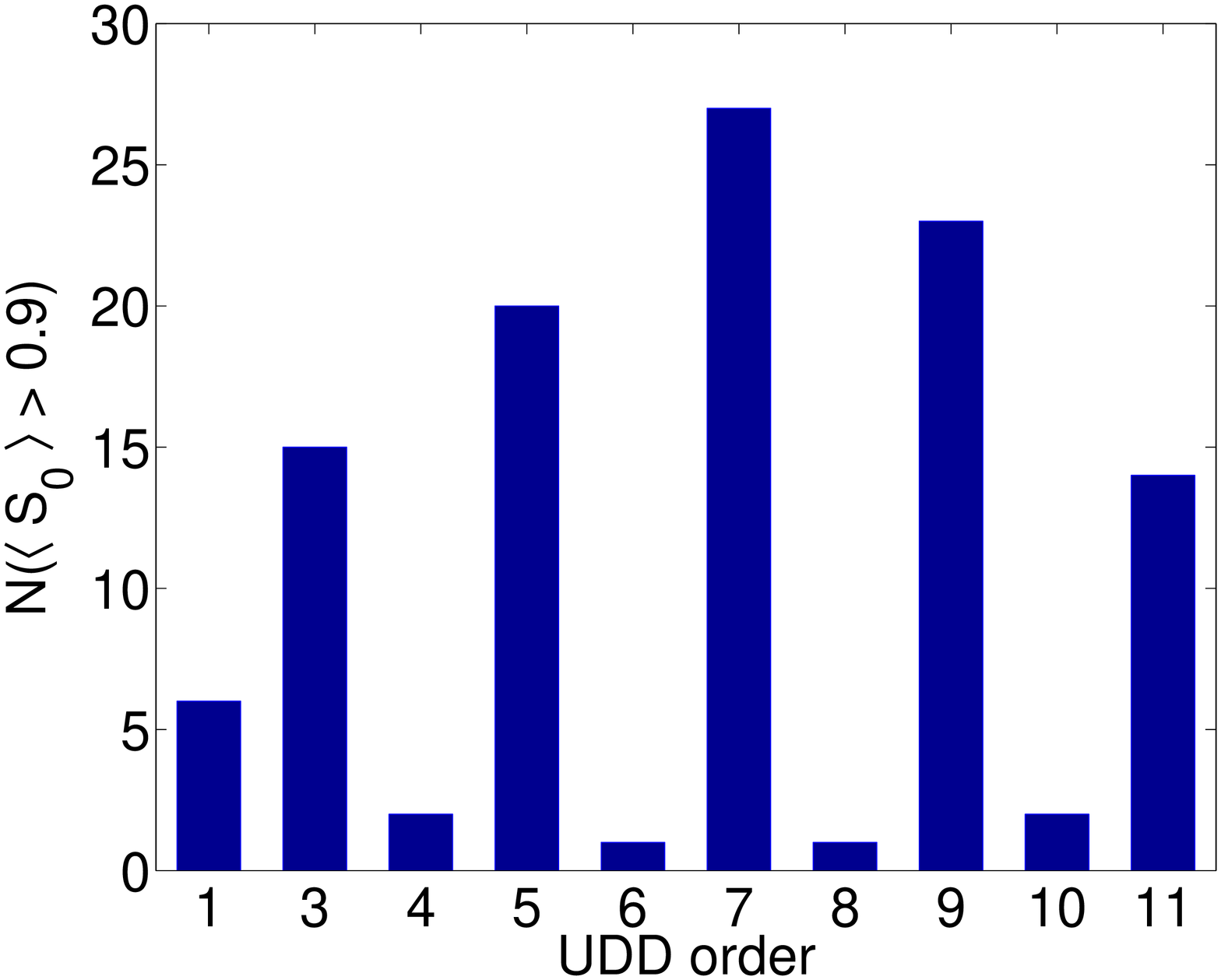} 
\caption[correlation exceeding 0.9 for]{The number of time instants at which the correlation 
exceeded 0.9 for various orders of UDD-N.
}
\label{bar1} 
\end{figure}

One way to quantify the efficiency 
of dynamical decoupling under various orders of UDD 
in figure \ref{uhrigall}, is by counting the number
of time instants in which the correlation of the preserved state
exceeds a given threshold.
The bar plot in Figure \ref{bar1} compares the number of time
instants during decoupling under various orders of UDD 
in which the correlation of the singlet state exceeded 0.9.
It can be seen that there exists an optimal order of
UDD (for a given $\tau_\mathrm{CPMG}$ and $\tau_\pi$), which performs the 
most efficient decoupling.  
The optimality may be because of the finite width of the $\pi$ pulse.
In a CPMG sequence the $\pi$ pulses are uniformly dispersed, 
while in Uhrig
sequence  the $\pi$ pulses are more crowded at the terminals 
(beginning and ending) of the sequence.
For example, if there are too many
$\pi$ pulses, Uhrig's formula will lead to an overlap of 
pulses.  Experimentally, the overcrowding of $\pi$ pulses 
may also lead to RF heating of the sample and the probe.
Thus the performance of the UDD sequence does not grow indefinitely
with the order of the sequence, but instead will fall beyond
a certain order.
In our experimental setting, we find that UDD-7 is the optimal
sequence for storing the singlet state.
There are recent suggestions for decoupling using finite pulses, 
however these are yet to be studied experimentally
\cite{uhrignjp10,uhrigcomm}.

\subsection{Decay of magnetization during various dynamical decouplings}
It can be noticed that the attenuated correlation
(expression (\ref{corr})) is insensitive
to the decay of the overall magnetization ($\epsilon$
in (\ref{rhodev})), 
but simply measures the overlap between $\rho_\Delta$
and the theoretical density matrix $\vert \psi \rangle \langle \psi \vert$.  
An alternate method is to monitor the decay of magnetization
(i.e., $\epsilon$) under dynamical decoupling.

\begin{figure}
\centering
\includegraphics[width=10cm]{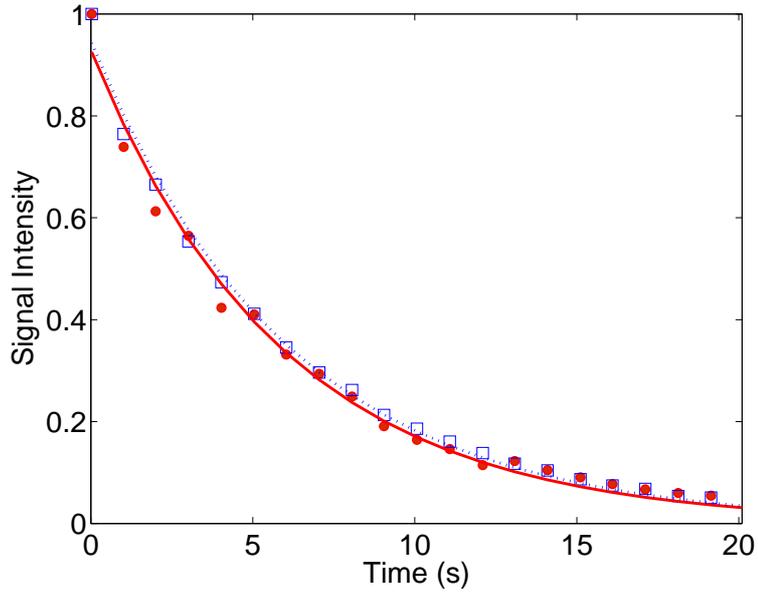} 
\caption[The decay of the singlet spin-order measured by]
{The decay of the singlet spin-order
measured by converting it into observable
single quantum magnetizations.  The decay was
studied under CPMG sequence (squares) as well as under
Uhrig sequence (filled circles).
The dashed and the solid line correspond to
the exponential fits for CPMG and UDD-7 data points
respectively.
}
\label{uhfig3} 
\end{figure}

As already
mentioned in Chapter 2, singlet state itself can not be measured directly,
but can be converted to observable magnetization by using
a chemical shift evolution for a duration $1/(4\Delta \nu)$ 
followed by a $\left( \frac{\pi}{2} \right)_{x(y)}$ pulse. 
Intensity of the resulting signal  as a function of the 
duration of dynamical decoupling is shown in Figure \ref{uhfig3}.
As can be seen, UDD-7 is no better than CPMG in preserving the
overall spin-order.  In fact the decay constant for CPMG 
and UDD-7 are 6.1 s and 5.9 s respectively.

\subsection{Efficiency of UDD over CPMG for a non-entangled state and various Bell states}
\begin{figure}
\centering
\hspace*{-.5cm}
\includegraphics[width=13.5cm]{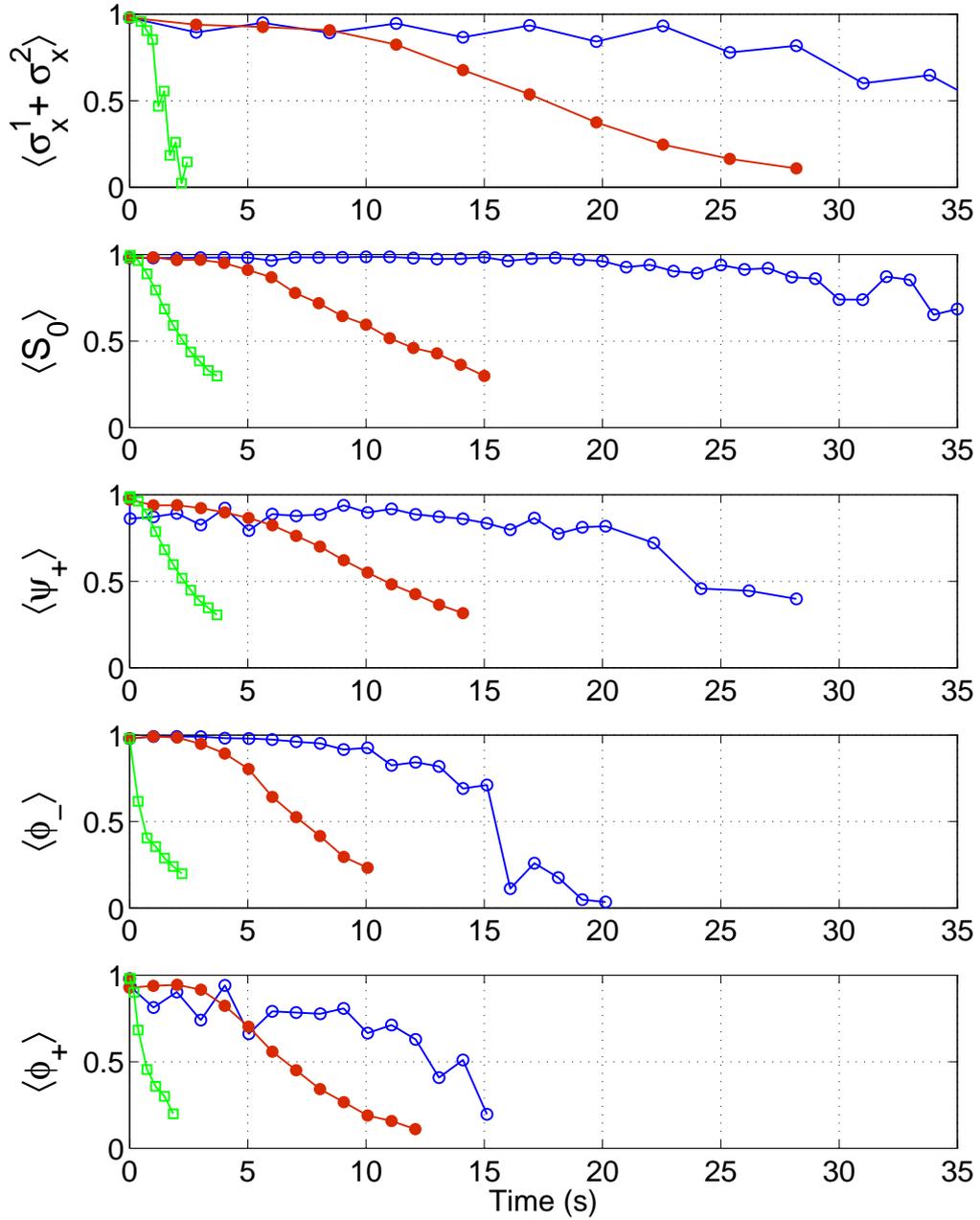} 
\caption[Experimental correlations of the product state and various Bell states]
{Experimental correlations of the product state
and various Bell states as a function of duration under
(i) no decoupling (open squares), (ii) CPMG sequence (filled
circles), and (iii) UDD-7 (open circles).
}
\label{uhfig4} 
\end{figure}

Now we compare the efficiency of the optimal sequence UDD-7 
with UDD-1 (CPMG) for preserving product state ($\sigma_x^1 + \sigma_x^2$)
and other Bell states.
Figure \ref{uhfig4} shows the variation of correlation of 
product states and the Bell states as a function of the 
decoupling duration \cite{soumyauhrig}.  Here, after preparing each of the initial state,
the dynamical decoupling was applied for a fixed duration of
time.  To monitor the correlation, we have carried out the
density matrix tomography as described earlier \cite{maheshjmr10}.
In the case of no decoupling,
we observe a rapid decay of the correlation.  The
UDD-1 (CPMG) sequence shows some improvement in the storage time.  
However, UDD-7 clearly exhibits much longer storage times than the CPMG sequence.
The superior performance of UDD-7 on the singlet
state compared to other Bell states is
presumably because of its antisymmetric property described
in section II.

\section{Conclusions}
In this chapter, we have described the effect of decoherence on the quantum system
and shows a method to tackle it in practical situations. Dynamical decoupling 
is a method by which we can reduce the effect of environment on the system and 
ultimately increase the coherence time scale of the system \cite{soumyauhrig}. CPMG is known to be 
the best known dynamical decoupling sequence both theoretically and experimentally 
for more than 50 years until 2007. In 2007, Uhrig introduced a new sequence where instead of
applying the $\pi$ pulses at regular intervals, one needs to apply $\pi$ pulses at
irregular intervals synchronizing with a sine-square bell. Theoretically it has been 
well proved that Uhrig dynamical sequence (UDD) performs better than CPMG sequence
for saving coherence orders of a quantum system. The coherence order for an entangled
state is also proved to be elongated by the application of UUD sequence compared to 
CPMG sequence. Stroboscopic spin flips have already been shown to prolong the coherence times of
quantum systems under noisy environments.  Uhrig's dynamical decoupling
scheme provides an optimal sequence for a quantum system interacting
with a dephasing bath.  Several experimental demonstrations have already been
verified the efficiency of such dynamical decoupling schemes
in preserving single qubit coherences.

  Here we have shown the first experimental study of UDD sequence on an NMR system.   
We have studied the efficiencies of CPMG and UDD sequences on 2-qubit Bell states both in terms of
magnetization as well as in terms of correlation decay  \cite{soumyauhrig}.  While
the Uhrig sequence is no better than CPMG sequence in terms of preserving
the overall magnetization (or spin order), it clearly outperforms the CPMG sequence in
preserving the correlation of the entangled as well as non-entangled 
states.  We summarize three important features:
\begin{enumerate}[(i)] \itemsep-.5em
  \item The even-order UDD sequences result is fluctuations in correlations.\\
  \item The odd-order UDD sequences out-perform the CPMG sequence.\\
  \item There exists an optimal length for the odd order UDD sequence which exhibits the most efficient decoupling.\\
\end{enumerate}  
In our case, UDD-7 of 28.2 ms duration appeared to outperform all other sequences
of both lower and higher orders.  
Further understanding on the subject can be achieved by carrying out   
investigations into the effects of other experimental issues like
RF inhomogeneity, resonance off-set, errors in calibration of
pulse angle etc.
These considerations may help in the theoretical and practical 
understanding of the optimal decoupling schemes.
           
\thispagestyle{empty}
\chapter{Violation of Leggett-Garg Inequality}

  In this chapter, we have performed experimental implementation of a protocol for testing the Leggett-Garg inequality (LGI) for nuclear spins in a NMR setup. The motivation and importance of this work is given in the introduction section \ref{lgiintro}.
In section \ref{seclgi}, we have laid out the mathematical formulation of Leggett-Garg inequality for a spin-1/2 nucleus in external magnetic field. In section \ref{secmoussa}, we have presented the Moussa protocol for evaluating the expectation values of a target operator using an ancilla qubit. Section \ref{secexp} shows the experimental results for the 3-qubit and 4-qubit measurements respectively. The conclusion is given in section \ref{secconclu}.

\section{Introduction}
\label{lgiintro}
Distinguishing quantum from classical behavior has been an important issue since the development of quantum theory \cite{einstein, bell, kochen, mermin, leggett}. This issue is also at the heart of physical realizations of quantum information processing (QIP) \cite{chuangbook}. Experimental tests for confirming quantumness in physical systems are usually guided by the Bell-type inequalities (BI) \cite{bell} and the Leggett-Garg inequality (LGI) \cite{leggett}. BI places bounds on certain combinations of correlation coefficients corresponding to measurement outcomes for space-like separated systems which are assumed unable to influence one another (\textit{local realism}). LGI, on the other hand, places bounds on combinations of temporal correlation coefficients between successive measurement outcomes for a system.  Here the system at any instant of time is assumed to be in one or the other of many possible states, and each measurement is assumed to be perfectly non-invasive, in the sense that it has no effect on system's subsequent dynamics  (\textit{macrorealism}). In other words, violation of LGI indicates that the system's dynamics cannot be understood in classical terms. In recent years various protocols for implementing LGI and its refined versions have been proposed and experimentally demonstrated \cite{ruskov, jordan, williams, goggin, wilde, laloy, dressel}. \\

 Here we have implemented the LGI protocol for individual spin-1/2 nuclei (from a liquid NMR sample) precessing in magnetic field and interacting with their local environments. A typical spin-1/2 system is genuinely `microscopic' and exhibits quantum behavior.  However, it is well-known that, due to decoherence, microscopic quantum systems appear to behave classically and as a consequence QIP tasks relying on such candidate systems tend to fail \cite{shlosshauer}. Nuclear spins from an NMR sample are examples of microscopic quantum systems that are in constant interaction with their local environment and are also candidate systems for QIP tasks. 
The interactions such as dipole-dipole and chemical-shift anisotropy are known to be leading to decoherence, dissipation and relaxation processes within the spin ensemble \cite{Kowalewski}. In experimental set-ups such as NMR, successful QIP implementation therefore demands confirmation of `survival' of and determination of `durability' of quantumness in candidate systems. While an LGI test was originally proposed for addressing the fundamental question about the ability of a \textit {macroscopic} system to behave quantum mechanically, considering its basic mathematical framework, we extend such a test to investigate survival and durability of quantumness within individual nuclear spins interacting with their environments. The investigation also sheds light on the possible consistency of the assumptions of macrorealism with the `decoherence perspective' \cite{kofler}. 

Although individual nuclear spins from an NMR sample are not directly addressable, the sample provides an easily accessible ensemble of nuclear spins from a large number of molecules. Therefore the experimental evaluation of a particular temporal correlation involves \textit {simultaneous} implementations of the LGI protocol on a large number of nuclei (identical `targets'). Further, an NMR read-out is an `ensemble average' obtained in terms of magnetization signal. One thus needs to relate the required temporal correlation from an LG string with the NMR signal. A quantum network for encoding correlation between measurement outcomes of a target system in the phase of a probe system  has recently been proposed by Moussa {\it et al} \cite{moussa}. With this network they were able to demonstrate quantum contextuality using nuclear spins from a solid state NMR sample. In this chapter, we exploit this network for testing LGI.

Experimental results shown for values of LG-strings containing three and four temporal correlations as 
functions of delay between successive measurements \cite{soumyalgi}. We have found good agreement between the quantum mechanically expected and experimentally observed values of the strings for short timescales over which the decay in correlations due to typical NMR relaxation processes are ineffective. Further, to demonstrate effect of decoherence on the state of individual target nuclei which leads to relaxation of the entire ensemble, we have also measured the values of LG strings over longer timescales and found that the LG strings gradually decay and ultimately fall within the classical bounds.

\section{Leggett-Garg inequality}
\label{seclgi}
Consider a system (the `target') whose state-evolution in time is governed by a particular Hamiltonian. To perform an LGI test for the system, a particular system-observable (say $\mathbbm{Q}$) that can be taken as `dichotomic', i.e. having two possible states with measurement outcomes $Q = \pm{1}$, requires to be identified. Next, from a set of `$n$' measurement instants $\{t_1,t_2,t_3,...,t_n \}$, pairs of instants $t_i$ and $t_j$,  such that $j = i+1$, and a pair containing the first $(i = 1)$ and the last $(j = n)$ instants are to be chosen. For each such pair, one is then required to perform measurements of $\mathbbm{Q}$ on the target system at the corresponding two instants and obtain outcomes $Q(t_i)$ and $Q(t_j)$. After repeating these two-time measurements over a large number of trials (say, $N$), one can obtain the two-time correlation coefficient (TTCC)  $C_{ij}$ for each pair given by the formula:
\begin{eqnarray}
\label{cij1}
C_{ij} = \frac{1}{N} \sum _{r=1}^{N}Q_{r} \left( t_i \right)\cdot Q_r \left(t_j \right).
\end{eqnarray}
where, $r$ is the trial number. Finally, the values of these coefficients are to be substituted in the \textit {n}-measurement LG string given by:
\begin{eqnarray}
K_{n} = C_{12}+ C_{23} + C_{34} +....+ C_{(n-1)n} - C_{1n}.
\end{eqnarray}
Each coefficient from the r.h.s. of the above LG string would have a maximum value of $+1$ corresponding to perfect correlation, a minimum value of $-1$ corresponding to perfect anti-correlation, and $0$ for no correlation. Thus, the upper bound for $K_{n}$ consistent with \textit{macrorealism} comes out to be $(n-2)$, the lower bound is $-n$ for $odd$ $n$, and $-(n-2)$ for $even$ $n$. With these considerations the LGI reads $-n \le K_n \le (n-2) \;\; \mathrm{for \; odd} \;n \mathrm{,\; and} 
-(n-2) \le K_n \le (n-2) \;\; \mathrm{for \; even} \; n$. For example, $-3 \le K_3 \le 1$ and $-2 \le K_4 \le 2$.
\begin{figure}
\centering
\includegraphics[width=14cm]{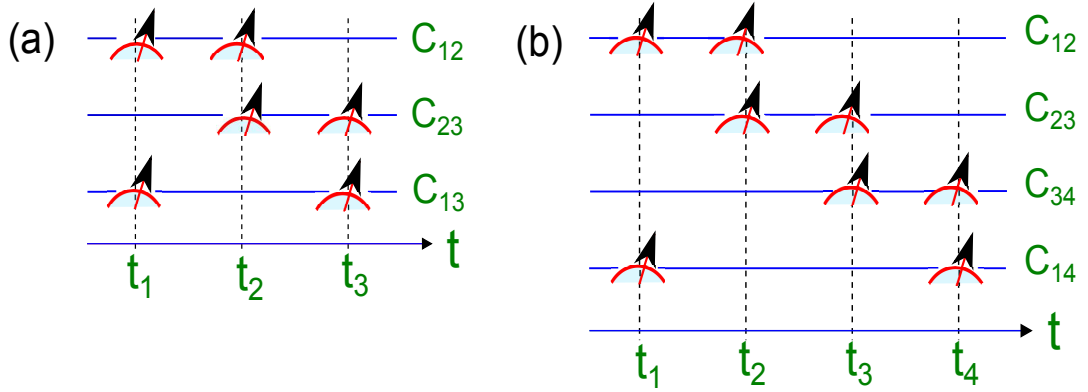} 
\caption[The protocols for evaluating $K_3 = C_{12}+C_{23}-C_{13}$]
{The protocols for evaluating $K_3 = C_{12}+C_{23}-C_{13}$ (a) and $K_4 = C_{12}+C_{23}+C_{34}-C_{14}$ (b).
In (a) three independent pairs of measurements are used to
evaluate TTCCs $C_{12}$, $C_{23}$, and $C_{13}$.  
Similarly (b) uses four pairs of independent measurements 
to evaluate $C_{12}$, $C_{23}$, $C_{34}$, and $C_{14}$.
}
\label{lgifig1} 
\end{figure}

\subsection{Spin-1/2 precession}
The Zeeman Hamiltonian for the precession of a spin-1/2 nucleus in a magnetic field about $z$-axis, 
is given by $\hat{H}=\frac{1}{2}\omega \hat{\sigma }_z$, with $\omega$ 
being the angular precession frequency and $\hat{\sigma}_z$ the Pauli-z operator. For the present work we choose the Pauli-x operator, i.e. $\hat{\sigma}_x$, as the dichotomic observable. The quantum mechanical 
expression of ${C}_{ij}$ for  $\hat{\sigma}_x$ measurements on the nucleus is given by  \cite{kofler}
\begin{eqnarray}
C_{ij}=\left 
\langle \hat{\sigma}_x\left(t_i\right) \hat{\sigma}_x\left(t_j\right) \right\rangle \approx \cos \left\{ \omega (t_j-t_i) \right\}.
\end{eqnarray} 
In Heisenberg representation one can obtain this relation from: 
\begin{eqnarray}
{C}_{ij}\approx \frac{1}{2}\sum _{k} \left[_z \langle {k}\vert\hat{\sigma}_x(t_i)\hat{\sigma}_x(t_j)\vert {k}\rangle_z \right] .
\end{eqnarray}
Here, $ \vert {k}\rangle_z \in \{\vert {0}\rangle, \vert {1}\rangle\}$, 
is an eigenstate of the Pauli-z operator. If we divide the total duration
from $t_{1}$ to $t_{n}$ into $(n - 1)$ equal intervals of duration $\Delta t$, we can express the LG string consistent with equation (3) as
\begin{eqnarray}
K_{n}=(n-1)\cos\{\omega\Delta t\} - \cos\{(n-1)\omega\Delta t\}.
\label{kn}
\end{eqnarray}
The protocols for evaluating $K_3$ and $K_4$ are
illustrated in Fig. \ref{lgifig1}.  It can be seen that
the quantum bounds for $K_3$ and $K_4$ are $[-3,+1.5]$ and $[-2\sqrt{2},+2\sqrt{2}]$
respectively.

\section{Evaluating TTCCs using network proposed by Moussa et al}
\label{secmoussa}
Suppose that we wish to evaluate correlations between the outcomes of repeated measurements of two commuting dichotomic unitary observables $S_1$ and $S_2$ for a target system (T). 
Consider an ancilla qubit (called `probe' P) and 
a unitary transformation for the joint system `T + P' ,
\begin{eqnarray}
\label{Us}
U_{S} = \mathbb{I}_\mathrm{P} \otimes ({P}_{+})_\mathrm{T} + ({\hat{\sigma}_z})_\mathrm{P} \otimes ({P}_{-})_\mathrm{T}.
\end{eqnarray}
Here ${P}_{+}$ and ${P}_{-}$ are the projectors onto the eigenspace of $S \in \{S_1,S_2\}$, such that
$ S = ({P}_{+})_\mathrm{T} - ({P}_{-})_\mathrm{T} $.

\begin{figure}
\centering
\includegraphics[width=14cm]{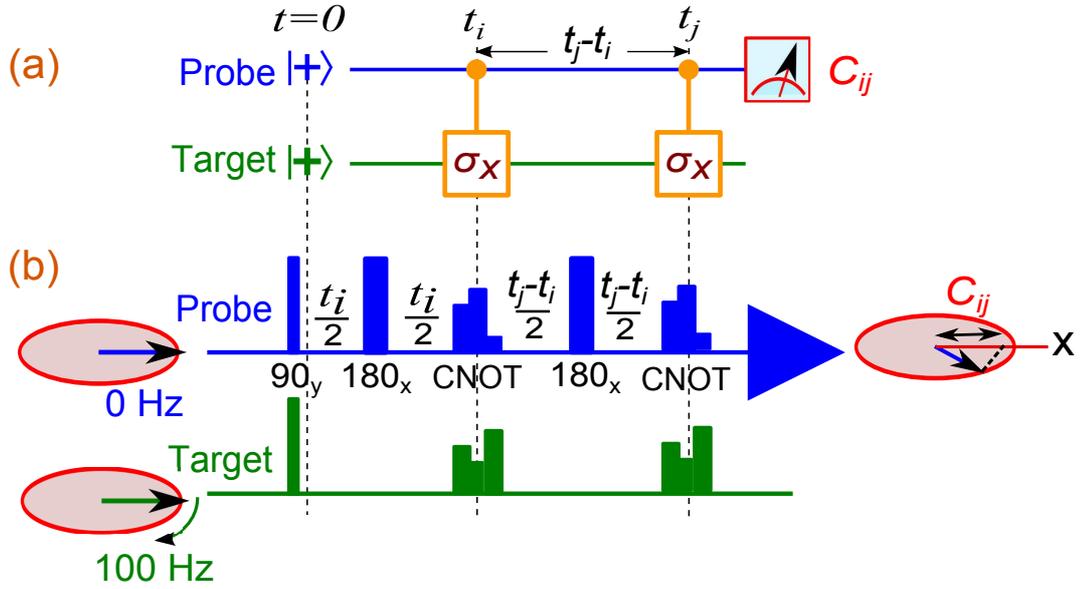} 
\caption[Quantum network for the evaluation of TTCCs]
{Quantum network for the evaluation of TTCCs (a) and
the corresponding NMR pulse sequence (b).
The ensemble was initially prepared according to
$(\rho)_\mathrm{P} \otimes (\rho)_\mathrm{T}$, where
$(\rho)_\mathrm{P} = 
(1 - \epsilon_P)\mathbb{I}/2+\epsilon_P \vert + \rangle \langle + \vert$,
and 
$(\rho)_\mathrm{T} = 
(1 - \epsilon_T)\mathbb{I}/2+\epsilon_T \vert + \rangle \langle + \vert$.
Here $\epsilon_{P/T}$ is a dimensionless quantity which represents the
purity of the initial states.}
\label{lgifig2} 
\end{figure}

Using equation \ref{Us}, it can be shown that the ensemble measurement of the `probe' gives correlation between successively measured commuting observables of the `target'. For evaluating TTCC's from an LG string, the observable-set for the target qubit is $ \left\{\hat{\sigma}_x (t_{i}), \hat{\sigma}_x (t_{j})\right\} $ and the corresponding unitaries to be applied to the joint (P + T) system at different time instants $ t_{i}<t_{j} $ are
\begin{eqnarray}
\label{Usx}
U_{\hat{\sigma}_x }(t_{q}) = \mathbb{I}_P \otimes {P}_{+} (t_{q}) + (\hat{\sigma}_z)_P \otimes {P}_{-} (t_{q}).
\end{eqnarray}
Here $ \hat{\sigma}_x (t_{q}) = {P}_{+} (t_{q}) - {P}_{-} (t_{q})$ and $q = i, j$ for time instants 
$t_{i}$ and $t_{j}$.
The quantum network for implementing these unitaries is shown
in Fig. \ref{lgifig2}(a).  

Let the target qubit `T' be initially prepared according to $ \rho $. If the probe qubit `P' is initially in one of the eigenstates of the $\hat{\sigma}_x$ operator, say $ \vert +\rangle = (\vert 0\rangle + \vert 1\rangle)/\sqrt{2} $, the density matrix of the joint system is given by
\begin{eqnarray}
(\rho)_\mathrm{P + T} = (\vert +\rangle \langle +\vert)_P \otimes (\rho)_T.
\end{eqnarray}
Due to the application of the unitaries (7) the joint density matrix evolves according to:
\begin{eqnarray}
\label{rp}
(\rho)_\mathrm{P + T} \longrightarrow  U(t_{j}, t_{i})  (\rho)_\mathrm{P + T}  {U{^{\dagger}}}(t_{i}, t_{j}) = (\rho')_\mathrm{P + T},
\end{eqnarray}

where $ U(t_{j}, t_{i}) = U_{\hat{\sigma}_x }(t_{j}) U_{\hat{\sigma}_x }(t_{i}) $.
In terms of the evolved joint density matrix, the probabilities of obtaining $ {\pm1} $ outcomes for the Pauli-x measurements on the probe are given by:
\begin{eqnarray}
\label{p}
p({\pm1}) = tr_\mathrm{P + T}[{(\rho')_\mathrm{P + T}} \{(\vert {\pm}\rangle \langle {\pm}\vert)_\mathrm{P} \otimes \mathbb{I}_\mathrm{T} \}].
\end{eqnarray}
By tracing over the probe states and using eqns. (\ref{Usx} - \ref{rp}) in eqn. \ref{p}, one obtains:
\begin{eqnarray}
\label{p2}
p({\pm1}) = tr_\mathrm{T}[\left\{{P}_{+} (t_{i}) {P}_{\pm} (t_{j}) + {P}_{-} (t_{i}) {P}_{\mp} (t_{j})\right\} (\rho)_T].
\end{eqnarray}
The ensemble average of the measurement outcome of joint (P + T) observable is given by:
\begin{eqnarray}
\label{sxp}
{\left\langle ({\hat{\sigma}_x })_\mathrm{P} \otimes {\mathbb{I}_\mathrm{T}} \right\rangle} = {+ p({+1}) - p({-1})}.
\end{eqnarray}
Substitution of results \ref{p2} in equation \ref{sxp} gives:
\begin{eqnarray}
\label{cij}
{\left\langle ({\hat{\sigma}_x })_\mathrm{P} \otimes {\mathbb{I}_\mathrm{T}} \right\rangle} & = & 
tr_\mathrm{T} [\hat{\sigma}_x (t_{i}) \hat{\sigma}_x (t_{j})(\rho)_T] \nonumber \\
 & = & {\left\langle {\hat{\sigma}_x (t_{i}) \hat{\sigma}_x (t_{j})}\right\rangle} = C_{ij}.
\end{eqnarray}
Comparing equations \ref{cij1} and \ref{cij}, it is clear that each TTCC in an LG string can be evaluated by applying unitaries (7) to the joint (probe + target) system followed by an ensemble measurement of Pauli-x operator on the probe. 

\section{Experiment}
\label{secexp}
NMR sample consisted of 2 mg of 
$^{13}$C labeled chloroform ($^{13}$CHCl$_3$) 
dissolved in 0.7 ml of deuterated dimethyl sulphoxide (DMSO).  
To implement the protocol described above, the spin-1/2 nuclei of
$^{13}$C and $^1$H atoms are treated as the target spin and the probe spin
respectively.  
All the
experiments are carried out on a Bruker 500 MHz spectrometer at an ambient
temperature of 300 K.
The carbon RF offset was chosen such that the $^{13}$C spin
precesses at an angular frequency of $\omega = 2\pi \times 100$ rad/s under the 
effective longitudinal field in the rotating frame of the RF. The proton
RF offset was chosen at the resonance frequency of $^1$H spin.  The indirect
spin-spin coupling constant (J) for these two spins is 217.6 Hz.
The spin-lattice (T$_1$) and spin-spin (T$_2$) relaxation
time constants for $^1$H spin are respectively 4.1s and 4.0 s.  The 
corresponding time constants for $^{13}$C are 5.5 s and 0.8 s.

The NMR pulse sequence for evaluating TTCCs is described in Fig. \ref{lgifig2}(b).
Initial 90 degree y-pulses on both probe and target prepares them in $\hat{\sigma}_x$ 
states.
All the spin manipulations including the C-NOT gates 
corresponding to $U_{\hat{\sigma}_x }$ operation are 
realized by specially designed strongly modulated pulses \cite{fortunato, MaheshNGE} 
having Hilbert-Schmidt fidelity of over 0.995.
These RF pulses are designed to be robust against the RF field inhomogeneity
in the range of 90\% to 110\% and static field inhomogeneity in the range of
$-5$ Hz to $+5$ Hz. 
The evolution of J-coupling during the intervals between the measurements are
refocused using $\pi$ pulses on $^1$H spin.  
Collective transverse magnetization of the probe spins induce an observable
emf on a resonant Helmholtz-type coil which is amplified, digitized and stored as 
the probe signal. Quadrature detection of the probe signal enables us to measure 
the x-component of the probe magnetization as the real part of the complex signal.
After Fourier transform, the probe signal is fitted to a mixed Lorentzian line shape
to extract the absorptive content.
A reference signal was obtained by an identical experiment with $\Delta t = 0$.
The correlation $C_{ij}(\Delta t)$ was measured  at each value of $\Delta t$ 
by normalizing the real part of the probe signal with the reference signal.
Below, first we will prove the dichotomic nature of nuclear spin observable
which is a requisite for the experimental verification of LGI violation. 
Later subsections shows the experimental results corresponding to LGI violations. 

\subsection{Confirmation of dichotomic nature of x-component of nuclear spin observable}
As the first step towards the implementation of any LGI protocol, one needs to identify a \textit {dichotomic} observable for the target system - i.e. having only two possible outcomes scalable as ${\pm}{1}$ - for measurements of which temporal correlations are to be evaluated. Although Pauli-spin operators (relevant to systems such as spin-1/2 nuclei) are routinely taken as dichotomic observables in NMR-QIP implementations, LGI test requires ensuring that this indeed is the case \textit {experimentally}, despite the presence of dominant couplings of the target nucleus with its environment.  

\begin{figure}[t]
\centering
\includegraphics[width=13cm]{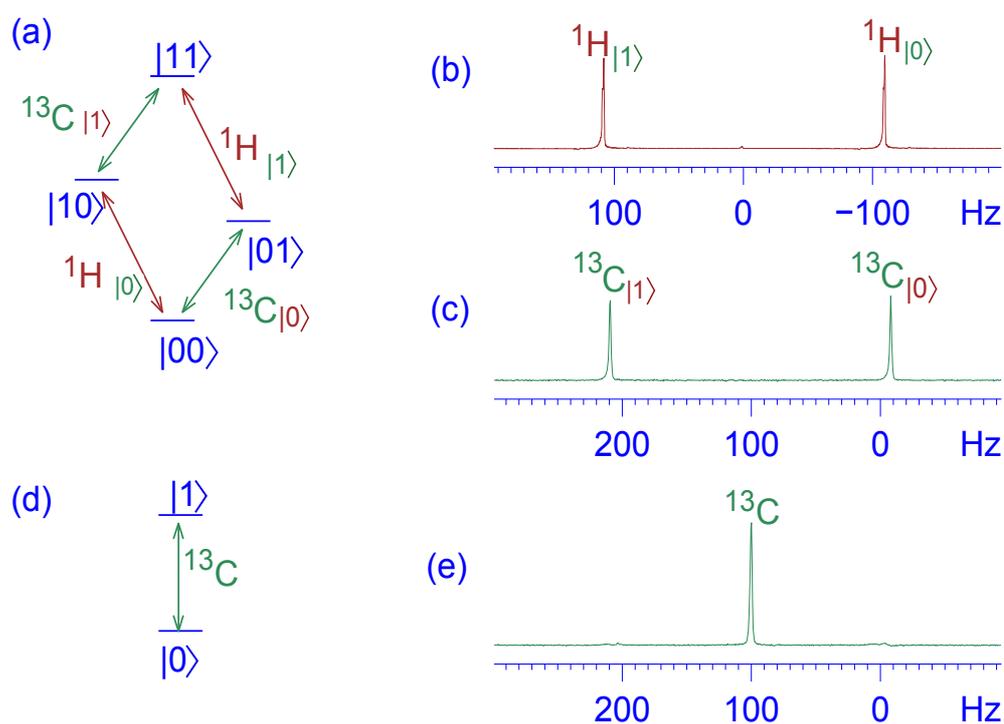}
\caption[The energy level diagram NMR spectra of $^1$H and $^{13}$C]
{The energy level diagram of $^1$H-$^{13}$C system (a) displaying four levels corresponding to two coupled spin-1/2 particles, NMR spectra of $^1$H (b) and $^{13}$C (c) nuclei showing splitting due to mutual interactions.  
The energy levels of $^{13}$C spin system after decoupling $^{1}$H spin (d), and the corresponding
$^{13}$C spectrum (e).
}
\label{dicho1} 
\end{figure}

\begin{figure}
\centering
\includegraphics[width=8cm,angle=-90]{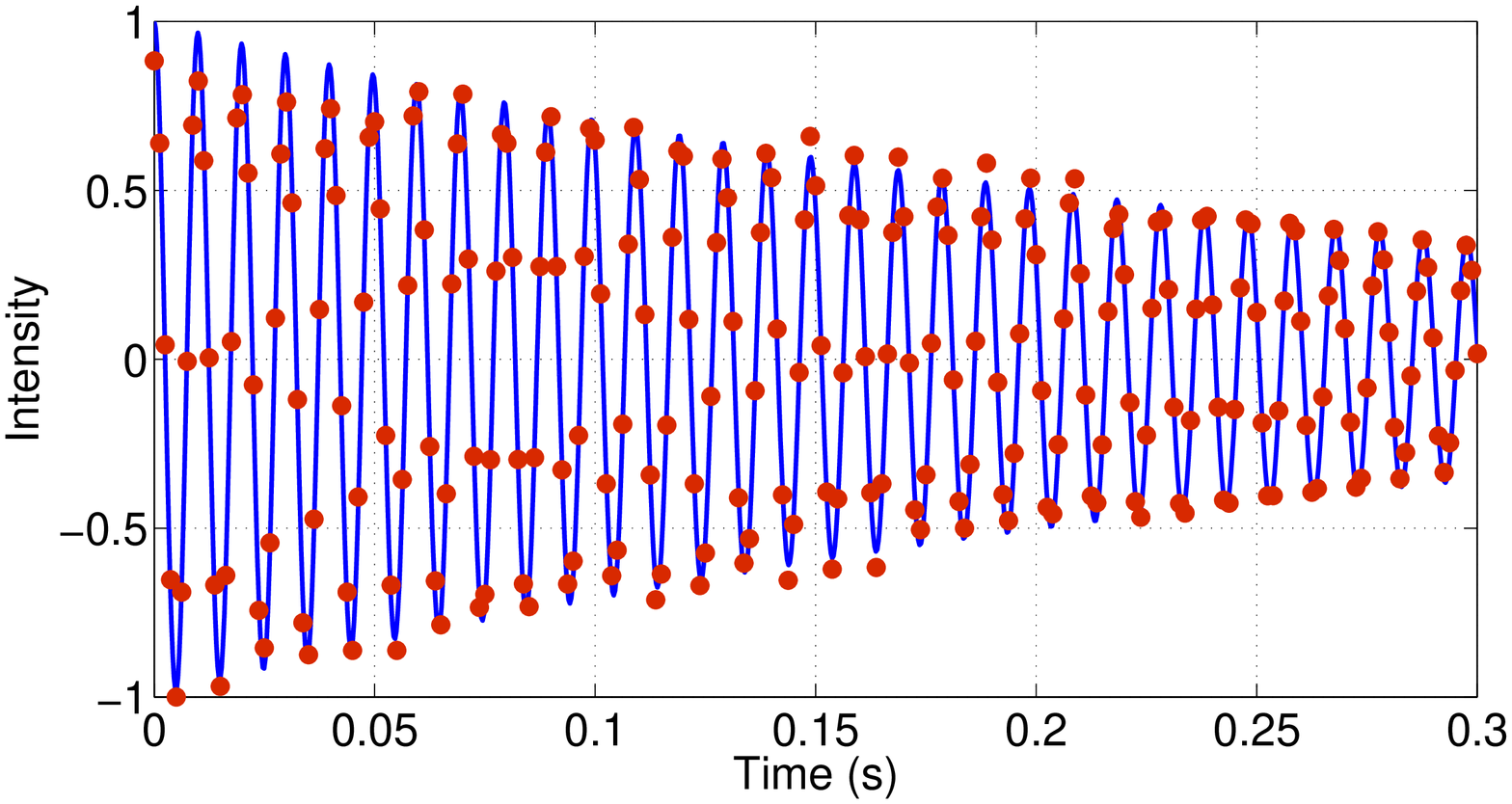}
\caption[Intensity of $^1$H decoupled $^{13}$C spectrum as a function]
{Intensity of $^1$H decoupled $^{13}$C spectrum as a function of time.  
Offsets of the rotating frame are adjusted such that $^1$H has zero 
precession frequency and $^{13}$C has a precession frequency of 100 Hz.
The continuous line is the cosine fit to the experimental data points (dots).
}
\label{dicho2} 
\end{figure}

The $^1$H and $^{13}$C spins in chloroform are coupled by indirect spin-spin interaction (J) with a strength of
217 Hz.  The Hamiltonian for such a two-spin system in a doubly rotating interaction frame can be written as 
\begin{eqnarray}
{\cal H} = h\nu_H \sigma_z^H/2 + h\nu_C \sigma_z^C/2 + hJ \sigma_z^H  \sigma_z^C /4,
\nonumber
\end{eqnarray}
where $\nu_H$ and $\nu_C$ are the precession frequencies of the two nuclei \cite{LevBook}.  In the present experiment
$\nu_H = 0$ Hz and $\nu_C = 100$ Hz. 
The energy level diagram of such a system is shown in Figure \ref{dicho1}a.  The experimental 
spectrum of $^1$H spin consists of two lines corresponding to the two eigenstates of the $^{13}$C spin
(and vice-versa) (Figure \ref{dicho1}b-c).  The effect of the probe spin (i.e., $^1$H) on $^{13}$C can be removed by
spin-decoupling.  Under decoupling, the $^{13}$C spectrum displays just a single line (Figure \ref{dicho1}e)
corresponding to a two-level system (Figure \ref{dicho1}d).

We have also recorded the real part of the intensity of signal corresponding to x-magnetization of $^{13}$C spin  (proportional to $\langle \sigma_x^C \rangle$), under $^1$H decoupling, as a function of precession duration (Figure \ref{dicho2}). The data
clearly fits to a cosine oscillation of single frequency.\\

 Thus, given the above confirmations that $^{13}$C spin is indeed a two-level system and the 
intensity of its signal corresponding x-magnetization has a cosine oscillation with a single 
frequency, we can say that $\sigma_x$ observable used for testing LGI in the present work
is dichotomic.\\

\begin{figure}[t]
\centering
\includegraphics[width=13cm]{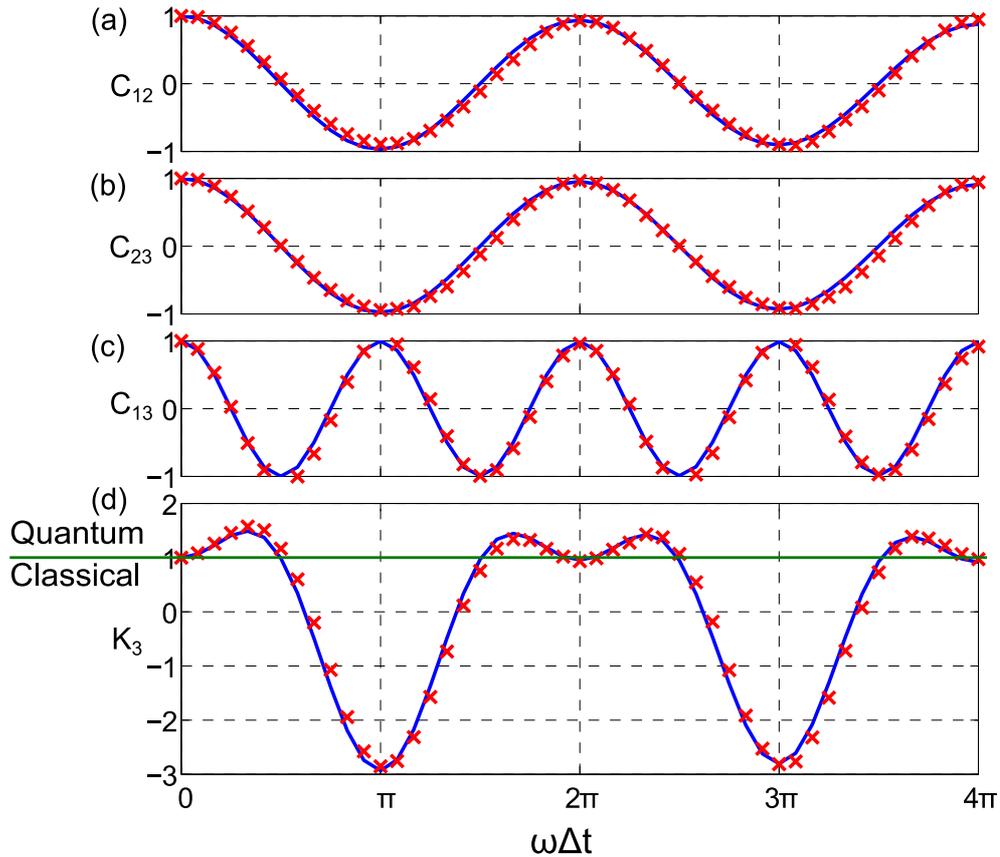} 
\caption[Correlations versus $\Delta t$]
{Correlations versus $\Delta t$: $C_{12}$ (a), $C_{23}$ (b), and $C_{13}$ (c).
$K_3$ is plotted for the range $\omega \Delta t \in [0,4\pi]$ (d). Continuous lines are theoretically
expecting plots with an exponential decay constant and crosses are experimentally achieved results at various time points.
The horizontal line in (d) demarcate the boundary between
the classical and the quantum regimes.
}
\label{lgifig3} 
\end{figure}

\begin{figure}
\centering
\includegraphics[width=14cm]{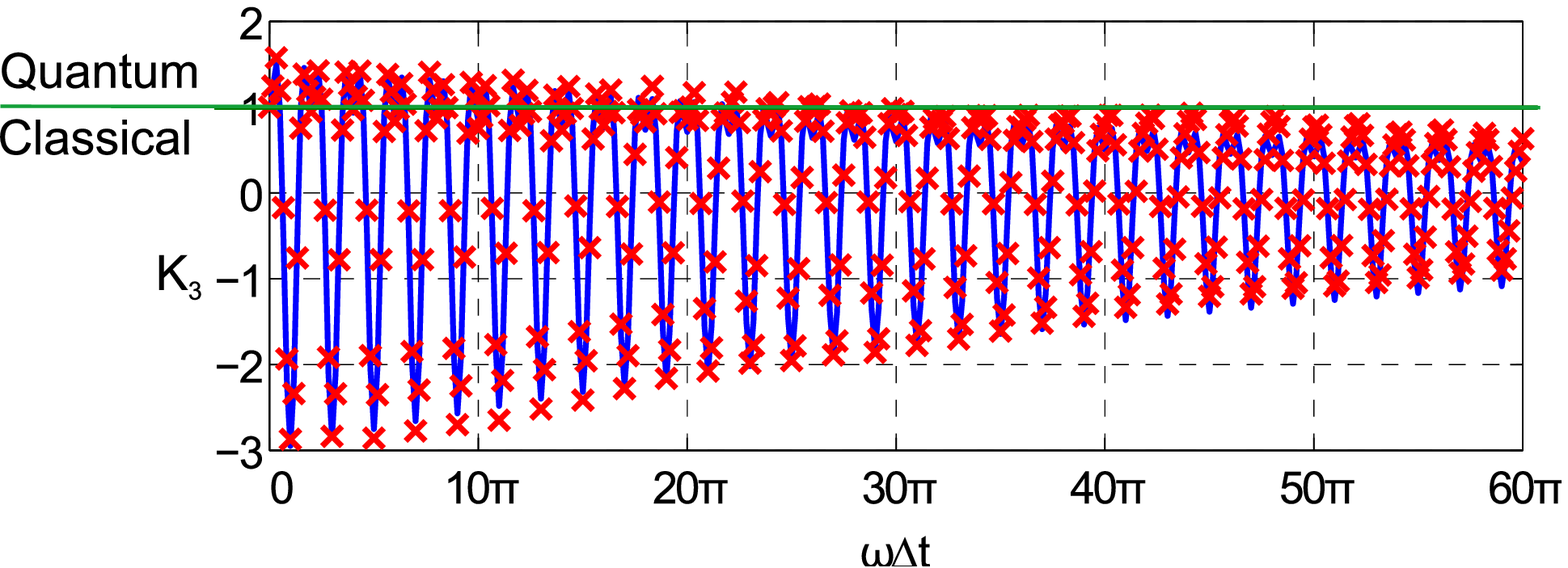} 
\caption[Decay of $K_3$ w. r. t. time]
{Decay of $K_3$ w. r. t. time: $K_3$ is plotted for the range $\omega \Delta t \in [0,60\pi]$ . 
Continuous lines and crosses are used for theoretical
($K_3$ with decay) and experimental values respectively.
The theoretical line was obtained by numerically fitting the 
$K_3$ function given in (\ref{kn}) with an exponential decay
to the experimental data.
The horizontal line demarcate the boundary between
the classical and the quantum regimes.
}
\label{lgifig3b} 
\end{figure}

 \subsection{Violation of LGI for 3 measurement case}
The 3-measurement LG string $K_3 = C_{12}+C_{23}-C_{13}$ was evaluated 
for $\omega \Delta t$ varying from 0 to 60$\pi$, with $\Delta t$ incremented from 0 to 300 ms
in 360 equal steps. 
The results of the experiment are shown in Fig. \ref{lgifig3}.  
The maximum random errors in these experiments were found to be about 0.5\%.
It is clearly seen that the experimental $K_3$ data points
violate the classical limit and hence macrorealism.
Fig. \ref{lgifig3}e shows the $K_3$ plot for an extended duration consisting of
30 periods.  It can be observed that the experimental values of $K_3$ gradually
decay at a time constant of about 288 ms predominantly due to
$T_1$ and $T_2$ relaxations and due to inhomogeneities in the 
magnetic field, thus eventually falling within the classical limit
for $\omega \Delta t > 26\pi$ ($\approx$ 42 ms).

\begin{figure}
\centering
\includegraphics[width=14cm]{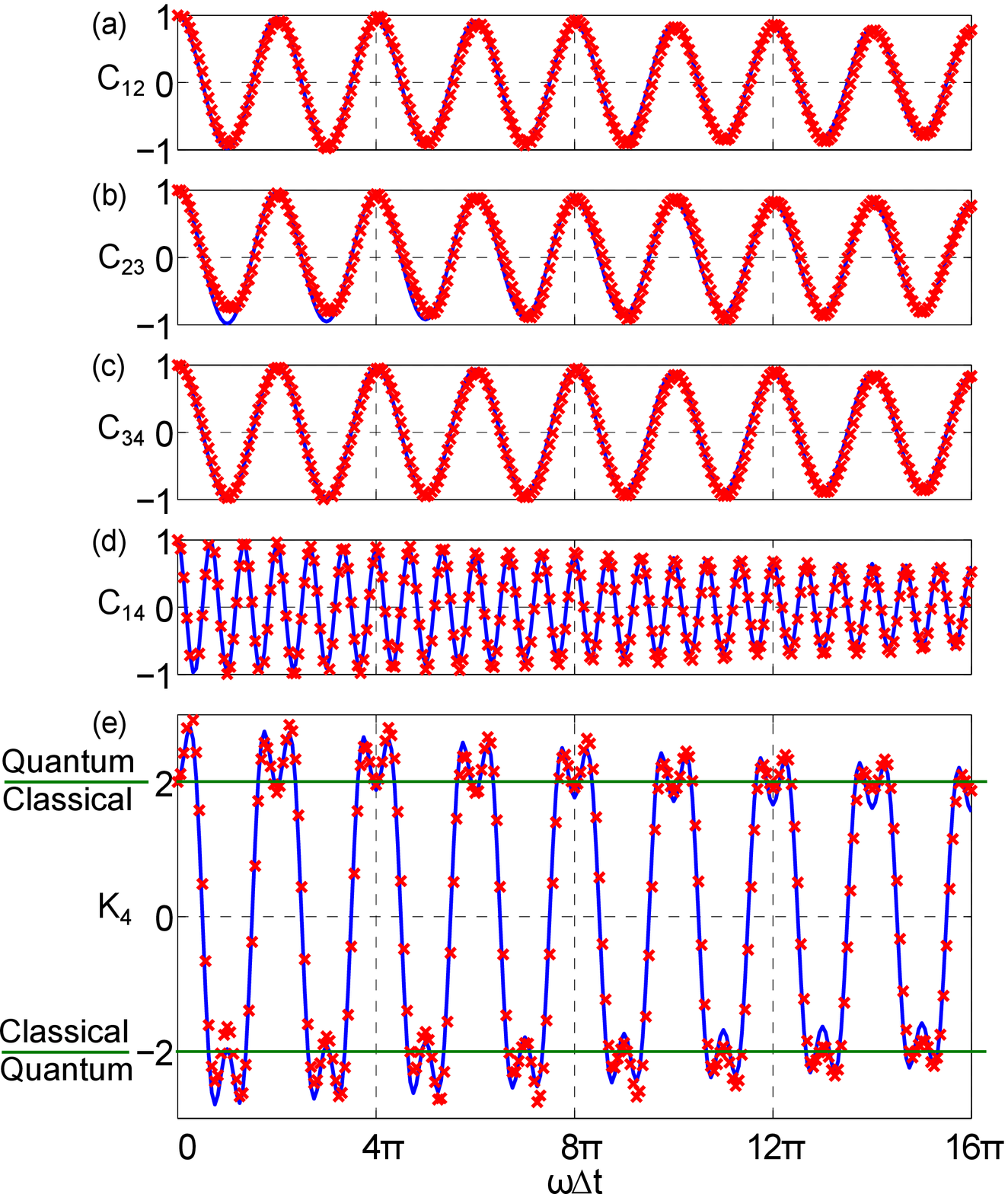} 
\caption[The individual correlations $C_{12}$, $C_{23}$, $C_{34}$, and]
{The individual correlations $C_{12}$, $C_{23}$, $C_{34}$, and $C_{14}$  
are plotted in (a-d) and $K_4 = C_{12}+C_{23}+C_{34}-C_{14}$ is plotted in (e) for the range $\omega \Delta t \in [0,16\pi]$.
Continuous lines and crosses are used for theoretical
($K_4$ with decay) and experimental values respectively.
The theoretical line was obtained by numerically fitting the 
$K_4$ function given in (\ref{kn}) with an exponential decay
to the experimental data.
The horizontal line demarcate the boundary between
the classical and the quantum regimes.}
\label{lgifig4} 
\end{figure}

\subsection{Violation of LGI for 4 measurement case}
Similarly, the 4-measurement LG string $K_4$ was measured for $\omega \Delta t$
varying from 0 to 16$\pi$ (i.e., for 8 periods), with $\Delta t$ varying from 0 to 80 ms. 
The results of the experiment are shown in Fig. \ref{lgifig4}.  Unlike the 3-measurement case,
where the classical and quantum mechanical lower limits for $K_3$ values match (i.e., $-3$),
the 4-measurement case displays 
violation of the classical limit both in the positive as well as
in the negative sides.  Similar to the previous case, we observe an
exponential decay of $K_4$ with a time constant of 
about 324 ms.  
Decay of LG strings is faster than the measured $T_2$ values of either spins
mainly because $T_2$'s have been measured using CPMG sequences which suppress 
the effects of static field inhomogeneity and local fluctuating fields. 

\section{Conclusion}
\label{secconclu}
The present investigation of LGI employs an ensemble of nuclear spins and alleviates the need for repeated experiments on single isolated systems \cite{moussa}. Simultaneous implementation of controlled operations on target-probe pairs enables evaluation of TTCCs and hence plotting of LG strings as functions of two-time measurement delays. 
The plots exhibit both violation and satisfaction of LGI respectively for delays shorter than and comparable to the relaxation timescales \cite{soumyalgi}. 
 we qualitatively interpret them as follows: For time scales, over which environmental effects on spin states are negligible, individual target spins can be taken as isolated quantum systems. The plots do reflect this fact in terms of violation of LGI. However, the spin-environment interaction tends to destroy phase relationship characterizing superposition of quantum states of the target nuclear spin. As a result, each member from the ensemble, with its respective environment traced out, begins to appear as if pre-existing in either one of the two states (of a spin observable chosen for performing measurements, which is Pauli-x in the present work) but not in their \textit {superposition}. Such a gradual transition from quantum to \textit {macrorealistic} behavior of individual \textit {microscopic} systems manifests itself in terms of decay of  TTCCs. 
This ultimately leads to the satisfaction of LGI. Our experimental results thus not only demonstrate initial macrorealism-violating dynamics in genuine microscopic systems such as individual nuclear spins, but also bring forward their environment-induced emergent macrorealistic behavior, captured in terms of satisfaction of LGI and consistent with decoherence mechanism. 
               
\thispagestyle{empty}
\chapter{Quantum Delayed-Choice Experiment}

 In this chapter, we have discussed Bohr's complementary principle and its implication
on light quanta and subsequently on quantum systems. After giving a short introduction of wave-particle duality
in section \ref{dcintro}, we discussed the various interferometer that is been used to study this strange property in section \ref{wpd}. Then we described the theory of recently 
proposed quantum delayed choice experiment in section \ref{dctheory}. In section \ref{dcexp},
we have shown the experimental approach for the implementation of quantum delayed choice circuit
in an NMR quantum information processor. The conclusion is given in section \ref{dcconclu}.

\section{Introduction}
\label{dcintro}
``Is light made up of waves or particles?" has been an intriguing question 
over past many centuries, and the answer remains a mystery even today.
The first comprehensive wave theory of light was advanced by Huygens
\cite{Huygens}. 
He demonstrated how waves might interfere to form a wavefront propagating 
in a straight line, and he could also explain reflection and refraction of
light.  Soon Newton could explain these
properties of light using corpuscular theory, in which light 
was made up of discrete particles \cite{newton}.  The corpuscular theory held over a
century till the much celebrated Young's double slit experiment clearly established
the wave theory of light \cite{young}.
In the Young's experiment, a monochromatic beam of light passing through an obstacle
with two closely separated narrow slits produced an interference pattern
with troughs and crests just like one would expect if waves from two
different sources would interfere.  Other properties of light like 
diffraction and polarization could also be explained easily using the
wave theory. 
The 20th century developments such as Plank's theory of 
black-body radiation and Einstein's theory of photoelectric effects required 
quantization of light into photons \cite{planck,einsteinphoto}.  But the question remained whether
individual photons are waves or particles. Subsequent development of 
quantum mechanics was based on the notion of wave-particle duality \cite{greiner}, which was
essential to explain the behavior not only of the light quanta, but also 
of atomic and sub-atomic entities \cite{bohr}.  

\section{Studying wave-particle duality by interferometers}
\label{wpd}
\subsection{Mach-Zhender Interferometer}
\begin{figure}
\centering
\includegraphics[width=13cm]{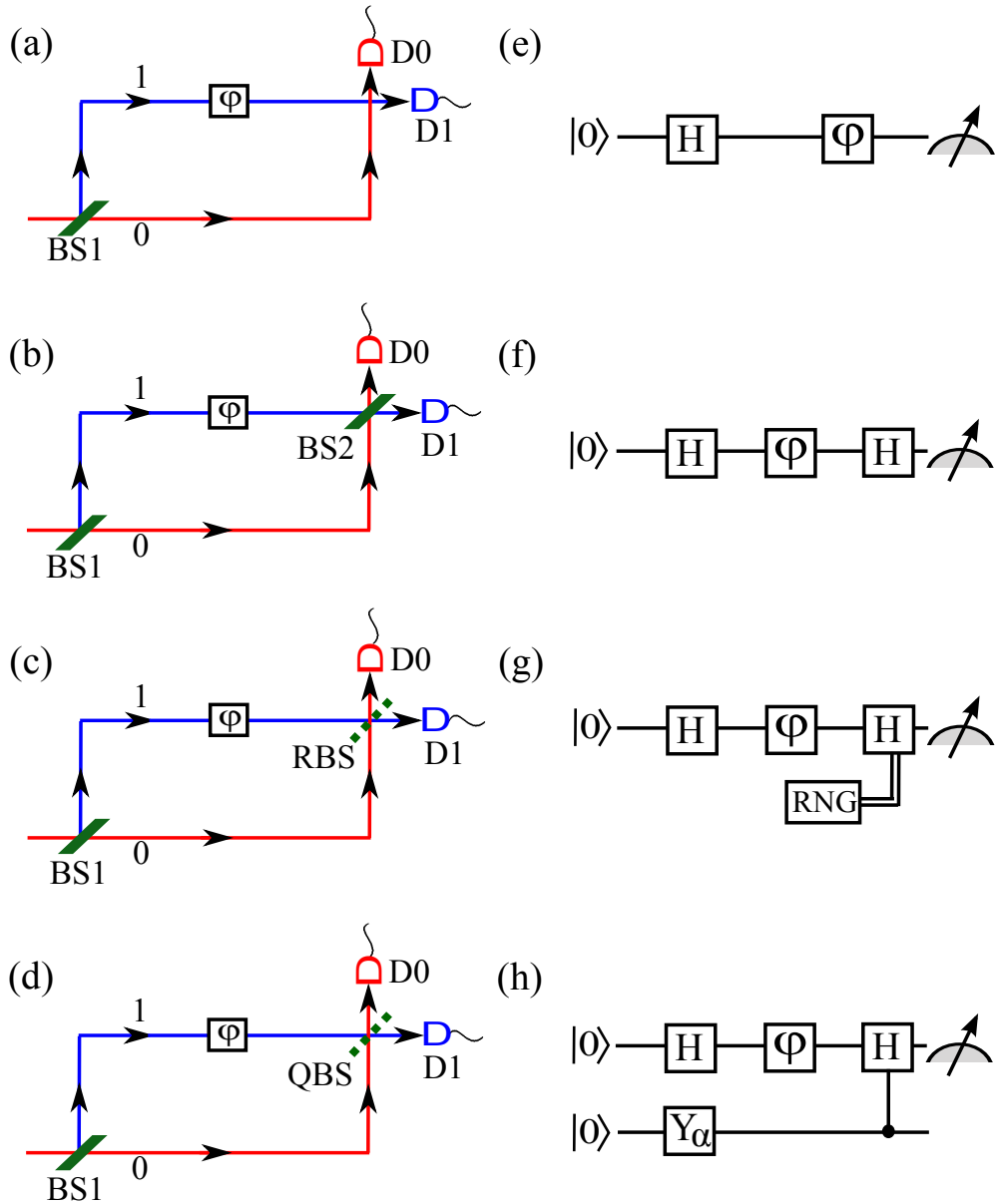}
\caption[Different types of Mach-Zehnder interferometer setups]
{Different types of Mach-Zehnder interferometer setups (a-d)
and equivalent quantum circuits (e-h). BS1 and BS2 are beam splitters,
$\phi$ is phase shifter, D0 and D1 are detectors.  RBS is a beam-splitter
switched ON or OFF by a random number generator (RNG) and  QBS is a beam-splitter
which is controlled by a quantum system in superposition.  In the quantum
circuits, $H$ is the Hadamard gate and $Y_\alpha = e^{-i \alpha \sigma_y}$
is used to prepare the state of ancilla qubit.
}
\label{dcfig1}
\end{figure}

The wave-particle duality of quantum systems is nicely illustrated
by a Mach-Zehnder interferometer (MZI) (see Fig. \ref{dcfig1})
\cite{zehnder,mach}.  The intensity
of the incident light is kept sufficiently weak so that photons enter
the interferometer one by one.
In the open-setup (Fig. \ref{dcfig1}a), it consists
of a beam-splitter BS1, providing each incoming photon with two possible
paths, named 0 and 1.  A phase-shifter in path-1 introduces a relative
phase $\phi$ between the two paths.  The two detectors D0 and D1 help to 
identify the path traveled by the incident photon. Experimental 
results show that only one of the detectors
clicks at a time  \cite{aspect}.  Each click can then be correlated
with one of the two possible paths by attributing particle nature to the
photons.
Here the phase-shifter has no effect on the intensity of
the photons measured by either detector, and therefore no 
interference is observed in this setup.

In the closed-setup (Fig. \ref{dcfig1}b), 
the interferometer consists of a second beam-splitter BS2, which 
allows the two paths to meet before the detection.  Experimental results again show
that only one detector clicks at a time. But much to the astonishment
of  common intuition, the results after many clicks do show  
an interference pattern, i.e., the intensities recorded by each 
detector oscillates with $\phi$ \cite{aspect}.  Since only one photon is present
inside the interferometer at a time, each photon must
have taken both paths in the interferometer and therefore this setup 
clearly establishes the wave property of photons.

The naive question by the classical mindset is ``whether the photon
entering the interferometer decides to take one of the paths or 
both the paths depending on the experimental setup?".  
Scientists who believed 
in a deterministic nature had proposed that, unknown to the 
current experimentalist, there exists some extra information about state 
of the quantum system, which in principle dictates whether the
photon should take either path, or both the paths \cite{bohm}.  In other words,
they assumed some hidden information availed by the photon coming
out of BS1 about the existence or non-existence of BS2.  

\subsection{Wheeler's delayed-choice experiments}
 In order to break this causal link between the two beam-splitters,
Wheeler proposed a modification in the MZI setup (Fig. \ref{dcfig1}c), 
in which the decision to introduce 
or not to introduce BS2 is to be made after the photon has already
passed through BS1 \cite{wheeler1,wheeler2,leggett}.  
This way, there is no causal connection 
between the selection of the paths by the photon and the presence of 
BS2.  Although initially considered as a `thought-experiment', this
proposal has recently been demonstrated by Jacques et al \cite{jacques}.  
In their experimental setup, the second beam-splitter (RBS) was controlled
by a random number generator (RNG), that choose to switch the beam-splitter
ON or OFF after the photon has already passed through BS1.
The results of this delayed-choice experiment was in agreement with 
Bohr's complementarity principle \cite{bohr}.  That is, the behavior of
the photon in the interferometer depends on the choice of the observable that is
measured, even when that choice is made at a
position and a time such that it is separated from
the entrance of the photon into the interferometer
by a space-like interval.
Breaking the causal link had no effect on
the results of the wave-particle duality, thus ruling out the existence
of hidden information \cite{jacques}. 

\subsection{Quantum delayed-choice experiments}
Recently, Ionicioiu and Terno have proposed a modified version
(Fig. \ref{dcfig1}d) of the
Wheeler's experiment which not only demonstrates the intrinsic duality, but
also shows that a photon can have a morphing behavior between particle and 
wave \cite{ionicioiu}. In their setup, BS2 is replaced with a beam splitter which is
switched OFF or ON depending on $\vert 0 \rangle$ or $\vert 1 \rangle$ state 
of a two-level quantum system.
Using this modification, Ionicioiu and Terno have been able to discard hidden 
variable theories which attempt to assign intrinsic wave or particle nature to individual 
photons even before the final measurement.  This proposed experiment is named
as `Quantum Delayed-Choice Experiment' \cite{ionicioiu}.

Using nuclear magnetic resonance (NMR) techniques we study the behavior of a target
spin-1/2 nucleus going through a similar situation as that of a photon going through an
interferometer \cite{soumyaqdc}.  Another spin-1/2 nucleus acts as an ancilla controlling 
the second beam-splitter.
In section \ref{dctheory} we briefly explain the theory and in section \ref{dcexp}
we describe the experimental results.

\section{Theory}
\label{dctheory}
In the following we shall use the terminology of quantum information. 
The two possible paths of the interferometer are assigned with the orthogonal
states $\vert 0 \rangle$ and $\vert 1 \rangle$ of a quantum bit.  
The equivalent quantum circuits for the different setups of MZI
are shown in Figs. \ref{dcfig1}(e-h).  
Similar circuits have previously been used in `duality computers' \cite{long1,long2,long3}.
In these circuits the Hadamard operator H has the function of the beam splitter BS1.
It transforms the initial state $\vert 0 \rangle$  to the superposition
$(\vert 0 \rangle + \vert 1 \rangle)/\sqrt{2}$ such that both $\vert 0 \rangle$
and $\vert 1 \rangle$ states are now
equally probable. 
The detection operators for the two detectors are $D_0 = \vert 0 \rangle \langle 0 \vert$ and
$D_1 = \vert 1 \rangle \langle 1 \vert$.

In the open setup (Fig. \ref{dcfig1}e), the state after the phase shift 
becomes, $\vert \psi_p \rangle = (\vert 0 \rangle + e^{i\phi} \vert 1 \rangle)/\sqrt{2}$.    
The intensities recorded by the two detectors are given by the 
expectation values,
\begin{eqnarray}
S_{\mathrm{p},0} & = & \langle \psi_\mathrm{p} \vert D_0 \vert \psi_\mathrm{p} \rangle = \frac{1}{2} \;\; \mathrm{and} \nonumber \\
S_{\mathrm{p},1} & = & \langle \psi_\mathrm{p} \vert D_1 \vert \psi_\mathrm{p} \rangle = \frac{1}{2},
\label{sp0}
\end{eqnarray}
independent of the phase introduced.  Therefore no interference can  be observed
and accordingly this setup demonstrates the particle nature of the quantum system.
The visibility of the interference
\begin{eqnarray}
\nu = \frac{\mathrm{max}(S) - \mathrm{min}(S)}{\mathrm{max}(S) + \mathrm{min}(S)},
\end{eqnarray}
is zero in this case.

The equivalent quantum circuit for the closed interferometer is shown in Fig.
\ref{dcfig1}f.  After the second Hadamard one obtains the state,
$\vert \psi_\mathrm{w} \rangle = \cos\frac{\phi}{2} \vert 0 \rangle - i\sin\frac{\phi}{2} \vert 1 \rangle,$
up to a global phase. The intensities recorded by the two detectors are now,
\begin{eqnarray}
S_{\mathrm{w},0} & = & \langle \psi_\mathrm{w} \vert D_0 \vert \psi_\mathrm{w} \rangle = \cos^2\frac{\phi}{2} \;\; \mathrm{and} \nonumber \\
S_{\mathrm{w},1} & = & \langle \psi_\mathrm{w} \vert D_1 \vert \psi_\mathrm{w} \rangle = \sin^2\frac{\phi}{2}.
\label{sw0}
\end{eqnarray}
Thus as a function of $\phi$, each detector obtains an interference pattern 
with visibility $\nu = 1$.  This setup clearly demonstrates the wave nature of the target qubit.

In the circuit corresponding to 
the Wheeler's experiment (Fig. \ref{dcfig1}g), the decision to insert or not to
insert the second Hadamard gate is to be made after the first Hadamard gate has been applied.

Here, we focus on the next modification, that is the quantum delayed-choice
experiment \cite{ionicioiu}.  
In the equivalent quantum circuit (Fig. \ref{dcfig1}h), the second Hadamard gate is to be
decided in a quantum way.  This involves an ancilla spin prepared in a
superposition state $\cos \alpha \vert 0 \rangle + \sin \alpha \vert 1 \rangle $.
This state can be prepared by rotating the initial $\vert 0 \rangle$ state of ancilla 
by an angle $2\alpha$ about $y$-axis (using operator $Y_\alpha = e^{-i \alpha \sigma_y}$).
The second Hadamard gate is set to be controlled by the ancilla qubit.  If the ancilla is in state
$\vert 0 \rangle$, no Hadamard gate is applied, else if the ancilla is in state 
$\vert 1 \rangle$, Hadamard gate is applied.  The combined state of the two-qubit system
after the control-Hadamard gate is
\begin{eqnarray}
\vert \psi_\mathrm{wp,\alpha} \rangle = \cos \alpha \vert \psi_\mathrm{p} \rangle \vert 0 \rangle 
+ \sin \alpha \vert \psi_\mathrm{w} \rangle \vert 1 \rangle,
\end{eqnarray}
wherein the second ket denotes the state of ancilla.
After tracing out the ancilla, the reduced density operator for the system becomes,
\begin{eqnarray}
\rho_\mathrm{wp} = \cos^2 \alpha \vert \psi_\mathrm{p} \rangle \langle \psi_\mathrm{p} \vert + 
\sin^2 \alpha \vert \psi_\mathrm{w} \rangle \langle \psi_\mathrm{w} \vert.
\label{psiwp}
\end{eqnarray}
Again, the intensity recorded by each detector can be obtained by calculating the expectation values.
For example, the intensity at the detector D0 is,
\begin{eqnarray}
S_\mathrm{wp,0} (\alpha,\phi) &=& \mathrm{tr}[D_0 \; \rho_\mathrm{wp}] \nonumber \\
 &=&  \mathrm{tr}[D_0 \vert \psi_\mathrm{p} \rangle \langle \psi_\mathrm{p} \vert] \cos^2 \alpha
 + \nonumber \\
 && \mathrm{tr}[D_0 \vert \psi_\mathrm{w} \rangle \langle \psi_\mathrm{w} \vert] \sin^2 \alpha \nonumber \\
  & = &  S_{\mathrm{p},0} \cos^2 \alpha +  S_{\mathrm{w},0} \sin^2 \alpha \nonumber \\
  & = & \frac{1}{2} \cos^2 \alpha + \cos^2\frac{\phi}{2} \; \sin^2 \alpha.
\label{swp0}
\end{eqnarray}
It can be immediately seen that the visibility $\nu$ for the above interference varies as $\sin^2 \alpha$.
When $\alpha = 0$, the quantum system has a particle nature and when $\alpha = \pi/2$, it
has a wave nature.  In the intermediate values of $\alpha$, the quantum system is morphed 
in between the particle and the wave nature.  
In the following section we describe the experimental demonstration of morphing of a quantum system between wave and particle
behaviors.

\section{Experiment}
\label{dcexp}
\begin{figure}
\centering
\includegraphics[width=12.5cm]{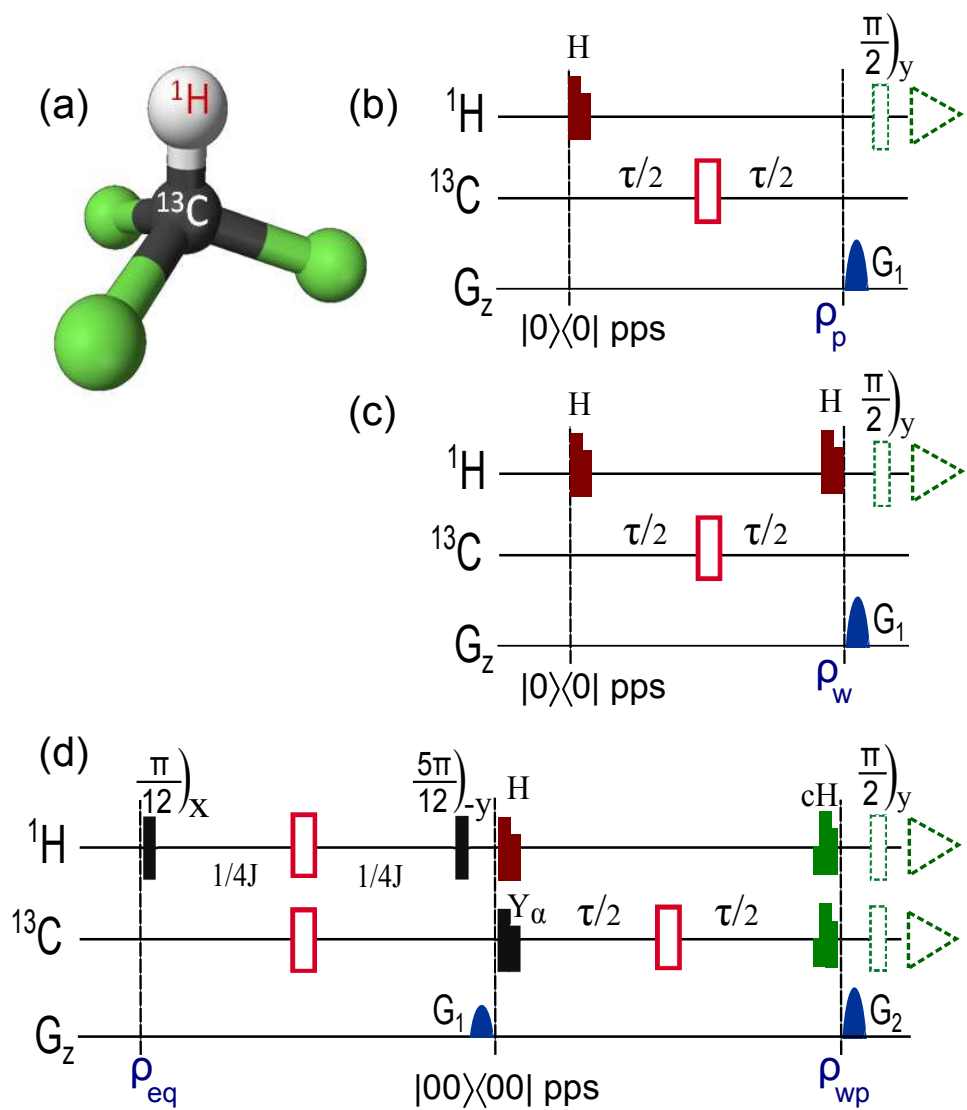}
\caption[Molecular structure of chloroform (a) and pulse-sequences]
{Molecular structure of chloroform (a) and pulse-sequences (b-d)
for different setups of MZI.
Figs. (b) and (c) correspond to the open and closed setups respectively,
and (d) corresponds to the quantum delayed-choice experiment.
The unfilled rectangles are $\pi$ pulses.  Shaped pulses are strongly modulated pulses corresponding
to Hadamard gate (H), Y$_\alpha$ gate, and control-Hadamard (cH) gate. 
$\pi/2$ detection pulses are shown in dotted rectangles.
$J$ is the coupling constant and $\tau$ is the phase-shifting delay.
G$_1$ and G$_2$ are two pulsed-field-gradients for destroying coherences.
In (d) two separate experiments for $^1$H and $^{13}$C are recorded
after applying respective $\pi/2$ detection pulses.  
$\rho_\mathrm{eq}$, 
$\rho_\mathrm{p} = \vert \psi_\mathrm{p} \rangle \langle \psi_\mathrm{p} \vert$, 
$\rho_\mathrm{w} = \vert \psi_\mathrm{w} \rangle \langle \psi_\mathrm{w} \vert$, 
and 
$\rho_\mathrm{wp} = \vert \psi_\mathrm{wp} \rangle \langle \psi_\mathrm{wp} \vert$ represent
the states at different time instants. 
}
\label{dcfig2}
\end{figure}

The sample consisted of $^{13}$CHCl$_3$ (Fig. \ref{dcfig2}a) 
dissolved in CDCl$_3$.    
Here $^1$H and $^{13}$C spins are used as the target and the ancilla qubits respectively.
The two spins are coupled by indirect spin-spin interaction with a coupling constant of $J=209$ Hz.
All the experiments
were carried out at an ambient temperature of 300 K in a 500 MHz Bruker NMR spectrometer.

\subsection{Open and closed interferometers}
The pulse-sequences corresponding to open and closed setups of MZI
are shown in Fig. \ref{dcfig2}(b-c).  In these cases, the circuits (Fig. \ref{dcfig1}(e-f)) need
only a single target qubit and no ancilla qubit. Here $^1$H spin
is used as the target qubit, and its interaction with $^{13}$C spin is refocused 
during the MZI experiments.
Ideally both of these
setups need initializing the target qubit to $\vert 0 \rangle$ state.
In thermal equilibrium at temperature $T$ and magnetic field $B_0$, an ensemble of 
isolated spin-1/2 nuclei 
 exists in a Boltzmann mixture,
\begin{eqnarray}
\rho_\mathrm{eq} =\frac{1}{2} e^{\epsilon/2} \vert 0 \rangle \langle 0 \vert
+ \frac{1}{2} e^{-\epsilon/2} \vert 1 \rangle \langle 1 \vert,
\end{eqnarray}
$\epsilon = \gamma \hbar B_0/kT$ is a dimensionless constant 
which depends on the magnetogyric ratio $\gamma$ of the spin.  
At ordinary NMR conditions $\epsilon \sim 10^{-5}$
and therefore $\rho_\mathrm{eq}$ is a highly mixed state.  Since preparing a pure 
$\vert 0 \rangle$ state requires extreme conditions, one can alleviate this problem
by rewriting the equilibrium state as the pseudopure state
\begin{eqnarray}
\rho_\mathrm{eq} =
\vert 0 \rangle \langle 0 \vert_\mathrm{pps} \approx
\frac{1}{2}\left(1-\frac{\epsilon}{2} \right)\mathbbm{1} + 
\frac{\epsilon}{2}\vert 0 \rangle \langle 0 \vert.
\end{eqnarray}
The identity part does neither evolve under the Hamiltonians, nor does it give raise
to NMR signals, and is therefore ignored.  
Thus the single qubit equilibrium state effectively mimics the state $\vert 0 \rangle$.

In all the cases (Fig. \ref{dcfig2}(b-d)), the first Hadamard gate on the target qubit is followed
by the phase shift. 
A 100 Hz resonance off-set of $^1$H spin was used to introduce the desired phase shift $\phi(\tau) = 200 \pi \tau$,
with the net free-precession delay $\tau$.  Experiments were carried out at 21 linearly spaced 
values of $\phi$ in the range $[0,2\pi]$.
The  $^{13}$C spin was set on-resonance and the 
$J$-evolution during $\tau$ was refocused with a $\pi$ pulse on $^{13}$C. 

\begin{figure}[bottom]
\centering
\includegraphics[width=9cm,angle=-90]{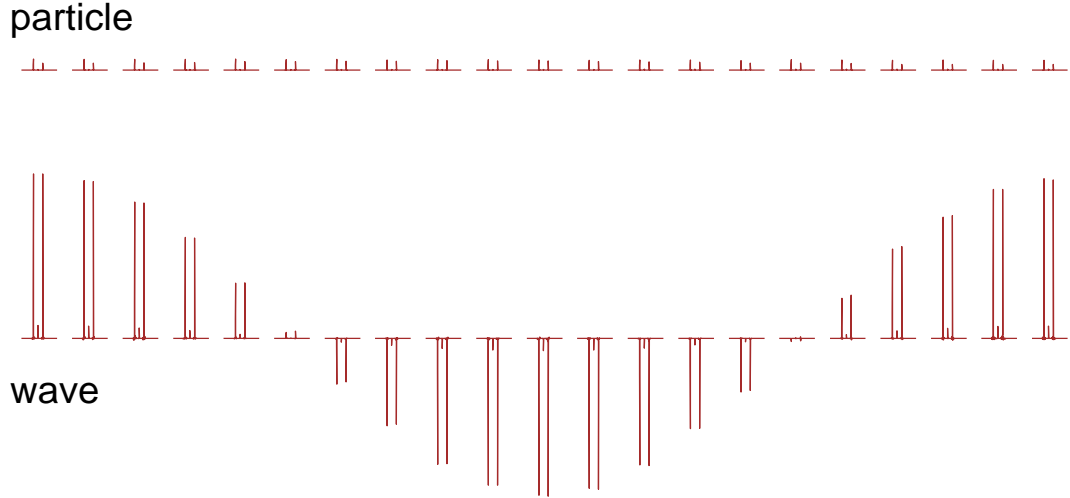}
\caption[The experimental spectra obtained after the open]
{The experimental spectra obtained after the open (top trace) and 
closed (bottom trace) setups of MZI.  
Each spectrum (pair of lines) corresponds to one of the 21 linearly spaced 
values of $\phi$ in the range $[0,2\pi]$.
}
\label{dcfig3}
\end{figure}

Unlike the open interferometer (Fig. \ref{dcfig2}b), the closed interferometer (Fig. \ref{dcfig2}c) has a second Hadamard 
gate.  In both of these cases, the intensity recorded by D1 detector corresponds to
the expectation value of $D_0 = \vert 0 \rangle \langle 0 \vert$ operator, which is a 
diagonal element of the density operator.  To measure this element, we destroy all the
off-diagonal elements (coherences) using a pulsed field gradient (PFG) G$_1$, followed by a $(\pi/2)_y$
detection pulse.  The most general diagonal density operator for a single qubit 
is
$\rho = \frac{1}{2} \mathbbm{1} + c \sigma_z$,
where $c$ is the unknown constant to be determined. After applying the $(\pi/2)_y$
detection pulse, we obtain $\frac{1}{2} \mathbbm{1} + c \sigma_x$.  The
corresponding NMR signal is proportional to $c$. The experimental NMR
spectra for the open and closed setups are shown in Fig. 
\ref{dcfig3}.  These spectra are normalized w.r.t. equilibrium detection.
Since both the pathways created by BS1 are equally probable in the open
MZI, $c=0$ and therefore spectrum vanishes.  On the other hand, 
because of the second beam-splitter (BS2) in closed MZI, $c$ becomes 
$\phi$ dependent, and hence the interference pattern.

\begin{figure}[t]
\centering
\includegraphics[width=12cm]{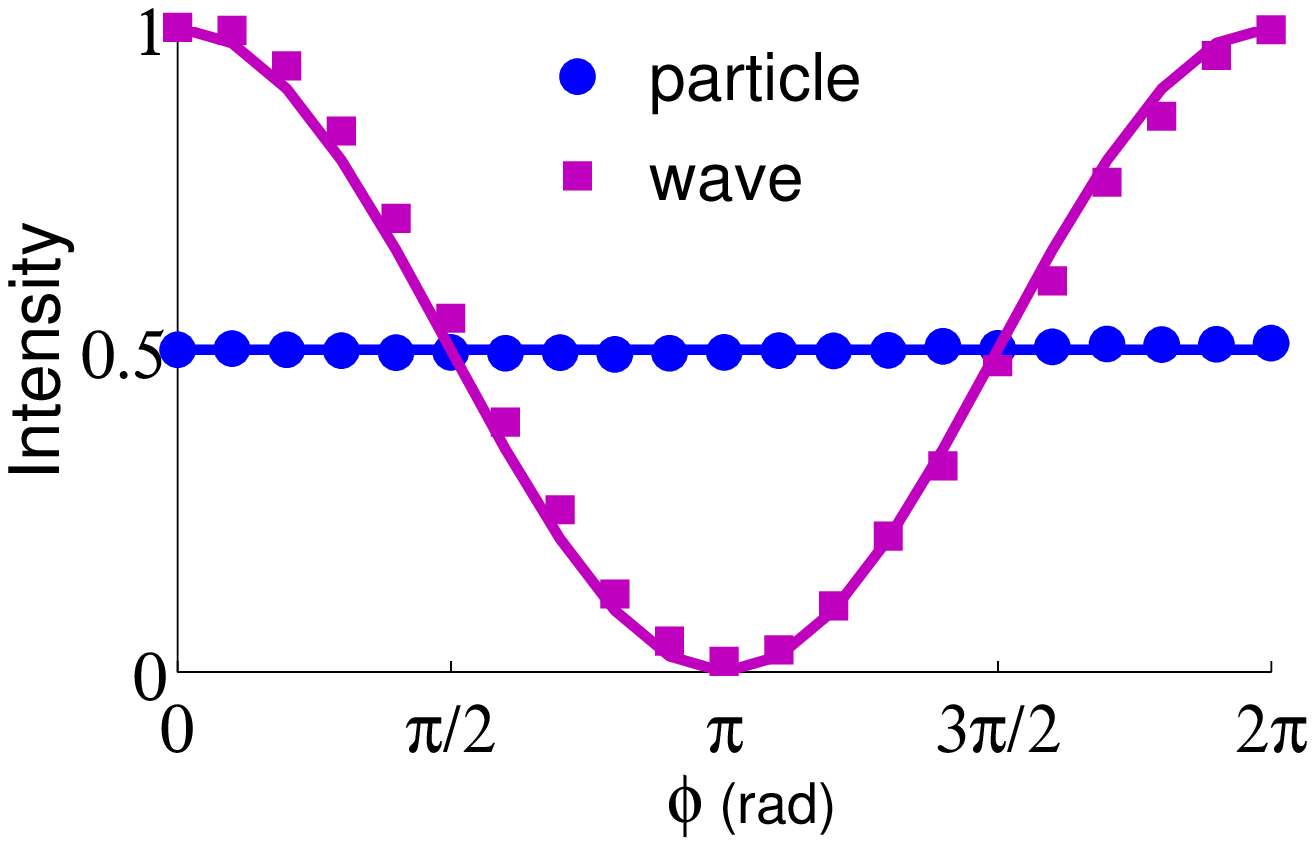}
\caption[The experimental intensities $S_\mathrm{p,0}$ (particle) and 
$S_\mathrm{w,0}$ (wave)]
{The experimental intensities $S_\mathrm{p,0}$ (particle) and 
$S_\mathrm{w,0}$ (wave) at various values of $\phi$.
}
\label{dcfig4}
\end{figure}

The corresponding intensities $S_\mathrm{p(w),0} = c+1/2$ are shown in
Fig. \ref{dcfig4}.  
The theoretical values from expressions (\ref{sp0}) and (\ref{sw0}) are also shown
in solid lines. 
The experimental visibility of interference in the 
particle case is 0.02 and that in the wave case is 0.97.
As explained in the previous section, the open setup demonstrates the particle nature
and the closed setup demonstrates the wave nature.

\subsection{Quantum delayed-choice experiment}
The circuit for quantum delayed-choice experiment is shown in Fig. \ref{dcfig1}h and
the corresponding NMR pulse-sequence is shown in Fig. \ref{dcfig2}d.  This circuit
requires one target qubit ($^1$H) and one ancilla qubit ($^{13}$C). 
The equilibrium state of the two-qubit system does not
correspond to a pseudopure state and therefore it is necessary to redistribute the 
populations to achieve the desired pseudopure state.
We used spatial averaging technique to prepare the pseudopure state \cite{corysp}
\begin{eqnarray}
\rho_\mathrm{pps} = \frac{1-\epsilon'}{4}\mathbbm{1} + \epsilon' \vert 00 \rangle \langle 00 \vert,
\end{eqnarray}
where $\epsilon'$ is the residual purity.

All the gates on the target and the ancilla were realized using
strongly modulated pulses (SMPs) \cite{fortunato, MaheshNGE}.
The SMPs were constructed to be robust against RF amplitude inhomogeneities,
which normally have a distribution of about 10 \% about the mean.
Robust pulses were achieved by calculating the Hilbert-Schmidt fidelity between the
desired operator and the experimental operator for different possible
RF amplitude distributions, and then maximizing the average fidelity \cite{dieterrev}.
An average fidelity of over 0.995 was achieved for each gate.
After the control-Hadamard gate, the state of the two-qubit system is expressed by
the density operator $\rho_\mathrm{wp}$ (eqn. \ref{psiwp}) up to the unit background.

\begin{figure}
\centering
\includegraphics[width=10cm,angle=-90]{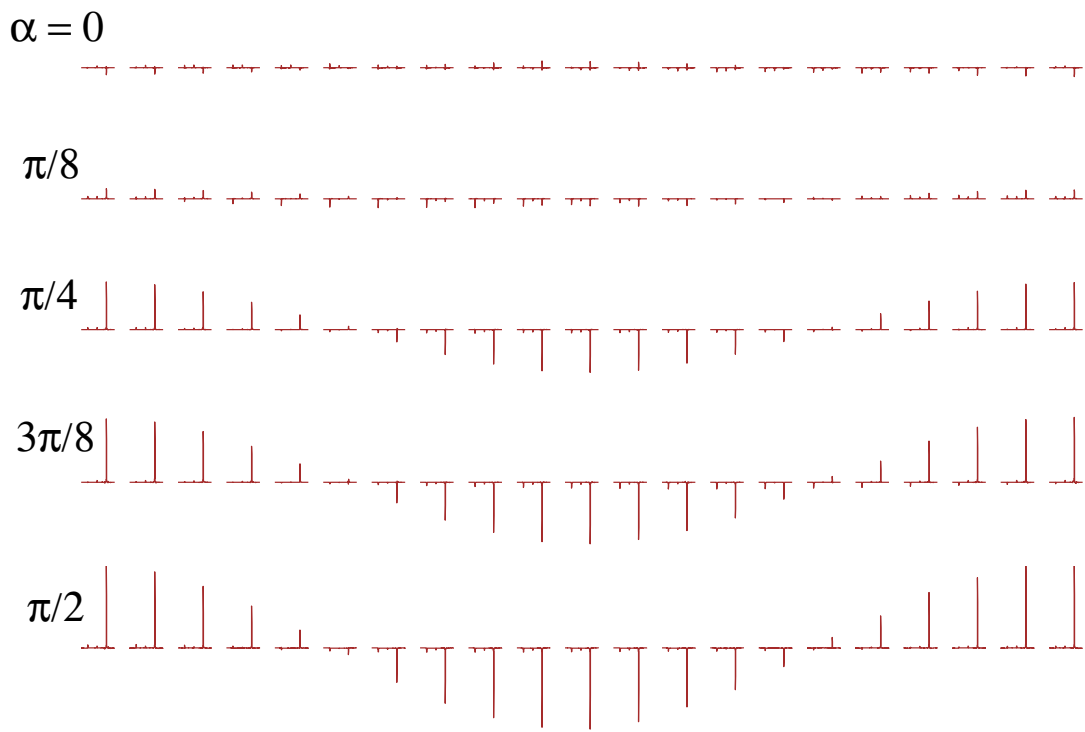}
\caption[The experimental spectra obtained after the quantum delayed choice experiment]
{The experimental spectra obtained after the quantum delayed choice
experiment with $(\pi/2)_y$ detection pulse on target ($^1$H) qubit.
These spectra are recorded with 21 equally spaced values of $\phi \in [0,2\pi]$ and
at different $\alpha$ values (as indicated).  In each spectrum, only one
line is expected due to the preparation of pseduopure state.
}
\label{dcfig5}
\end{figure}

The interference $S_\mathrm{wp,0}$ (in eqn. \ref{swp0}) due to the detection operator 
$D_0 = \vert 00 \rangle \langle 00 \vert$ 
can be obtained by measuring the first diagonal element of the density matrix, and
hence complete density matrix tomography is not necessary \cite{maheshjmr10}.
As in the single qubit case, we apply a PFG $G_2$ which averages out all the coherences and 
retains only the diagonal part of the density matrix.  The most general diagonal
density matrix of a two-qubit system is of the form
\begin{eqnarray}
\rho = \frac{1}{4}\mathbbm{1} \otimes \mathbbm{1} 
+ c_1 \sigma_z \otimes  \mathbbm{1}  
+ c_2 \mathbbm{1} \otimes  \sigma_z  
+ c_3 \sigma_z \otimes  \sigma_z,
\end{eqnarray}
with the unknown constants $c_1$, $c_2$, and $c_3$.

Recording the target spectrum after a $(\pi/2)_y$ pulse on the above state
gives two signals proportional to $c_1+c_3$ and $c_1-c_3$.  
The spectra of the target qubit at various values of $\phi$ and
$\alpha$ are shown in Fig. \ref{dcfig5}.  
The signals obtained after applying a $(\pi/2)_y$ pulse on either qubit after
preparing the $\vert 00 \rangle$ pseudopure state are used to normalize these
intensities.  In each spectrum, the left transition (corresponding to the $\vert 0 \rangle$
state of ancilla), vanishes because of the particle nature (similar to the top trace of Fig. \ref{dcfig3})
and the right transition (corresponding to the $\vert 1 \rangle$ state of ancilla) displays the 
interference pattern because of the wave nature (similar to the bottom trace of Fig. \ref{dcfig3}).

\begin{figure}
\centering
\includegraphics[width = 14cm]{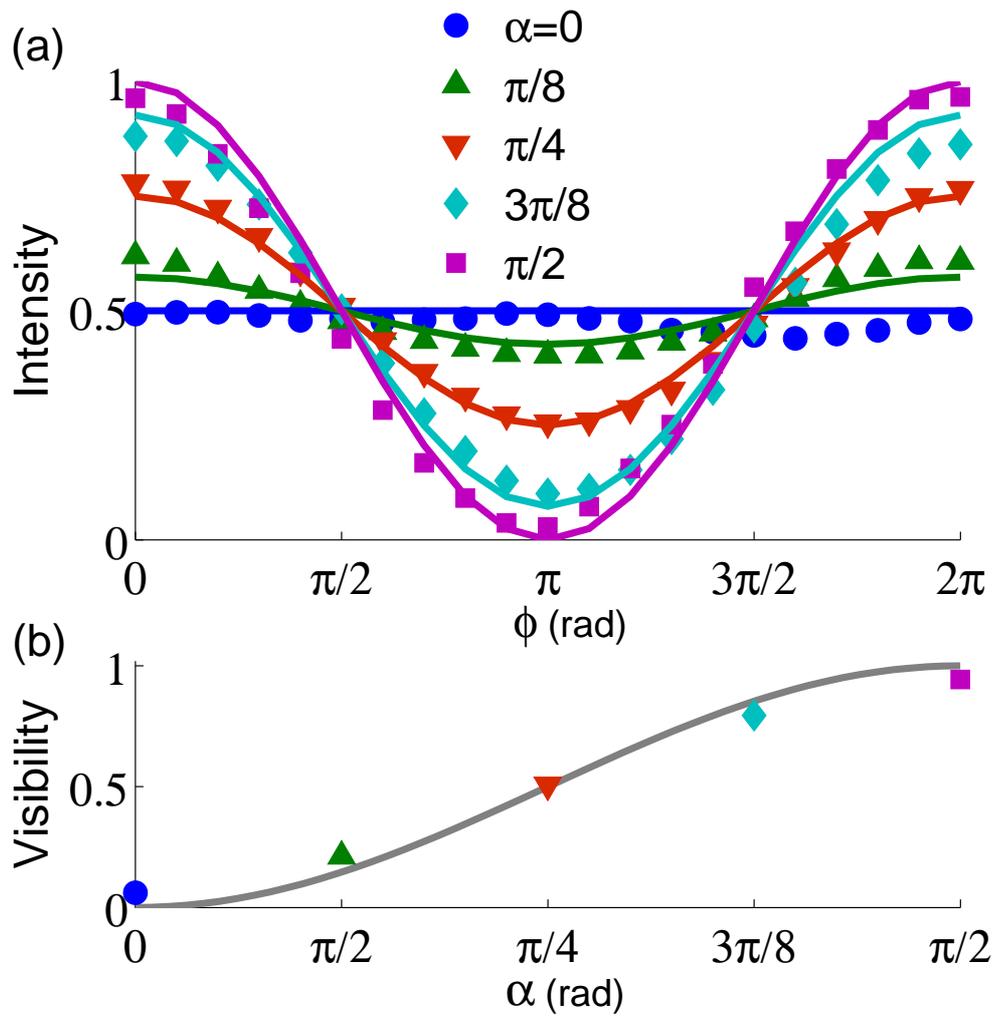}
\caption[The intensities $S_\mathrm{wp,0}(\alpha,\phi)$ versus phase $\phi$]
{The intensities $S_\mathrm{wp,0}(\alpha,\phi)$ versus phase $\phi$ for
different values of $\alpha$ (a) and the visibility $\nu$ versus $\alpha$ (b).
The theoretical values are shown in solid lines and the experimental results
are shown by symbols.  
}
\label{dcfig6}
\end{figure}

Similarly, recording the ancilla spectrum after a $(\pi/2)_y$ pulse 
gives two signals proportional to $c_2+c_3$ and $c_2-c_3$.  
From these four transitions one can precisely determine all the three unknowns $c_1$,
$c_2$, and $c_3$, and obtain the 
population $S_\mathrm{wp,0} = 1/4 + c_1+c_2+c_3$.
Calculated experimental intensities $S_\mathrm{wp,0}$ are shown in 
Fig. \ref{dcfig6}a. The intensities were measured for five
values of $\alpha$ in the range $[0,\pi/2]$, and for 21 values of $\phi$ in the range
$[0,2\pi]$.  The theoretical values from expression (\ref{swp0}) are also shown
in solid lines. 
The experimental values were found to have small random errors with a standard deviation less 
than 0.01.  
The significant systematic errors are due to experimental limitations
such as radio-frequency inhomogeneity and spectrometer non-linearities.

The visibility $\nu$ calculated at different values of $\alpha$ are plotted in 
Fig. \ref{dcfig6}b.  The theoretical visibility varies as $\sin^2\alpha$ as
explained in the section II.
There appears a general agreement between the
quantum mechanical predication (solid-line) and the experiments (symbols).

\section{Conclusions}
\label{dcconclu}
In this chapter, we have studied the open and closed setups of Mach-Zehnder interferometer using
nuclear spin qubits, and demonstrated the particle-like and wave-like behaviors
of the target qubit. 
Previously NMR interferometer has been used to study dipolar oscillations in solid state NMR \cite{stoll}
and to measure geometric phases in multi-level systems \cite{dietergp,arindamgp,dugp}. 
We have reported the first experimental demonstration of the quantum delayed-choice
experiment using NMR interferometry.  

Bohr's complementarity principle is based on mutually exclusive experimental arrangements.
However, the quantum delayed-choice experiment proposed by Ionicioiu and Terno 
\cite{ionicioiu}, 
suggests that we can study the complementary properties like particle and wave behavior
of a quantum system in a single experimental setup if the ancilla is prepared
in a quantum superposition.  This experiment is the quantum version of the 
Wheeler's delayed-choice experiment.
The quantum delayed-choice experiment suggests a reinterpretation of complementarity principle:
instead of complementary experimental setups, the new proposal suggests 
complementarity in the experimental data.  

The NMR systems provide perfect platforms for studying such phenomena \cite{soumyaqdc}.
In our experiments we found a general agreement between the intensities and
the visibilities of the interference with the theoretically expected values.
These experiments not only confirm the intrinsic wave-particle duality of
quantum systems, but also demonstrates continuous morphing of quantum systems
between wave and particle behavior of the target qubit depending on the quantum state 
of the ancilla qubit.

\appendix \chapter{Density Matrix tomography for a pair of spin-1/2 homonuclear system}


A density matrix describes the statistical state of a quantum system and is considered to be the most profound way of representing a quantum state. A pair of spin-1/2 system can be expressed fully by a density martix of $4 \times 4$ order. In general the size of the density matrix depends on the number of spins, such as a n-spin system can be represented fully by a $2^n \times 2^n$ density matrix. The density matrix of any quantum state is a Hermitian matrix with trace 1. In a NMR spin system, the diagonal elements of the density matrix represent populations, whereas the off-diagonal elements represent coherence orders. The density matrix for a typical two spin-1/2 system is given below:
\begin{equation}
 \bordermatrix{~ & \vert 00 \rangle & \vert 01 \rangle & \vert 10 \rangle & \vert 11 \rangle \cr
 \langle 00 \vert & P_{00} & SQ & SQ & DQ \cr
 \langle 01 \vert & & P_{01} & ZQ & SQ  \cr
 \langle 10 \vert & & & P_{10} & SQ  \cr
 \langle 11 \vert & & & & P_{11}   \cr}
\end{equation}
\paragraph\
   Here, $P_{00}, P_{01}, P_{10}$, and $P_{11}$ are representing the populations in $\vert 00 \rangle$, $\vert 01 \rangle$, $\vert 10 \rangle$, $\vert 11 \rangle$ states respectively. Single, double, and zero-quantum coherences are denoted by $SQ$, $DQ$, and $ZQ$ respectively. In NMR only single quantum coherences are directly observable. Hence to determine all these elements, we need to perform some unitary transformations by which we can effectivle transfer other non-SQ terms into a SQ coherence and hence can be readout from the output signal.
\paragraph\
   The general traceless deviation density matrix consists of 15 independent real numbers:
   \begin{eqnarray}
\rho = 
\left(
\begin{array}{cccc}
  p_0 \;\;\;\; &  r_3 + i s_3  & r_1 + i s_1 &  r_5 + i s_5  \\
  & p_1 &  r_6 + i s_6  &  r_2 + i s_2  \\
 & & p_2 &  r_4 + i s_4  \\
 & & & -\sum_{i=0}^2p_i \\
\end{array}
\right).
\end{eqnarray}
Each coherence element has a real ($r$) and a imaginary ($s$) part in it. The elements below the diagonal are determined by Hermitian condition : $\rho_{jk} = \rho_{kj}^{*}$.
Since, $^1H$ spin systems have smaller coupling constants, often it is difficult to readout each of the line seperately incase of a dispersive spectra. Here, we devoloped the tomographic technique where we need to do total integratation for one spin. This way integration errors for a dispersive spectra can be reduced significantly. Four combinations of different readout elements can be obtained from the integration of complex line shapes of spin 1 and 2 respectively.
\begin{itemize}
\item Real part of spin 1 :  R$_1$ = (r$_1$+r$_2$)
\item Imaginary part of spin 1 :  S$_1$ = (s$_1$+s$_2$)
\item Real part of spin 2 :  R$_2$ = (r$_3$+r$_4$)
\item Imaginary part of spin 2 :  S$_2$ = (s$_3$+s$_4$)
\end{itemize}   

Now consider an RF sequence with propagator representing by $U$, that 
transforms the original density matrix $\rho$ into $\rho' = U \rho U^\dagger$. 
Single quantum coherences of $\rho'$ will lead to different linear combinations 
of various elements in $\rho$.  Thus, by applying different propagators on $\rho$,
we can measure the values of different linear combinations of various elements of $\rho$.
The real and imaginary values of the integration of $j$th spin in $k$th experiment
will be labeled as $R_j^k$ and $S_j^k$ respectively.  Following six 
one-dimensional NMR experiments were found to be sufficient to tomograph a two-spin 
density matrix:
\begin{itemize}
\item[1.]{$\mathbbm{1}$
}
\item[2.]{$90_x$
}
\item[3.]{$\frac{1}{4J} \cdot 180_x \cdot \frac{1}{4J}$
}
\item[4.]{$45_x \frac{1}{4J} \cdot 180_x \cdot \frac{1}{4J}$
}
\item[5.]{$45_y \frac{1}{4J} \cdot 180_x \cdot \frac{1}{4J}$
}
\item[6.]{$\frac{1}{2\Delta \nu} \cdot 45_y \frac{1}{4J} \cdot 180_x \cdot \frac{1}{4J}$
}
\end{itemize}
Here $\mathbbm{1}$ is the identity i.e., direct observation without applying any extra pulses.
$\Delta \nu$ and $J$ are the chemical shift difference  and the scalar coupling 
respectively (both in Hz).  The offset
is assumed to be at the center of the two doublets and the RF amplitudes are assumed to 
be much stronger than $\Delta \nu$. 
\paragraph\
Now, we will operate each of this propagator one by one and read out the four detectable element orders. We will
follow the spin-operator formalism for mathemical treatments. $I_{j}^{k}$ defines as spin-operator, $j$ can be x, y, or z depending on the phase of a particular pulse and $k$ can be either 1 or 2 depending on the spin, which is going through the evolution. E. g. $I_{x}^{1}$ denotes evolution of spin-1 w. r. t. $x$ axis. The single spin operator can be written as:

\begin{align}
I_x &= \sigma_x/2 = \frac{1}{2}
  \begin{pmatrix}
       0 & 1   \\       
       1 & 0   \\       
 \end{pmatrix};
  & 
I_y &= \sigma_y/2 = \frac{1}{2}
  \begin{pmatrix}
       0 & -i   \\       
       i & 0    \\      
  \end{pmatrix};
   &  
I_z &= \sigma_z/2 = \frac{1}{2}
  \begin{pmatrix}
       1 & 0   \\       
       0 & -1  \\        
  \end{pmatrix}.
  & 
\end{align}  
Where $\sigma_x, \sigma_y, \sigma_z$ are Pauli matrices. Now, for a two spin system matrix operator is of the order of $4 \times 4$. Hence $I_{x}^{1} = I_x \otimes I_d$, where $I_d$ denotes identity matrix of $2 \times 2$. Again  $I_{x}^{2} = I_d \otimes I_x$.
Similarly $I_{y}^{1}$, $I_{y}^{2}$, $I_{z}^{1}$, $I_{z}^{2}$  can be written. Obviously for a two-spin system $I_x = I_{x}^{1}+I_{x}^{2}$, $I_y = I_{y}^{1}+I_{y}^{2}$, $I_z = I_{z}^{1}+I_{z}^{2}$
 
The spin-operator formalism for the 6 tomographic experiments described in detail.\\
  
\subsubsection{I.}
The first experiment is the identity operator and hence the density matrix will be same and the readout elements are as written above:
\begin{equation}
R_{1}^{1} = (r_1+r_2),\quad R_{2}^{1} = (r_3+r_4),\quad S_{1}^{1} = (s_1+s_2),\quad S_{2}^{1} = (s_1+s_2).
\end{equation}
The left hand side of each of these equations are known values (achieved by integrating the spectra) and righ hand side are unknowns to be determined.

\subsubsection{II.}
The second unitary operator is a  $90_x$ pulse. The matrix form of this operator can be written as:
\[
U_2 =e^{(-i.\frac{\pi}{2}.I_x)} = \frac{1}{2}
  \begin{bmatrix}
       0 & 1 & 1 & 0   \\       
       1 & 0 & 0 & 1   \\      
       1 & 0 & 0 & 1   \\      
       0 & 1 & 1 & 0
 \end{bmatrix}.
\]

Now operating $U_2$ on $\rho$ can be written as: $\rho_2 = U_2 \cdot \rho \cdot U_2^{\dagger}$. The readout elements from $\rho_2$ are :

\begin{equation}
R_{1}^{2} = (r_1+r_2),\quad R_{2}^{2} = (r_3+r_4),\quad S_{1}^{2} = (p_0+p_1),\quad S_{2}^{2} = (p_0+p_2).
\end{equation}

\subsubsection{III.}
The third experiment is a 1/2J evolution of J-coupling, whereas chemical shift is refocussed since the offset is set
in between the spins. Mathematically the operator form for this unitary evolution can be written as: 

\[
U_3 =e^{(-i.\frac{\pi}{4}.2.I_{z}^{1}I_{z}^{2})}.e^{(-i.\pi.I_x)}.e^{(-i.\frac{\pi}{4}.2.I_{z}^{1}I_{z}^{2})} = \frac{1}{\sqrt2}
  \begin{bmatrix}
       0 & 0 & 0 & -1+i   \\       
       0 & 0 & -1-i & 0   \\      
       0 & -1-i & 0 & 0   \\      
       -1+i & 0 & 0 & 0
 \end{bmatrix}.
\]
Similarly as above, $\rho_3 = U_3 \cdot \rho \cdot U_3^{\dagger}$. The readout elements from $\rho_3$ are :

\begin{equation}
R_{1}^{3} = (s_1-s_2),\quad R_{2}^{3} = (s_3-s_4),\quad S_{1}^{3} = (r_1-r_2),\quad S_{2}^{3} = (r_3-r_4).
\end{equation}
\subsubsection{IV.}
The fourth experiment comprises of a $45_x$ pulse and a 1/2J evolution similar to the previous one. 
Mathematically, the operator can be writen in a similar way as shown previously.
$U_4 =e^{(-i.\frac{\pi}{4}.2.I_{z}^{1}I_{z}^{2})}.e^{(-i.\pi.I_x)}.e^{(-i.\frac{\pi}{4}.2.I_{z}^{1}I_{z}^{2})}.e^{(-i.\pi/4.I_x)}$. Then, $\rho_4 = U_4 \cdot \rho \cdot U_4^{\dagger}$. The readout elements from $\rho_4$ are:

\begin{eqnarray}
R_{1}^{4} &=& \frac{1}{4}(p_0-p_1-p_2)+\frac{1}{2}(r_5-r_6+s_1-s_2-s_3+s_4),\nonumber \\
R_{2}^{4} &=& \frac{1}{4}(p_0-p_1-p_2)+\frac{1}{2}(r_5-r_6-s_1+s_2+s_3-s_4),\nonumber \\
S_{1}^{4} &=& \frac{1}{\sqrt 2}(r_1-r_2-s_5+s_6),\nonumber \\ 
S_{2}^{4} &=& \frac{1}{\sqrt 2}(r_3-r_4-s_5-s_6).
\end{eqnarray}

\subsubsection{V.}
The operator form of the fifth experiment can be written as,
$U_5 =e^{(-i.\frac{\pi}{4}.2.I_{z}^{1}I_{z}^{2})}.e^{(-i.\pi.I_x)}.e^{(-i.\frac{\pi}{4}.2.I_{z}^{1}I_{z}^{2})}.e^{(-i.\pi/4.I_y)}$.\\
 Then, $\rho_5 = U_5 \cdot \rho \cdot U_5^{\dagger}$. The readout elements from $\rho_5$ are:

\begin{eqnarray}
R_{1}^{5} &=& \frac{1}{\sqrt 2}(s_1-s_2-s_5-s_6),\nonumber \\ 
R_{2}^{5} &=& \frac{1}{\sqrt 2}(s_3-s_4-s_5+s_6), \nonumber \\
S_{1}^{5} &=& \frac{1}{4}(p_0-p_1-p_2)+\frac{1}{2}(r_1-r_2-r_3+r_4-r_5-r_6),\nonumber \\
S_{2}^{5} &=& \frac{1}{4}(p_0-p_1-p_2)+\frac{1}{2}(-r_1+r_2+r_3-r_4-r_5-r_6).
\end{eqnarray} 

\subsubsection{VI.}
The operator form of the sixth and last experiment can be written as,
$U_6 =e^{(-i.\frac{\pi}{4}.2.I_{z}^{1}I_{z}^{2})}.e^{(-i.\pi.I_x)}.e^{(-i.\frac{\pi}{4}.2.I_{z}^{1}I_{z}^{2})}.e^{(-i.\pi/4.I_x)}.e^{(-i.\frac{\pi}{2}.(I_{z}^{1}-I_{z}^{2}))}$.\\
 Then, $\rho_6 = U_6 \cdot \rho \cdot U_6^{\dagger}$. The readout elements from $\rho_6$ are:

\begin{eqnarray}
R_{1}^{6} &=& \frac{1}{4}(p_0-p_1-p_2)+ \frac{1}{2}(r_1-r_2+r_3-r_4+r_5+r_6),\nonumber \\ 
R_{2}^{6} &=& \frac{1}{4}(p_0-p_1-p_2)+ \frac{1}{2}(-r_1+r_2-r_3+r_4+r_5+r_6),\nonumber \\ 
S_{1}^{6} &=& \frac{1}{\sqrt 2}(-s_1+s_2-s_5+s_6),\nonumber \\
S_{2}^{6} &=& \frac{1}{\sqrt 2}(s_3-s_4-s_5+s_6).
\end{eqnarray} 

So, at the end of these 6 experiments, one should have a total 24 linear equations (Eqs. A.3-A.8) with 15 unknowns. All these equations can be written together in a matrix format, as written below: (see Eq. \ref{tomoeq}) ($AX = Y$, where A is a constant matrix, X is an unknown matrix to be calculated and Y is a known matrix).

\begin{eqnarray}
\left[
\begin{array}{ccc cccc cc cccc cc}
 0&            0&            0&            1&            1&            0&            0&            0&            0&            0&            0&            0&            0&            0&            0      \\
      
      0&            0&            0&            0&            0&            1&            1&            0&            0&            0&            0&            0&            0&            0&            0     \\

      0&            0&            0&            0&            0&            0&            0&            0&            0&            1&            1&            0&            0&            0&            0    \\
        
      0&            0&            0&            0&            0&            0&            0&            0&            0&            0&            0&            1&            1&            0&            0     \\


      0&            0&            0&            1&            1&            0&            0&            0&            0&            0&            0&            0&            0&            0&            0 \\
           
      0&            0&            0&            0&            0&            1&            1&            0&            0&            0&            0&            0&            0&            0&            0     \\

     \frac{1}{2}&          \frac{1}{2}&         \frac{-1}{2}&         0&            0&            0&            0&            0&            0&            0&            0&            0&            0&            0&            0      \\
     
     \frac{1}{2}&         \frac{-1}{2}&          \frac{1}{2}&          0&            0&            0&            0&            0&            0&            0&            0&            0&            0&            0&            0      \\


      0&            0&            0&            0&            0&            0&            0&            0&            0&            1&           -1&            0&            0&            0&            0     \\
       
      0&            0&            0&            0&            0&            0&            0&            0&            0&            0&            0&            1&           -1&            0&            0     \\

      0&            0&            0&            1&           -1&            0&            0&            0&            0&            0&            0&            0&            0&            0&            0     \\
       
      0&            0&            0&            0&            0&            1&           -1&            0&            0&            0&            0&            0&            0&            0&            0     \\


     \frac{1}{4}&         \frac{-1}{4}&         \frac{-1}{4}&          0&            0&            0&            0&           \frac{1}{2}&         \frac{-1}{2}&          \frac{1}{2}&         \frac{-1}{2}&         \frac{-1}{2}&          \frac{1}{2}&           0&            0      \\
     
     \frac{1}{4}&         \frac{-1}{4}&         \frac{-1}{4}&          0&            0&            0&            0&           \frac{1}{2}&         \frac{-1}{2}&         \frac{-1}{2}&          \frac{1}{2}&          \frac{1}{2}&         \frac{-1}{2}&           0&            0      \\

      0&            0&            0&            \frac{1}{\sqrt{2}}&      \frac{-1}{\sqrt{2}}&         0&            0&            0&            0&            0&            0&            0&            0&         \frac{-1}{\sqrt{2}}&       \frac{1}{\sqrt{2}}   \\
      
      0&            0&            0&            0&            0&          \frac{1}{\sqrt{2}}&      \frac{-1}{\sqrt{2}}&         0&            0&            0&            0&            0&            0&         \frac{-1}{\sqrt{2}}&      \frac{-1}{\sqrt{2}}   \\


      0&            0&            0&            0&            0&            0&            0&            0&            0&          \frac{1}{\sqrt{2}}&      \frac{-1}{\sqrt{2}}&         0&            0&         \frac{-1}{\sqrt{2}}&      \frac{-1}{\sqrt{2}}   \\
      
      0&            0&            0&            0&            0&            0&            0&            0&            0&            0&            0&          \frac{1}{\sqrt{2}}&      \frac{-1}{\sqrt{2}}&      \frac{-1}{\sqrt{2}}&       \frac{1}{\sqrt{2}}   \\

     \frac{1}{4}&         \frac{-1}{4}&         \frac{-1}{4}&          \frac{1}{2}&         \frac{-1}{2}&         \frac{-1}{2}&          \frac{1}{2}&         \frac{-1}{2}&         \frac{-1}{2}&           0&            0&            0&            0&            0&            0      \\
     
     \frac{1}{4}&         \frac{-1}{4}&         \frac{-1}{4}&          \frac{-1}{2}&          \frac{1}{2}&          \frac{1}{2}&         \frac{-1}{2}&         \frac{-1}{2}&         \frac{-1}{2}&           0&            0&            0&            0&            0&            0      \\


     \frac{1}{4}&         \frac{-1}{4}&         \frac{-1}{4}&          \frac{1}{2}&         \frac{-1}{2}&          \frac{1}{2}&         \frac{-1}{2}&          \frac{1}{2}&          \frac{1}{2}&           0&            0&            0&            0&            0&            0      \\
     
     \frac{1}{4}&         \frac{-1}{4}&         \frac{-1}{4}&          \frac{-1}{2}&          \frac{1}{2}&         \frac{-1}{2}&          \frac{1}{2}&          \frac{1}{2}&          \frac{1}{2}&           0&            0&            0&            0&            0&            0      \\

      0&            0&            0&            0&            0&            0&            0&            0&            0&         \frac{-1}{\sqrt{2}}&       \frac{1}{\sqrt{2}}&         0&            0&         \frac{-1}{\sqrt{2}}&      \frac{-1}{\sqrt{2}}   \\
      
      0&            0&            0&            0&            0&            0&            0&            0&            0&            0&            0&          \frac{1}{\sqrt{2}}&      \frac{-1}{\sqrt{2}}&      \frac{-1}{\sqrt{2}}&       \frac{1}{\sqrt{2}}   \\

\end{array}
\right]
\left[
\begin{array}{c}
p_0 \\
p_1 \\
p_2 \\
r_1 \\
r_2 \\
r_3 \\
r_4 \\
r_5 \\
r_6 \\
s_1 \\
s_2 \\
s_3 \\
s_4 \\
s_5 \\
s_6
\end{array}
\right]=
\left[
\begin{array}{c}
R_1^1 \\
R_2^1 \\
S_1^1 \\
S_2^1 \\
R_1^2 \\
R_2^2 \\
S_1^2 \\
S_2^2 \\
R_1^3 \\
R_2^3 \\
S_1^3 \\
S_2^3 \\
R_1^4 \\
R_2^4 \\
S_1^4 \\
S_2^4 \\
R_1^5 \\
R_2^5 \\
S_1^5 \\
S_2^5 \\
R_1^6 \\
R_2^6 \\
S_1^6 \\
S_2^6
\end{array}
\right].
\label{tomoeq}
\end{eqnarray}

Certainly, this is a over-determined problem since 24 linear equations are to be used for 15 unknowns.  
This redundancy however works to increase the precision of the solution by reducing the
condition number of the constraint matrix.  For the 24 $\times$ 15 matrix in equation \ref{tomoeq},
condition number is about 3.7, meaning the solutions are precise to 5 significant digits. 
The equation \ref{tomoeq} can be solved either by
singular value decomposition (SVD) or by Gaussian elimination method (both of which are implemented in 
MATLAB). 


\thispagestyle{empty}
\chapter{Density Matrix tomography for a three spin-1/2 homonuclear system}

 The method for three-spin tomography is similar to one that we have described for a two-spin system. Since the number of unknowns here for three-spin system is much higher (63 unknowns) than two-spin (15 unknowns) system, we need to have more number of experiments in order to find out all the unknowns faithfully. The $8 \times 8$ general density matrix ($\rho$) for a three-spin system can be written as follows:

\begin{equation}
\hspace{-1cm}
 \bordermatrix{~ & \vert 000 \rangle & \vert 001 \rangle & \vert 010 \rangle & \vert 011 \rangle & \vert 100 \rangle
                   & \vert 101 \rangle & \vert 110 \rangle & \vert 111 \rangle \cr
 \langle 000 \vert & P_{0} & r_9 + i s_9 & r_5 + i s_5 & r_{13} + i s_{13} & r_1 + i s_1 & r_{14} + i s_{14} & r_{15} + i s_{15} & r_{25} + i s_{25}  \cr
 \langle 001 \vert & & P_{1} & r_{16} + i s_{16} & r_6 + i s_6 & r_{17} + i s_{17} & r_2 + i s_2 & r_{26} + i s_{26} & r_{18} + i s_{18}  \cr
 \langle 010 \vert & & & P_{2} & r_{10} + i s_{10} & r_{19} + i s_{19} & r_{27} + i s_{27} & r_3 + i s_3 & r_{20} + i s_{20}  \cr
 \langle 011 \vert & & & & P_{3} & r_{28} + i s_{28} & r_{21} + i s_{21} & r_{22} + i s_{22} & r_4 + i s_4  \cr
 \langle 100 \vert & & & & & P_{4} & r_{11} + i s_{11} & r_7 + i s_7 & r_{23} + i s_{23} \cr
 \langle 101 \vert & & & & & & P_{5} & r_{24} + i s_{24} & r_8 + i s_8 \cr
 \langle 110 \vert & & & & & & & P_{6} & r_{12} + i s_{12} \cr
 \langle 111 \vert & & & & & & & & \sum_{j=0}^{6}-P_{j} \cr}
\end{equation}

 The lower triangle of the density matrix can be filled by applying the Hermitian property of it ($\rho_{jk}=\rho_{kj}^{*}$). The diagonal elements ($P_j ,j=0 \to 6$) are representing population distributions of the density matrix. Applying the traceless property (or identity trace property), one of the unknowns can be reduced.  All other off-diagonal elements are representing the various coherence orders. Each coherence elements has a real ($r$) and imaginary ($s$) part in it. Elements $r_j$ and $s_j$, with $j=1 \to 12$ representing the real and imaginary part of single quantum coherences. Whereas $r_j$ and $s_j$, with $j=13 \to 28$ representing the real and imaginary part of double, triple, or zero quantum coherences. Only single quantum coherences are directly accessible in NMR. As described in the 2-spin tomography method, we have to find suitable unitary transformations which can transfer the double, triple, zero, and population orders into single quantum coherences. Consider a propagator $U$, that transforms the original density matrix $\rho$ into $\rho' = U \rho U^\dagger$. 
 Following 13  unitary transformations were found to be sufficient to tomograph a three-spin homonuclear system. 
\begin{itemize}
\item[(1).]{$\mathbbm{1}$
}
\item[(2).]{$\frac{1}{J_{13}}$
}
\item[(3).]{$\frac{1}{2J_{13}}$
}
\item[(4).]{$\frac{1}{J_{23}}$
}
\item[(5).]{$\frac{1}{2J_{13}} \cdot 60_{90}$
}
\item[(6).]{$\frac{1}{J_{13}} \cdot 90_{45}$
}
\item[(7).]{$\frac{1}{2J_{13}} \cdot 90_{135}$
}
\item[(8).]{$\frac{1}{2J_{13}} \cdot 45_{0}$
}
\item[(9).]{$\frac{1}{J_{23}} \cdot 60_{45}$
}
\item[(10).]{$\frac{1}{J_{13}} \cdot 45_{135}$
}
\item[(11).]{$\frac{1}{2J_{13}} \cdot 30_{45}$
}
\item[(12).]{$\frac{1}{J_{13}} \cdot 90_{0} \cdot \frac{1}{2J_{13}} \cdot 90_{0}$
}
\item[(13).]{$\frac{1}{2J_{13}} \cdot 60_{90} \cdot \frac{1}{J_{13}} \cdot 90_{135}$
}
\end{itemize}
Here $\mathbbm{1}$ represents the identity operator i.e., direct observation without applying any extra pulses.
$J_{12}$, $J_{23}$, and $J_{13}$ are the scalar couplings between spin 1 \& 2, spin 2 \& 3, and spin 1 \& 3 
respectively (in Hz). The offset is assumed to be at the center of the spin-1 and spin-2 and the RF amplitudes are assumed to 
be much stronger than $\Delta \nu$. Hence, all the pulses used are non-selective RF pulses. For this particular case $J_{12}$ is not used since it has very small coupling constant (a small coupling constant leads to larger duration of evolution which inturn makes the result more error prone). 

 Interms of unitary operator the delays ($\frac{1}{J_{13}}$ or $\frac{1}{J_{23}}$) can be written as bellow. For example, let us take the $2^{nd}$ experiment ($\frac{1}{J_{13}}$),
\begin{equation}
U_2 = e^{-i(H_j + H_{cs}).\frac{1}{J_{13}}},
\end{equation}  
where,  $H_{cs}$ and $H_j$ denoting Hamiltonian due to chemical shifts and J-couplings. 
\begin{equation}
H_{cs} = \displaystyle\sum_{i=1}^{3}\nu_i I_{z}^{i} ;\hspace{1.5cm}
H_{j} = \underset{i\neq j}{\displaystyle\sum_{i=1}^{3}\sum_{j=1}^{3}} 2\pi J_{ij} I_z^i I_z^j.
\end{equation}
 A combination of pulses and delays can also be seen as a required tomographic experiments. A pulse can be easily be written 
 as a unitary transformation as shown in detail in previous appendix. For example, we can take the experiment named $60_{90}$ (experiment no. 5). The unitary operator for this pulse can be written as,
\begin{equation}
U_5 = e^{-i\frac{\pi}{3} (I_x^1+I_x^2+I_x^3)}.
\end{equation} 
 
We need to apply this unitary operator one by one on the primitive density matrix (B. 1). By doing individual 
integration on each of the transition (12 transition for each experiment) and taking the imaginary values as well, we can
get a total 312 linearly dependent equations. These equations can be solved by singular value decomposition (SVD) method and
all the 63 unknowns can be find out.


\bibliographystyle{plain}

\end{document}